\documentclass{aa}
\usepackage{graphicx}
\usepackage{xcolor}
\usepackage{txfonts}

\definecolor{myblue}{rgb}{0.00, 0.0, 0.9}
\definecolor{myred}{rgb}{0.90, 0.0, 0.0}

\usepackage[breaklinks,  citecolor=myblue, linkcolor=myred, urlcolor=purple, colorlinks=true, debug, baseurl=' ']{hyperref}

\begin{document} 

        \title{Modeling the magnetized Local Bubble from dust data}

\titlerunning{Modeling the magnetized Local Bubble}

   \author{V. Pelgrims \inst{1} \fnmsep \inst{2} \fnmsep \thanks{pelgrims@physics.uoc.gr},
   K. Ferri{\`e}re \inst{3}, F. Boulanger \inst{4}, R. Lallement \inst{5}, L. Montier \inst{3}
          }
\institute{Institute of Astrophysics, Foundation for Research and Technology-Hellas, GR-71110 Heraklion, Greece
\and
Department of Physics, Univ. of Crete, GR-70013 Heraklion, Greece
\and
Institut de Recherche en Astrophysique et Plan{\'e}tologie (IRAP), CNRS, Universit{\'e} de Toulouse, CNRS, 9 Avenue du Colonel
Roche, BP 44346, 31028, Toulouse Cedex 4, France
\and
{\'E}cole normale sup{\'e}rieure/LERMA, Observatoire de Paris, Sorbonne Universit{\'e}, Universit{\'e} PSL, CNRS, Paris, France
\and
GEPI, Observatoire de Paris, PSL University, CNRS, 5 Place Jules Janssen, 92190 Meudon, France
}

\authorrunning{V. Pelgrims et al.}
             
   \date{Received November 21, 2019; accepted February 17, 2020}

  \abstract{
The Sun is embedded in the so-called Local Bubble (LB) -- a cavity of hot plasma created by supernova explosions and surrounded by a shell of cold, dusty gas. 
Knowing the local distortion of the Galactic magnetic field associated with the LB is critical for the modeling of interstellar polarization data at high Galactic latitudes. In this his paper, we relate the structure of the Galactic magnetic field on the LB scale to three-dimensional (3D) maps of the local interstellar medium (ISM).
First, we extracted the geometry of the LB shell,  its inner surface, in particular from 3D dust extinction maps of the local ISM. We expanded the shell inner surface in spherical harmonics, up to a variable maximum multipole degree, which enabled us to control the level of complexity for the modeled surface. 
Next, we applied an analytical model for the ordered magnetic field in the shell to the modeled shell surface.
This magnetic field model was successfully fitted to the \textit{Planck} 353~GHz dust polarized emission maps over the Galactic polar caps.
For each polar cap, the direction of the mean magnetic field derived from dust polarization
(together with the prior that the field points toward longitude $90^\circ \pm 90^\circ$)
is found to be consistent with the Faraday spectra of the nearby diffuse synchrotron emission.
Our work presents a new approach to modeling the local 
structure of the Galactic magnetic field.
We expect our methodology and our results to be useful both in modeling the local ISM as traced by its different components and in modeling the dust polarized emission, which is a long-awaited input
for studies of the polarized foregrounds for  cosmic microwave background.
}

                       
   \maketitle
%

\section{Introduction}
\label{sec:intro}

The interstellar medium (ISM) that surrounds the Sun out to a radius on the order of 100$-$300~pc  is known to have an unusually low
atomic gas density of $n_{\rm HI} \lesssim 0.1$ cm$^{-3}$ (\citealt{Cox1987}). This rarefied interstellar region is filled with a soft X-ray emitting plasma, as confirmed by the latest measurements and by recent analyses that take the heliospheric contribution to the soft X-ray background into account (\citealt{Pus2014}; \citealt{Gal2014}; \citealt{Sno2015}; \citealt{Liu2017}).
This so-called Local Cavity, also known as the Local Bubble (LB), is bounded by a shell of cold neutral gas and dust.

The LB was most likely created by supernova explosions that occurred
over the past 10$-$15 Myrs (\citealt{Mai2001,Bre2016}). 
According to these authors,
the progenitors of these supernovae belonged to stellar currents moving near the Galactic plane (within about 50 pc) and whose surviving members are probably part of the Scorpius-Centaurus (Sco-Cen) OB association.
As discussed by \cite{Mai2001}, backwards extrapolations of the trajectories of Sco-Cen OB association members in the
Local Standard of Rest (LSR) show that the positions of the supernovae that exploded in the past 10~Myrs fall outside, albeit very close to, the present boundary of the LB.
If these supernovae are indeed located at the origin of the LB, one would have expected the weighted mean of their positions to be close to the center of the LB.
However, this expectation implicitly relies on the assumptions that the local ISM (including the LB) moves exactly at the same velocity as the LSR and that the expansion motions driven by the explosions are isotropic, neither condition of which is actually satisfied.
In reality, the local ISM is believed to move at a velocity of $\simeq 15~{\rm km~s}^{-1}$ with respect to the LSR (e.g., \citealt{Gry2014}), corresponding to a displacement of $\simeq 150$~pc in 10~Myrs.
In addition, large-scale density and pressure gradients in the local ISM must have favored expansion in certain directions,
typically, away from the Galactic plane and towards the outer Galaxy.
For instance, \cite{Mai2001} suggested that pressure imbalance between a large molecular complex in the Galactic Center direction and a pre-existent rarefied volume in the opposite direction may have shifted the LB center away from the mean explosion center;
a realistic pressure-driven drift at a few km~s$^{-1}$ would be large enough to create a displacement $\sim$ 50~pc in 10~Myrs.
Altogether, the present LB center could be offset by as much as $\sim 200$~pc from the sites of the explosions that occurred 10~Myrs ago.

Global hydrodynamic and magneto-hydrodynamic models of the Galactic disk subjected to the effect of supernova explosions have been developed and a fraction of the computed cavities can match, at some stage, the characteristics (size, temperature, density range, and ion abundances) of the LB (see, e.g., \citealt{deA2009}).
More directly, the present shape and size of the LB can be extracted from three-dimensional (3D) maps of the dusty ISM surrounding the Sun (e.g., \citealt{Gre2019}; \citealt{Lal2019}, hereafter L19; and \citealt{Lei2019}, hereafter LE19).
In addition to being interesting in its own right, determining and modeling the geometry of the LB is expected to be useful to model the interstellar density distribution in our Galactic vicinity, to constrain the expansion motions driven by the supernova explosions that created the LB, and to model the local Galactic magnetic field.

Several studies have demonstrated that the magnetic field in the local ISM does not follow the large-scale Galactic magnetic field (e.g. \citealt{Hei1998}; \citealt{Ler1999}; \citealt{San2011}; \citealt{Fri2012}; \citealt{Ber2014}; \citealt{Gon2019}). For the first time, \cite{Alv2018} (hereafter, A18) quantified the association between the LB and the local magnetic field distortion. 
They developed an analytical model for the ordered magnetic field in the LB shell, which they assumed to be very thin and to result from purely radial expansion motions.
Approximating the shape of the shell as a spheroid, A18 fitted their magnetic field model to the measured \textit{Planck} dust polarized emission in the Galactic polar caps ($|b| > 60^\circ$), where the contribution from the magnetized LB shell was expected to be dominant compared to the contribution from the large-scale Galactic magnetic field. 
Recently, \cite{Ska2019} were able to confirm this expectation: by comparing the dust polarized emission at 353~GHz with starlight optical polarization, they showed that the 353~GHz polarized sky is dominated at high Galactic latitudes by a dusty and magnetized structure extending from about 200 to 300~pc from the Sun.
Thus, an accurate modeling of the magnetic field in the LB shell becomes an important milestone towards a comprehensive 3D modeling of the large-scale Galactic magnetic field, which, in turn, is critical for the physical characterization of the Galactic polarized foregrounds to the Cosmic Microwave Background (CMB).

\smallskip

In this paper, we develop a physically motivated approach to model the Galactic dust polarized emission in the Galactic polar caps. 
We infer the geometry of the LB shell directly from 
observational data.
We then describe the shell geometry in mathematical terms in order to be able to study the local perturbation of the Galactic magnetic field associated with the formation of the LB.
In that sense, our paper follows up and improves on the modeling of the magnetized LB shell proposed by A18.
We also satisfactorily provide the first self-consistent physical model of the dust polarized sky at high Galactic latitudes using actual 3D data.

Our work contains two main parts, structured as follows.
In Sect.~\ref{sec:LB_shape}, we extract the location and shape of the LB shell from 3D extinction maps and we provide a mathematical model, in terms of spherical harmonics, for the shell inner surface.
We quantitatively compare the modeled surfaces
obtained from the L19 and LE19 extinction maps both with each other and with the shape of the Local Hot Bubble derived from X-ray emission data.
In Sect.~\ref{sec:GMF_LB}, we apply the magnetic field model of A18 to our shell inner surface, and we constrain this model by fitting it to the \textit{Planck} dust polarized emission in the Galactic polar caps.
We also test the stability of our results for the magnetic field and, therefore, for the dust polarized emission against several sources of uncertainty.
Finally, we compare the best-fit magnetic fields derived with our approach to those from
simpler models, and we
confront our results to those of Faraday tomographic studies of the local ISM.
Section~\ref{sec:end} summarizes the work carried in this paper and presents some perspectives.

\section{Geometry of the LB shell}
\label{sec:LB_shape}

\subsection{Data set}
In recent years, a growing number of 3D maps of the dusty Galactic space surrounding the Sun have been produced (see, e.g., the introduction of L19 for an exhaustive review).
These data sets were made possible thanks to large photometric and spectroscopic surveys such as 2MASS, Pan-STARRS, SDSS/APOGEE, accurate parallax measurements from Gaia, for example, and elaborate inversion techniques (e.g., see \citealt{Gre2019}; L19; LE19 and references therein).
Spectroscopic data improve the quality of the maps in different ways. Stellar spectral features constrain stellar types, and, in turn, extinction estimates. Absorption by gaseous species (e.g., Diffuse Interstellar Bands) can be used directly for mapping
(see, e.g., \citealt{Far2019} or used as proxies for dust extinction and merged with photometric determinations (\citealt{Cap2017}).
To date, the latest products corresponding to large Galactic volumes are those presented in the first three aforementioned papers. Every 3D map comes with its own set of characteristics (covered volume, resolution, etc.) and with the strengths and weaknesses from either or both of the applied data sets and inversion methods.

\smallskip

In this work, 
we rely on 3D dust density maps to model the geometry of the LB shell.
As discussed further in Sect.~\ref{subsec:comp3Dmap}, we find that the most suitable available 3D map to perform our analysis is the map of L19, which is
 the only publicly available map that covers, in all directions, a volume large enough to contain the entire LB.

L19 constructed a 3D map of dust reddening based on Gaia DR2 photometric data combined with 2MASS measurements to derive extinction towards stars that possess accurate photometry and relative uncertainties in DR2 parallaxes smaller than 20\%. They applied a hierarchical inversion algorithm which includes spatial correlation and which is adapted to large datasets and to an inhomogeneous target distribution.
The resulting map is delivered on a Cartesian grid with voxel size of (5 pc)${}^3$.
It covers a volume of $\left[6.0 \times 6.0 \times 0.8\right]$ kpc$^3$ centered on the Sun with the largest extent in the Galactic disk. The maximal spatial resolution achieved in that iterative inversion process is 25 pc.
We refer the reader to the aforementioned paper for further details regarding the map-making process, the map itself and the description of the different data sets it relies on.

\begin{figure*}[t]
\centering
\includegraphics[trim={0cm 7.2cm 0cm 9cm},clip,width=.98\linewidth]{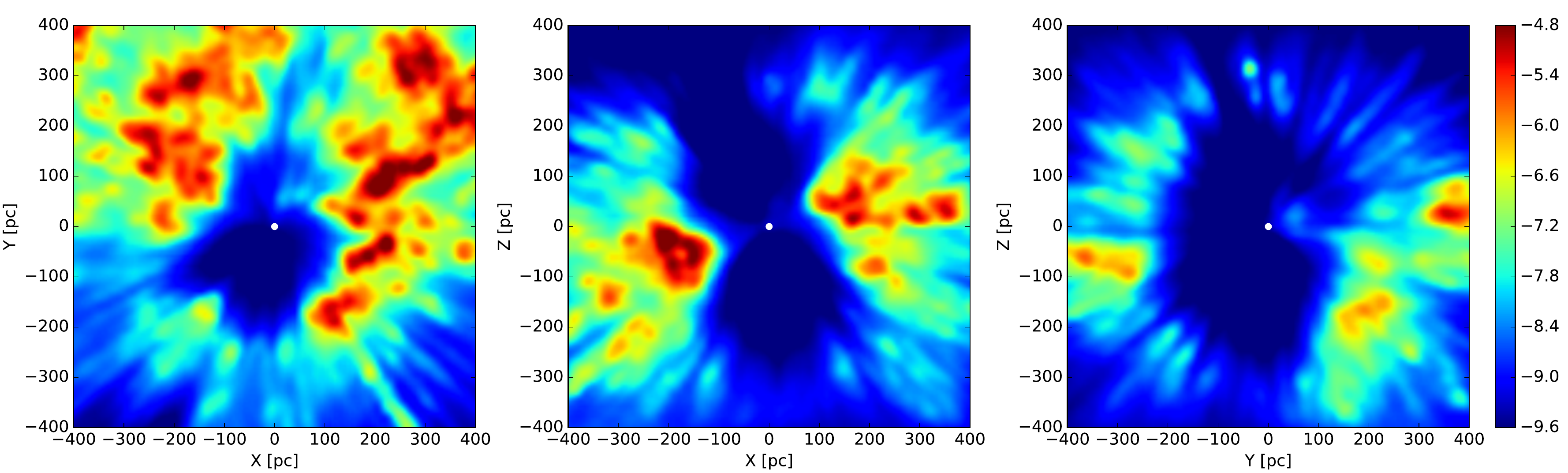}
\caption{Crosscuts along the planes XY, XZ, and YZ in the L19 3D dust extinction map. 
The Sun is at the center.
The X axis points from the Sun to the Galactic center at Galactic longitude $l=0^\circ$, the Y axis points towards $l=90^\circ$ and the Z axis points to the North Galactic pole.
The Galactic center is to the right in the left and middle panels and back to the reader in the right panel. 
The color scale shows
$\log(A'_{\rm{v}})$, where $A'_{\rm{v}}$ is the differential extinction,
in units of magnitude per parsec.
}
\label{fig:CC_L19}
\end{figure*}

In Fig.~\ref{fig:CC_L19}, we show three crosscuts of the
solar neighborhood according to the 3D dust extinction map of L19.
This map gives the differential extinction, $A'_{\rm{v}}(r) \equiv {\rm{d}} A_{\rm{v}}(r) / {\rm{d}}r$ (with $r$ the distance to the Sun), which we implicitly assume to be a proxy for the gas density.
These crosscuts show the XY, XZ and YZ planes, where the X axis points from the Sun to the Galactic center at Galactic longitude $l=0^\circ$, the Y axis points towards $l=90^\circ$ and the Z axis points to the North Galactic pole at Galactic latitude $b = 90^\circ$. The LB cavity clearly stands out in this triptych.

\smallskip

Throughout this paper, we assume that the LB cavity results
from supernova explosions, which shocked and swept up the ambient interstellar matter together with the frozen-in magnetic field.
It is the layer of swept-up matter between the cavity and the surrounding ISM that we call the shell of the LB. 
In this section, we provide simple, but realistic, models of the shape of this shell.

\begin{figure}
\centering
\includegraphics[trim={.4cm 0cm .4cm .8cm},clip,width=\linewidth]{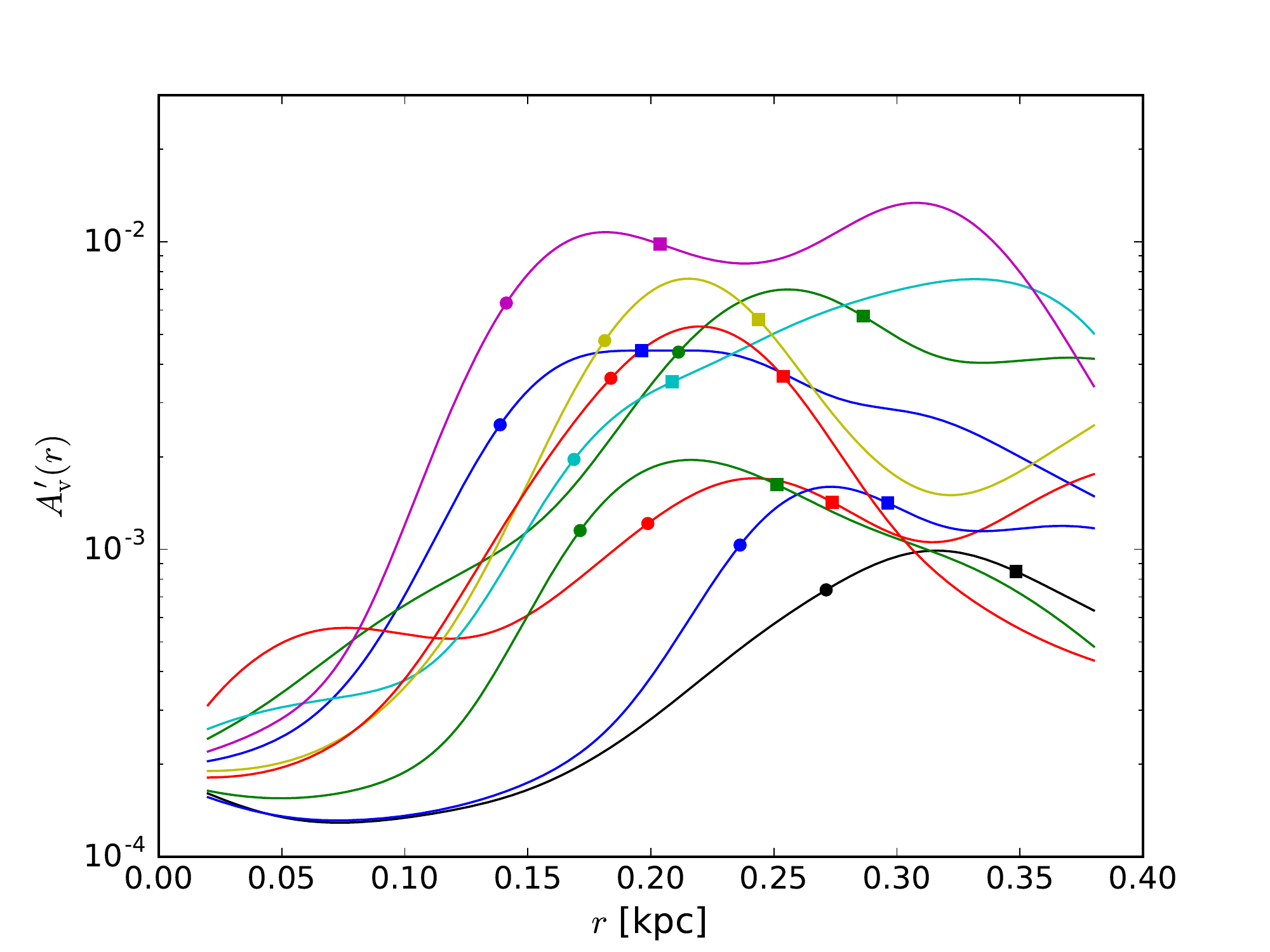}
\caption{Example of radial profiles of differential extinction
($A'_{\rm{v}}(r) \equiv {\rm{d}} A_{\rm{v}}(r) / {\rm{d}}r$)
as a function of distance to the Sun ($r$).
Ten lines of sight were randomly chosen in the XY plane of the Galaxy.
For each profile, the inner and outer radii of the LB shell, as determined in Sect.~\ref{subsubsec:def_LBedges}, are marked with filled circle and square, respectively.}
\label{fig:L19_los_expl}
\end{figure}

\subsection{Method}
\label{subsec:method}
\subsubsection{Inner and outer surfaces of the LB shell}
\label{subsubsec:def_LBedges}
To determine the geometry of the LB shell, we chose to rely on a criterion that is based on relative, rather than absolute, values of the reconstructed dust density. The procedure should be carried out automatically through the full data set. Our method can be described as follows.

\begin{figure*}[t]
\centering
\includegraphics[trim={0cm 7.2cm 0cm 9cm},clip,width=.98\linewidth]{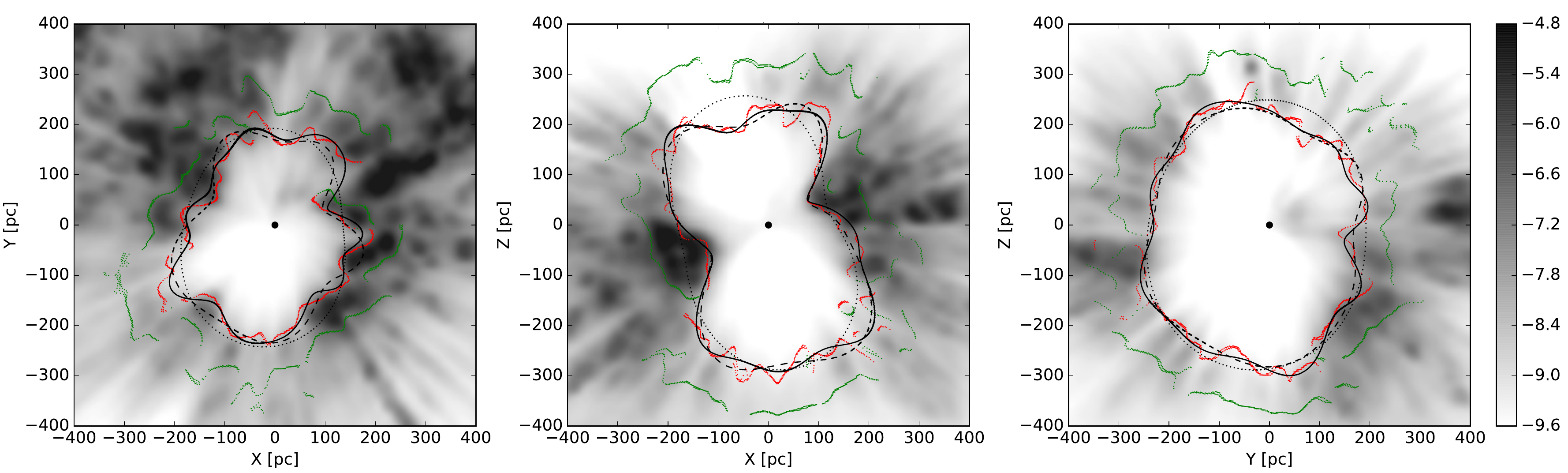}
\caption{
Crosscuts along the planes XY, XZ, and YZ in the L19 3D dust extinction map, with the same conventions as in Fig.~\ref{fig:CC_L19}.
The (common) gray scale shows $\log(A'_{\rm{v}})$, with $A'_{\rm{v}}$ in units of magnitude per parsec.
The red and green lines mark the inner and outer surfaces of the LB shell ($r_{\rm{inner}}$ and $r_{\rm{outer}}$, respectively), as extracted from the L19 map (Sect.~\ref{subsubsec:def_LBedges}).
The black dotted, dashed, and solid lines trace our models of the inner surface ($r_{\rm{LB}}$), as obtained through a spherical harmonic expansion up to $l_{\rm{max}} = 2,\,6$ and $10$, respectively (Sect.~\ref{subsubsec:mod_LBedges}).
}
\label{fig:CC_L19_mod}
\end{figure*}
To begin with, we draw lines of sight originating from the Sun and running outwards with a radial sampling step of 2.5~pc. We perform the angular sampling according to a HEALPix tessellation of the celestial sphere (\citealt{Gor2005}). We set the $N_{\rm{side}}$ parameter to 128, providing an angular resolution of about
27\arcmin.5.
Out to the 400 pc radial distance that we probe, the 3D extinction map is therefore well over-sampled and we do not miss material in the line-of-sight cones. 
To each node of our spherical grid, we assign a value derived from the 3D extinction map. 
Because the latter utilizes a uniform Cartesian grid, 
we need to convert from Cartesian to spherical coordinates. Here, we rely on a linear interpolation over the nearest neighbors of the Cartesian grid.
For each line of sight, we thus obtain a radial profile of the differential extinction, $A'_{\rm{v}}(r)$.
The interpolation process induces spurious noise in the differential extinction curves. To eliminate this noise, we smooth these curves using a one-dimensional Gaussian smoothing kernel, with a standard deviation of 25~pc.
This value corresponds to the maximum resolution of the 3D extinction map of L19.
In Fig.~\ref{fig:L19_los_expl}, we show ten differential extinction curves randomly chosen in the XY plane of the Galaxy.

For each line of sight, we compute the first and second derivatives of $A'_{\rm{v}}(r)$ with respect to $r$. We then define the radius of the inner surface of the LB shell, $r_{\rm{inner}}$, as the distance to
the first (closest to the Sun) inflection point, where the curve changes from convex to concave, that is, the first point at which ${\rm{d}}^2 A'_{\rm{v}}(r) / {\rm{d}}r^2 = 0$.
A cursory look at Fig.~\ref{fig:CC_L19}, particularly at the first quadrants of the left and right panels, reveals localized dust structures which are most likely unrelated to the LB shell and, therefore, should be ignored.
The iterative procedure described in Sect.~\ref{subsubsec:mod_LBedges} makes it possible to bypass these dust structures and prevent them from biasing the determination of the shell inner surface.

Similarly, we locate the radius of the outer surface of the LB shell, $r_{\rm{outer}}$, at the second inflection point, where the $A'_{\rm{v}}(r)$ curve changes from concave to convex. Because of the complex dust density distribution in the interstellar medium, especially in the Galactic disk, we find that our derived $r_{\rm{outer}}$ is not reliable in some places.\footnote{
About 2\,\% of the lines of sight do not present a second inflection point within the radial distance of 400 pc, which delineates the probed ISM volume. 22\,\% have their second inflection point closer than 50 pc from the edge, a distance at which we estimate that the second derivative of the radial profile might be biased by the applied smoothing.
}
For this reason, we focus on the modeling of the inner surface in the next subsection.

\smallskip

We apply the above method to the 3D extinction map of L19.
The result is shown in the triptych of Fig.~\ref{fig:CC_L19_mod}, where the inner and outer surfaces of the LB shell are plotted in red and green,
respectively.
Figure~\ref{fig:CC_L19_mod} conveys a good sense of the complex geometry of the LB shell.
An intervening cloud can be spotted towards $(l,\,b) \approx (0^\circ,\, -50^\circ)$
in the fourth quadrant of the middle panel.
It also emerges from Fig.~\ref{fig:CC_L19_mod} that the shell is
relatively thick (ranging from 50 to 150 pc) and, more crucially with regard to our study, present all around the cavity, including the area towards the Galactic polar caps ($|b|\geq60^\circ$).
This was not immediately obvious from the 3D extinction map alone, where the LB looks more like an open chimney.
Towards the polar caps, the LB shell extends roughly from 200 to 300 pc, in agreement with the conclusion reached by \cite{Ska2019}, who estimated the shell extent based on stellar distances and polarization data only.

\subsubsection{Mathematical model for the inner surface of the LB shell}
\label{subsubsec:mod_LBedges}

\begin{figure}
    \centering
    \includegraphics[trim={.0cm 2.cm 0cm 0cm},clip,width=\linewidth]{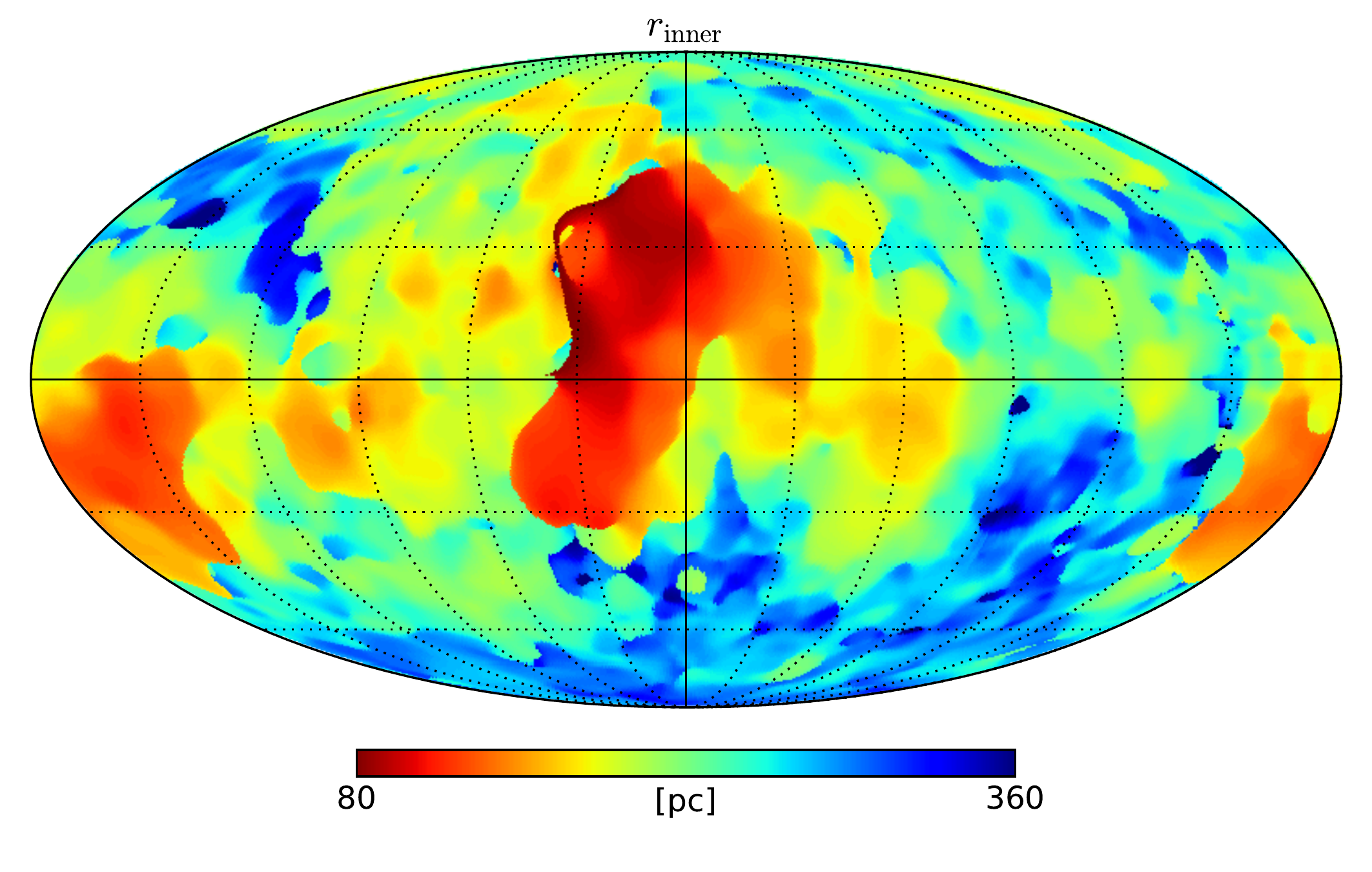} \\
    \includegraphics[trim={0cm .5cm 0cm 0cm},clip,width=\linewidth]{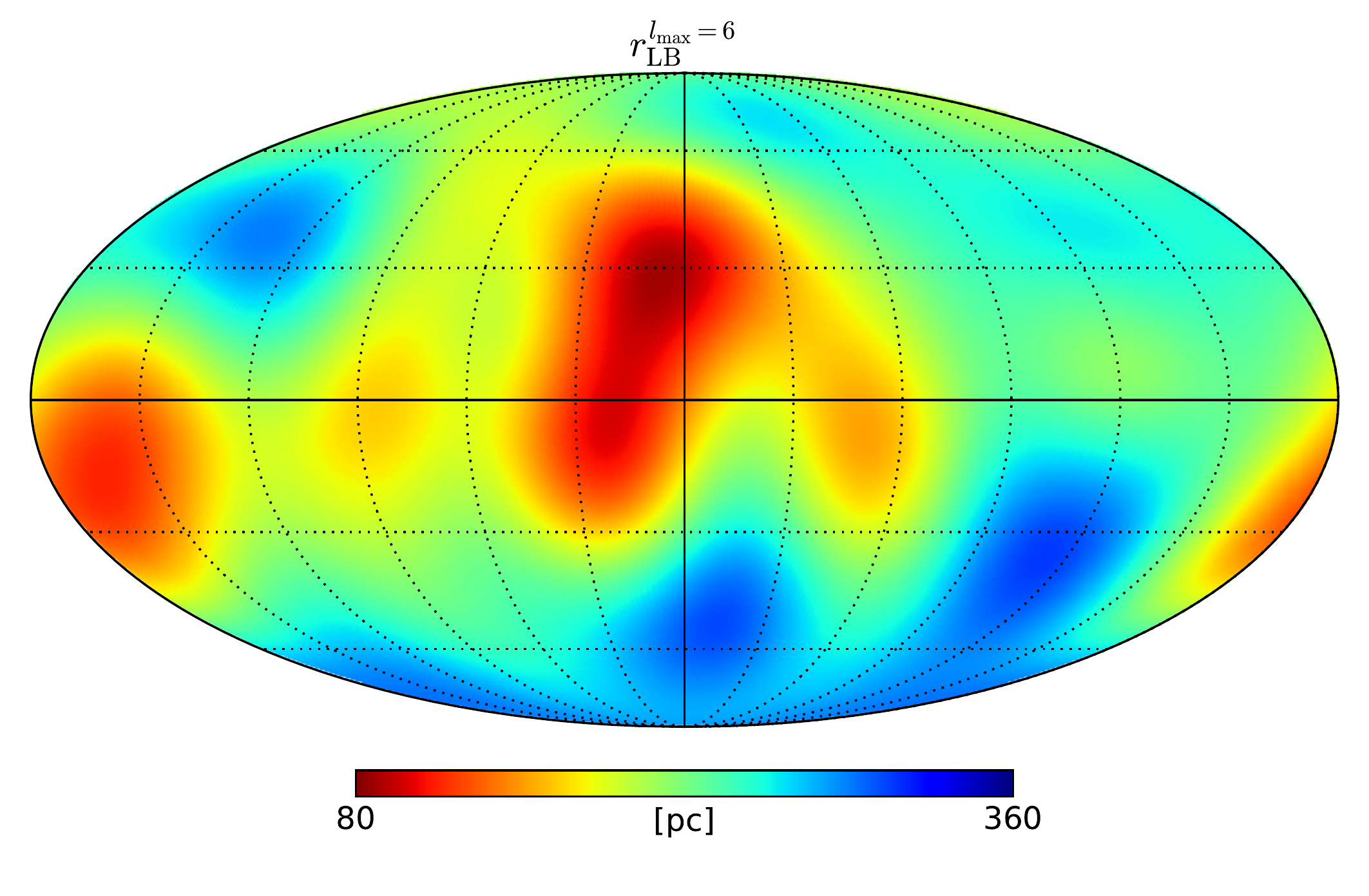} \\
    \caption{Full-sky map of (top) the inner surface of the LB shell ($r_{\rm{inner}}$) 
    as extracted from the L19 3D extinction map (Sect.~\ref{subsubsec:def_LBedges})
    and (bottom) our model of this surface through a spherical harmonic expansion up to $l_{\rm{max}} = 6$ (Sect.~\ref{subsubsec:mod_LBedges}).
    The maps are in Galactic coordinates, the center points towards the Galactic center and longitude increases to the left.
    }
    \label{fig:L19-map_rinnerand06}
\end{figure}

The inner surface of the LB shell can be visualized in 3D or in map format as shown in the top panel of Fig.~\ref{fig:L19-map_rinnerand06}.
In order to characterize the geometrical shape of this surface, to extract its main features, and to provide a good model of it with a small number of parameters, we rely on a spherical harmonic expansion.
By limiting the expansion to a maximum multipole degree, $l_{\rm{max}}$, we can easily adjust the level of complexity of the modeled surface.

We utilize the Python version of the HEALPix package to handle spherical harmonics. 
For a given $l_{\rm{max}}$, the routines return a set of positive spherical harmonic coefficients, from which we can 
build a model of the input surface smoothed out to the desired angular scale.

The expansion in spherical harmonics is meaningful to describe the shape of the inner surface of the LB shell.
Indeed, the coefficients decay rapidly with increasing $l$, which indicates that the spherical harmonic expansion converges for large $l_{\rm{max}}$. 
We find that the power spectrum of the 2D map of $r_{\rm{inner}}$ follows a power law with index $-2.95$ up to $l = 300$.

It is clear that the modeling of the shell inner surface described above can be biased by the presence of small dust clouds inside the cavity.
To correct for this bias, we proceed iteratively.
We start from the 2D map of $r_{\rm{inner}}$ extracted from the L19 3D dust extinction map in Sect.~\ref{subsubsec:def_LBedges}.
Then for any chosen value of $l_{\rm{max}}$,
we proceed as follows:

\begin{enumerate}[($i$)]
\item 
We expand the input map of the shell inner radius, 
$r_{\rm{inner}}$, in spherical harmonics up to $l_{\rm{max}}$.
\item With the retained spherical harmonic terms,
we approximate $r_{\rm{inner}}$ by a modeled inner radius, $r_{\rm{LB}}$.
\item For all lines of sight with
$r_{\rm{LB}} > r_{\rm{outer}}$,
we reset $r_{\rm{inner}}$ to $r_{\rm{LB}}$.
\item We repeat steps (\textit{i}) to (\textit{iii})
until the modeled surface does not change from the previous iteration.
\end{enumerate}

The reason why step (\textit{iii}) is needed is because the $r_{\rm{outer}}$ value of a line of sight that points towards an intervening cloud is smaller than the $r_{\rm{inner}}$ values of the neighboring lines of sight that avoid the intervening cloud. 
This iterative procedure should work as long as the intervening clouds are not too extended in the sky, such that statistically $r_{\rm{LB}}$ is indeed determined by  the inner surface of the LB shell.
It is, however, clear that this procedure might mistakenly erase abrupt changes in $r_{\rm{inner}}$.
This appears to happen for $l_{\rm{max}}=2,\,4$, because the shape of the modeled inner surface is too simple compared to the input surface.
For these values of $l_{\rm{max}}$, we find that a total of 10 iterations is a good compromise that enables us to skip over intervening clouds, without artificially scooping out the shell inner surface.
On the other hand, for $l_{\rm{max}}=6,\,8,\,10$, only 4, 8 and 10 iterations are required before the modeled inner surface becomes totally  stable. 

We visually check that the first and final models are very close to one another.
Moreover, for each $l_{\rm{max}}$, we quantify the difference between the first and final models by computing the mean Euclidean distance between the two sets of real-valued spherical harmonic coefficients,
$\tilde{a}_{lm}$:
\begin{equation}
d(s,s') = \frac{1}{(l_{\rm{max}} + 1)^2}\sqrt{\sum_{l=0}^{l_{\rm{max}}} \sum_{m=-l}^{m=l} \left( \tilde{a}_{lm}^s - \tilde{a}_{lm}^{s'} \right) ^2} \ ,
\label{eq:E_dist}
\end{equation}
where $s$ and $s'$ refer to two different (here, the first and final) models of the shell inner surface.
Real-valued spherical harmonics, which are better suited for describing real surface functions,
are related to the standard complex spherical harmonic coefficients, $a_{lm}$, through
\begin{equation}
\label{eq:E_norm_coef}
\tilde{a}_{lm} = \begin{cases}
        \begin{array}{ll}
                \frac{- i}{\sqrt{2}} \left( a_{l -|m|} - (-1)^m a_{l |m|}         \right) & \text{if}\, m<0 \\
                a_{l0}  & \text{if}\, m=0 \\
                \frac{1}{\sqrt{2}} \left( a_{l -|m|} + (-1)^m a_{l |m|} \right) & \text{if}\, m>0
                \end{array}
        \end{cases}
.\end{equation}
Here, we normalize the $a_{lm}$ coefficients to $a_{00}$ 
because the overall scale of the LB shell is irrelevant for our magnetic field modeling in Sect.~\ref{sec:GMF_LB}.

In Fig.~\ref{fig:Edist}, we plot the Euclidean distance between the first and final models of the shell inner surface, for several values of $l_{\rm{max}}$ (gray dotted line).
We consider this distance as a measure of the intrinsic accuracy of our model of the shell inner surface based on a given 3D extinction map.

\begin{figure*}
\centering
\includegraphics[trim={0cm 7.2cm 0cm 9cm},clip,width=.98\linewidth]{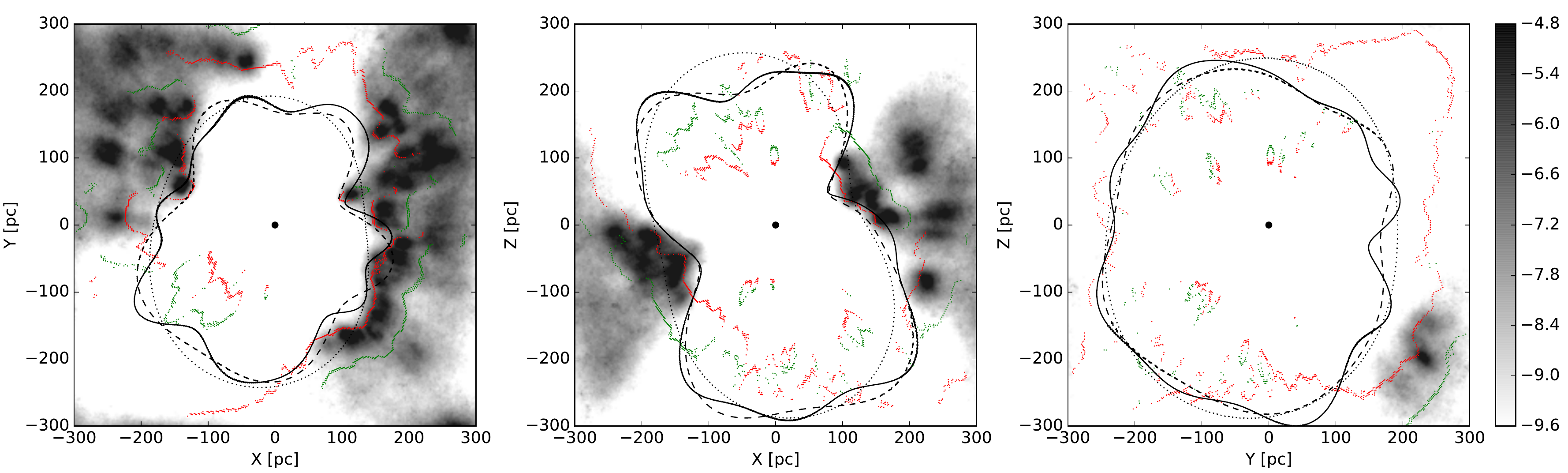}
\caption{Same as for Fig.~\ref{fig:CC_L19_mod}, except that the underlying dust density distribution and the corresponding inner and outer surfaces
of the LB shell (red and green lines, respectively)
are from the LE19 3D map.
Our models of the inner surface (black dotted, dashed and solid lines) are still based on the L19 map.
}
\label{fig:CC_L19_mod-on_LE19}
\end{figure*}

\subsection{Results}
In the triptych of Fig.~\ref{fig:CC_L19_mod}, we show 
our models of the inner surface of the LB shell as obtained through a spherical harmonic expansion
up to different values of $l_{\rm{max}}$ (black lines).
In the bottom panel of Fig.~\ref{fig:L19-map_rinnerand06}, we show a full-sky map of our model of the inner surface obtained for $l_{\rm{max}} =6$.
The model can be directly compared to the input map shown in the top panel of the same figure.

\subsection{Comparison with other 3D maps}
\label{subsec:comp3Dmap}

\begin{figure*}
\centering
\includegraphics[trim={0cm 7.2cm 0cm 9cm},clip,width=.98\linewidth]{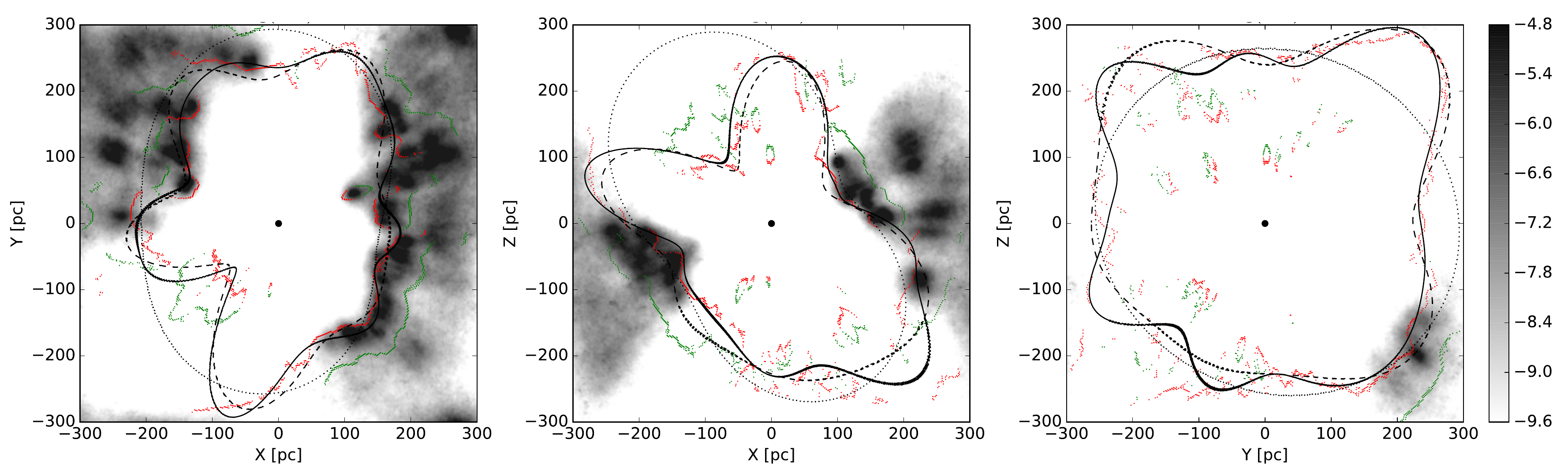}
\caption{Same as for Fig.~\ref{fig:CC_L19_mod-on_LE19}, except that our models of the inner surface of the LB shell (black dotted, dashed and solid lines) are now based on the LE19 map.
}
\label{fig:CC_LE19_mod}
\end{figure*}

In this subsection, we investigate the impact of the choice of the 3D extinction map on our modeling of the shape of the LB shell. 
It is important to test the stability of our results against other data sets. 
However, it is not our purpose to provide a comparison study between the different 3D maps of the dust density distribution that are available.
Future analyses should help pinpoint which 3D map is the most reliable and which is the best suited for the kind of analysis presented here.
The most advanced 3D maps that can compete with that of L19 are those from LE19, \cite{Gre2019} and \cite{Che2019}.

\cite{Gre2019} constructed a 3D map of dust reddening based on stellar parallaxes from Gaia DR2 and on stellar photometry from Pan-STARRS~1 and 2MASS.
Their map relies on 800 million objects, has unprecedented angular resolution and extends out to a distance of several kpc, but it is limited by the Pan-STARRS footprint. It covers (only) about 75 percent of the sky for declination $\delta > -30^\circ$.
Our model of the shape of the LB shell would suffer from this large hole in the sky, which would bias all the low $l_{\rm{max}}$ components.

\cite{Che2019} used stellar parallaxes from Gaia DR2 together with optical and near-infrared photometry from Gaia, WISE and 2MASS to trace dust reddening. 
Because they focused on the Galactic disk, they analyzed only lines of sight with Galactic latitudes $|b| \leq 10^\circ$.
As a result, their map is not at all suited for our study.

In contrast, LE19 constructed a 3D map of dust reddening that is full-sky once projected on the sky, but covers a smaller volume than the L19 map, namely a (600 pc)${}^3$ cube centered on the Sun.
Unlike L19, they constructed a statistical model with non-parametric kernel and applied a Bayesian variational scheme to Gaia DR2 distances and reddening estimates from \cite{And2018}, producing a set of fifty 3D maps. We refer the reader to their paper for further details on their inversion method and their results.

Focusing on the overlapping volume to compare with the 3D map of \cite{Lal2018}, LE19 showed that their mean reconstruction gives values of the dust density that range from a few orders of magnitude lower to one order of magnitude higher.
The latter corresponds to a cloud size that is one order of magnitude smaller, which is expected given that \cite{Lal2018} used a fixed minimum size for their two co-existent kernels.
A comparison with the L19 map leads to the same conclusions, which again is expected since, despite the new hierarchical technique, the L19 final step also has a 25 pc resolution limitation. On the other hand, unexpectedly, the close vicinity of the Sun in the LE19 map appears to have too low reddening, that is, to be too empty, compared to other maps.
As discussed in LE19, potential causes are the choice of the data sets used to reconstruct the 3D map, or an artifact of their reconstruction.
In addition, the authors cautioned against using the external parts of their reconstructed map, as periodic boundary conditions were assumed for algorithmic reasons, and the northern and southern tops of the "chimneys" fall in this category (see below). 
Finally, the authors noted a pronounced tension in the 3D positions of some dust clouds. 
Despite the higher angular resolution of LE19, we find a good agreement in the sky positions of the clouds, but we also detect some differences in their distances to the Sun.

Our method can be directly applied to the LE19 maps since they are full-sky once projected onto the sky.
Therefore, with the above caveats in mind, below we use the mean LE19 map to test the robustness of our model of the shape of the LB shell with respect to the adopted 3D extinction map.
As performed in Sect.~\ref{subsubsec:def_LBedges} with the L19 3D extinction map, we extract the radial profiles of differential extinction, $A'_{\rm{v}}(r)$, smooth them using a Gaussian kernel with a standard deviation of 25 pc in order to eliminate spurious high-frequency variations in $A'_{\rm{v}}(r)$, and define the inner and outer surfaces of the LB shell.

In the triptych of Fig.~\ref{fig:CC_L19_mod-on_LE19}, we show the modeled inner surface of the LB shell obtained in Sect.~\ref{subsubsec:mod_LBedges} with the L19 map, but over-plotted on the gray-scale dust density distribution with corresponding inner and outer surfaces of the LB shell from the LE19 map. An overall qualitative agreement is reached, but significant differences are observed. Additional structures, likely to be intervening clouds, appear in the LE19 map or are found to be closer to the Sun than in the L19 map.
For some lines of sight, the opposite trend is observed, as illustrated in the right panel of Fig.~\ref{fig:CC_L19_mod-on_LE19}.
In fact, in the LE19 map, the inner surface is often found far from the Sun and quite close to the boundary of the modeled interstellar volume.
In this region of space, we expect the distance to the shell to be biased in a non-trivial way by the presence of the boundary.
LE19 did indeed caution against the fact that the periodic boundary conditions used in their inversion process might produce artifacts up to about 15~pc from the sides of the modeled volume.
We estimate that these hardly quantifiable systematics are propagated further inside the volume, for example, up to about 50 pc, by the line-of-sight smoothing that we adopt to eliminate spurious noise in the differential extinction radial profiles.

\begin{figure}
\centering
        \includegraphics[trim={0.4cm 12cm 16.6cm 0cm},clip,width=\linewidth]{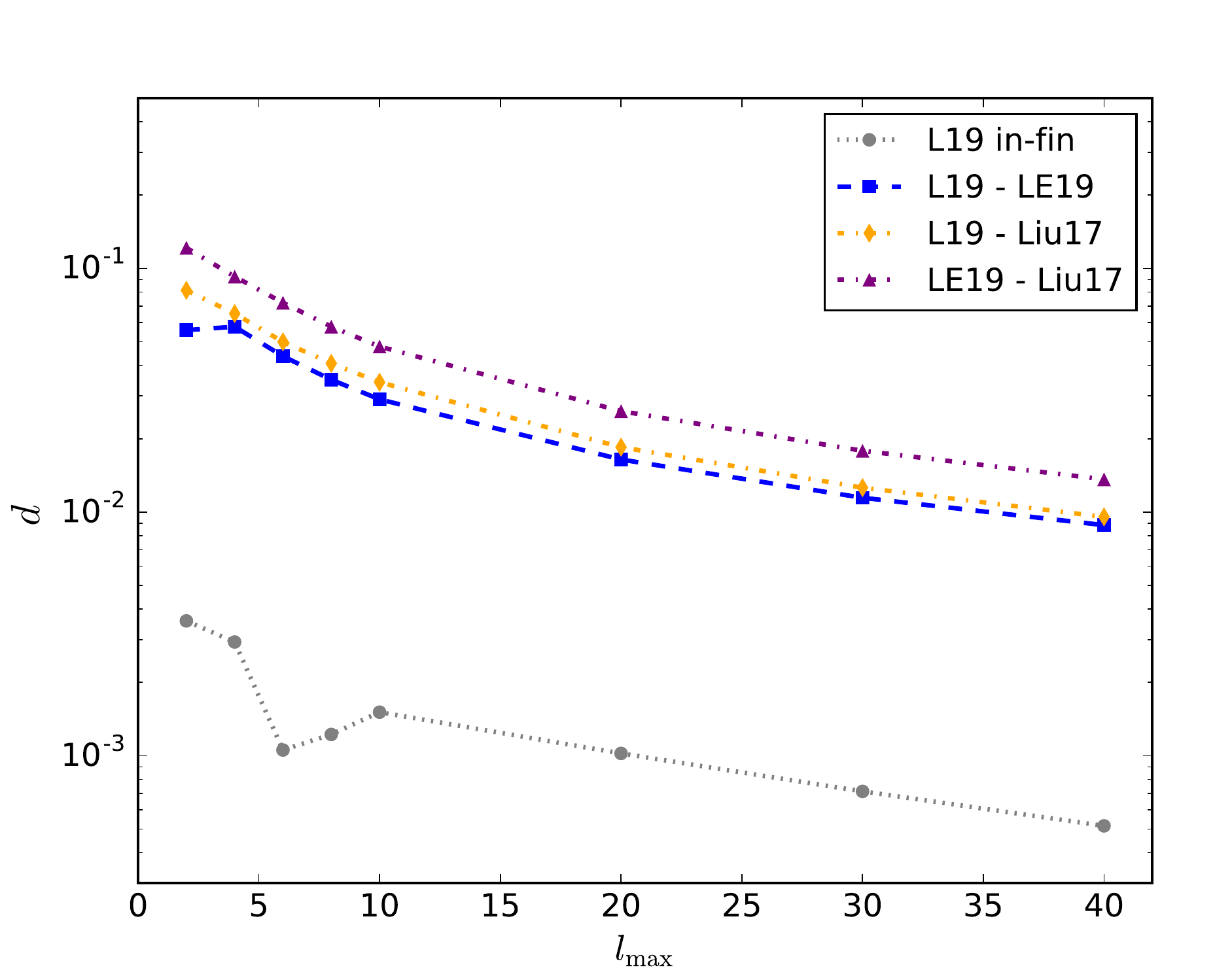}
\caption{Euclidean distances ($d$, given by Eq.~\ref{eq:E_dist}) between several models of the LB shell inner surface as a function of the maximum multipole degree ($l_{\rm{max}}$).
The gray dotted line compares the first and final iterations of the iterative procedure discussed in Sect.~\ref{subsubsec:mod_LBedges} and applied to the L19 data.
The blue dashed line compares the final models obtained with the L19 and LE19 data.
The orange and purple dot-dashed lines compare the final models obtained with the L19 and LE19 data, respectively, with the shape of the LHB from \cite{Liu2017} (see Sect.~2.5).
}
\label{fig:Edist}
\end{figure}

To go beyond the qualitative comparison given around Fig.~\ref{fig:CC_L19_mod-on_LE19}, we model the inner surface of the LB shell based on the LE19 3D map 
in the same manner as we model it in Sect.~\ref{subsubsec:mod_LBedges} based on the L19 map.
The results are displayed in the triptych of Fig.~\ref{fig:CC_LE19_mod}.
We also compute the Euclidean distances (Eq.~\ref{eq:E_dist}) between the real-valued spherical harmonic coefficients of the modeled inner surfaces derived from the L19 and LE19 maps, for several values of $l_{\rm{max}}$.
These distances are plotted in Fig.~\ref{fig:Edist}. They are about one order of magnitude larger than the distances between the first and last iterations in our modeling procedure (see Sect.~\ref{subsubsec:mod_LBedges}).

In conclusion, it appears that our models of the inner surface of the LB shell depend quite substantially on the underlying 3D extinction map.
In view of the above discussion, we prefer to rely on the L19 map; we consider the resulting models of the shell inner surface to be more suitable for our present purpose.

\subsection{Comparison with the shape of the Local Hot Bubble}
\label{subsec:comp3D-LHB}
Based on data from the \textit{DXL} sounding rocket mission, \cite{Liu2017} obtained a reliable map of the X-ray emission attributed to the Local Hot Bubble (LHB).
Their modeling of the X-ray emission allowed them to estimate the shape of the LHB, among other physical parameters, under the assumption of hot gas homogeneity.
The shape of the LHB was found to match qualitatively well the shape of the dust cavity in the 3D extinction map of \cite{Lal2014}.

Comparing X-ray data, which probe the hot ionized gas, with extinction data, which probe the dust, is a milestone in its own right towards a global understanding and physical modeling of the multi-phase ISM, in particular in the solar vicinity.
A detailed comparison between the physical properties inferred from both kinds of data is beyond the scope of our paper. 
In this subsection, we are content to compare the shape of the X-ray emitting LHB with the shape of the shell inner surface that we modeled from dust extinction maps.
To do so, we compute the Euclidean distances 
(Eq.~\ref{eq:E_dist}) between the real-valued spherical harmonic coefficients of both shapes, for several values of $l_{\rm{max}}$.

For the reasons explained below Eq.~\ref{eq:E_norm_coef}, we use the normalized coefficients.
This enables us to get rid of an overall scale difference 
between the LHB and the shell inner surface.
We note that only the shell, which is seen through dust extinction, has a reliable size; 
the LHB has an uncertain size that depends on the assumed electron density.
The computed distances are plotted in Fig.~\ref{fig:Edist}, which indicates that the shapes of the LB derived from different tracers compare as well as the shapes extracted from different extinction maps.
This suggests an overall consistency across the different phases of the ISM.

\section{Modeling the magnetic field in the LB shell}
\label{sec:GMF_LB}
A18 presented the first physical model of the magnetic field in the shell of the LB. 
Their model relies on the common assumption that the LB was created by supernova explosions, which swept out a cavity of hot ionized gas and pushed most of the evacuated matter, together with the frozen-in magnetic field, into a dense shell of cold neutral gas and dust around the cavity.
Their model further assumes that the initial magnetic field is uniform in strength and direction within the whole volume encompassed by the present-day LB, that the expansion motions driven by the explosions are purely radial, and that the shell is very thin 
(in the sense that it can be approximated by a 2D surface).
Adopting a spheroid to describe the shape of the shell, A18 constrained their magnetic field model by fitting it to the 2015 \textit{Planck} 353~GHz observations of the dust polarized emission towards the Galactic polar caps.

In this second part of the paper, we go one step further than what was done by A18:
we take up their magnetic field model, relax their simplifying assumption of a spheroidal shell, and adopt instead the more realistic geometry derived in Sect.~\ref{sec:LB_shape} on purely observational grounds.
To remain consistent with the very thin shell approximation, we replace the actual thick shell found in Sect.~\ref{sec:LB_shape} (see, e.g., Fig.~\ref{fig:CC_L19_mod}) 
by an idealized very thin shell.
A priori the best way of defining this very thin shell would be to identify it with the surface of maximum density inside the actual thick shell, that is, the surface at $r_{\rm{max}}$ where the differential extinction, $A'_{\rm{v}}(r)$, reaches its first maximum beyond $r_{\rm{inner}}$.
In practice, however, this criterion does not work, because some lines of sight turn out to have $r_{\rm{max}} > r_{\rm{outer}}$ (see, e.g., cyan radial profile in Fig.~\ref{fig:L19_los_expl}).
Moreover,  as already discussed in Sect.~\ref{subsubsec:def_LBedges}, $r_{\rm{outer}}$ is often unreliable.
Since neither $r_{\rm{outer}}$ nor $r_{\rm{max}}$ can be confidently determined in all directions, we take the inner surface of the actual thick shell to represent the very thin shell.
We note, however, that the outer surface appears to have a shape roughly similar to that of the inner surface.
Since our fit to the polarization data depends only on the shape of the shell, not on its size, the impact of substituting the inner surface of the shell for the shell itself is probably small.
In the remainder of this section, the word "shell" refers to the idealized very thin shell.

\subsection{Magnetic field model}
\label{subsec:GMF_model}
We start with the general magnetic field model derived in A18.
The present-day (ordered) magnetic field in the LB shell, $\mathbf{B}$, can be fully described in terms of the initial magnetic field, $\mathbf{B}_0$, the shape of the shell, and the position of the explosion center.
If several explosions took place, the explosion center is taken to be a point from which the expansion motions driven by the explosions can be considered to be purely radial.
In a spherical coordinate frame centered at the explosion center, the mathematical expression of $\mathbf{B}$ as a function of position $\mathbf{r}$ is given by Eq.~6 in A18:
\begin{equation}
\mathbf{B}(\mathbf{r}) = \frac{r_0}{r}\, \frac{\partial r_0}{\partial r}\, \frac{1}{\mathbf{n} \cdot \mathbf{e}_r} \left[ \mathbf{n} \times ( \mathbf{B}_0 \times  \mathbf{e}_r) \right] \ ,
\label{eq:Bfield_model}
\end{equation}
where
$\mathbf{n}$ is the unit vector normal to the surface of the shell,
$\mathbf{e}_r$ is the unit vector in the radial direction (from the explosion center),
and $r_0$ is the initial radial position of a particle presently at radial position $r$.

The measured polarized intensity of the thermal dust emission depends on the orientation of the magnetic field, but not on its strength.
Therefore, the prefactor in Eq.~\ref{eq:Bfield_model} is irrelevant and only the orientation of the vector within square brackets matters.
In the expression of this vector, the unit vector normal to the shell, $\mathbf{n}$, can be derived directly from the known shape of the shell;
the radial unit vector, $\mathbf{e}_r$,
is set by the position of the explosion center, which, in turn, is defined by its Cartesian coordinates ($\delta x$, $\delta y$, $\delta z$);
and the orientation of the initial magnetic field, $\mathbf{B}_0$, is given by its Galactic angular coordinates, ($l_0$, $b_0$).
Hence, we are left with a total of five free parameters: ($\delta x$, $\delta y$, $\delta z$, $l_0$, $b_0$).

\medskip

For the shape of the LB shell, we adopt the inner surface extracted from the L19 3D extinction map (in Sect.~\ref{subsubsec:def_LBedges}) and expanded in spherical harmonics up to $l_{\rm{max}}$ (in Sect.~\ref{subsubsec:mod_LBedges}).
We only consider values of $l_{\rm{max}} \le 10$.
Larger values of $l_{\rm{max}}$ would enable us to capture finer details of the original shell surface, but because of the low resolution of the L19 3D extinction map, these fine details are probably not physical.
More important, our simple magnetic field model is not suited for a very convoluted shell.
It is, therefore, legitimate to restrict our investigation to smooth shell models.

It directly emerges from Eq.~\ref{eq:Bfield_model} that for a given shell surface, that is, for given normal vector, $\mathbf{n}$, and for a given orientation of the initial magnetic field, $\mathbf{B}_0$, the orientation of the present-day magnetic field, $\mathbf{B}$, remains unchanged when the explosion center is displaced along a line parallel to $\mathbf{B}_0$.
It then follows that there will be a degeneracy between the three coordinates of  the explosion center, with the degeneracy line being parallel to $\mathbf{B}_0$.
In other words, our modeling will not enable us to determine the 3D location of the explosion center, but only to constrain its 2D position in a plane perpendicular to $\mathbf{B}_0$.

\subsection{Constraints from dust polarized emission}
To constrain the free parameters of our magnetic field model, we compute the associated Stokes parameters $Q$ and $U$ of the linearly polarized thermal dust emission, and we confront them to the observed Stokes parameters at 353~GHz from the 2018 \textit{Planck} data release (hereafter PR3).

We start from the integral equations for the Stokes parameters similar to those given in Appendix~B of \cite{PlaXX2015}
\footnote{Using the HEALPix convention (\url{https://healpix.jpl.nasa.gov/html/intronode12.htm)}}. 
For optically thin emission at frequency $\nu$:
\begin{equation}
\label{eq:simulated_I}
I=\int S_\nu\,\left[1-\mathfrak{p}_0\left(\cos^2\gamma-\frac{2}{3}\right)\right]\,
n_\mathrm{H} \, \sigma_\mathrm{H} \, \mathrm{d} s,
\end{equation}
\begin{equation}
\label{eq:simulated_Q}
Q=\int \mathfrak{p}_0\,S_\nu\,\cos\left(2\phi\right)\cos^2\gamma\,
n_\mathrm{H} \, \sigma_\mathrm{H} \, \mathrm{d}s,
\end{equation}
\begin{equation}
\label{eq:simulated_U}
U=\int \mathfrak{p}_0\,S_\nu\,\sin\left(2\phi\right)\cos^2\gamma\,
n_\mathrm{H} \, \sigma_\mathrm{H} \, \mathrm{d} s,
\end{equation}
where the integrals are computed along the line of sight over the emitting region (here, the LB shell); $S_\nu$ is the source function, $\mathfrak{p}_0$ a parameter related to dust polarization properties combining grain cross sections and the degree of alignment with the magnetic field, $n_\mathrm{H}$ the gas density, $\sigma_\mathrm{H}$ the dust cross-section
per hydrogen atom averaged over angles, $\gamma$ the angle of the local magnetic field to the plane of the sky, and $\phi$ the local polarization angle (see Fig.~14 in \cite{PlaXX2015}).

As in \cite{Lee85} and \cite{PlaXLIV2016}, we account for variations in the magnetic field orientation along the line of sight by introducing an effective depolarization factor $F$ that includes turbulence effects as well as small departures of our ideal model from reality.
Within this approximation, the Stokes parameters $Q$ and $U$ may be written as:
\begin{equation}
Q = I_{\rm{d}} \, p_0 \, \frac{\left(B_\theta^2 - B_\phi^2  \right)} {|\mathbf{B} |^2} \\ ; \\
U = - 2\, I_{\rm{d}} \, p_0 \, \frac{\left(B_\theta \, B_\phi \right)}{|\mathbf{B} |^2} \;,
\label{eq:QandU_model}
\end{equation}
where $p_0 = F\, \mathfrak{p}_0$ is an effective polarization fraction, 
$B_\theta$ and $B_\phi$ are the plane-of-sky components of the ordered magnetic field expressed in the spherical coordinate system centered on the observer ($\mathbf{e}_\theta$ points southwards and $\mathbf{e}_\phi$ eastwards).
Equations~\ref{eq:QandU_model} introduce the sky map $I_{\rm{d}}$ defined as
\begin{equation}
I_{\rm{d}} = \int S_\nu\,n_\mathrm{H} \, \sigma_\mathrm{H} \, \mathrm{d} s = 
\frac{I + P}{1 + \frac{2}{3} p_0}.
\label{eq:Id_vs_I}
\end{equation}
Following \cite{PlaXLIV2016} and \cite{Van2017}, we approximate $I_{\rm{d}}$ with $I$.  We note that the mean values of $I_{\rm{d}}$ and $I$ are equal when averaging over angles.
Hereafter, we assume that $p_0$ is constant across the fitted sky region. This assumption is supported by the tight scaling observed between the amplitude of the dust polarization power spectra and the dust total intensity, with no systematic difference between the northern and southern Galactic latitudes (\citealt{PlaXXX2016}; \citealt{PlaXI2018}).

\smallskip

In principle, the model of the magnetic field in the LB shell and its contribution to the dust polarized sky can be evaluated for the full-sky, corresponding to a first layer of the Galactic dust polarized foregrounds.
However, because our model does not include any component from the large-scale Galactic magnetic field, we follow A18 in restricting the fitted area to the Galactic polar caps, $|b| \geq 60^\circ$.
Using star optical polarization measurements and star distances to estimate the line-of-sight distance of the region responsible for the 353~GHz polarized emission, 
\cite{Ska2019} provided statistical evidence that in the Galactic polar caps the 353~GHz polarized emission is dominated by a dusty and magnetized structure extending from about 200 to 300~pc from the Sun. We naturally identify this structure with the LB shell, as also suggested by the triptych of Fig.~\ref{fig:CC_L19_mod}.

\medskip

We constrain the free parameters of our magnetic field model by maximizing the profiled log-likelihood function,
\begin{equation}
\mathcal{L}(d | m) = - \frac{1}{2} (d - m)^\dagger \, C^{-1}\, (d - m) \ ,
\label{eq:log_LL}
\end{equation}
where $d$ is the concatenation of the observed Stokes $Q$ and $U$ maps and $m$ is the concatenation of the modeled $Q$ and $U$ maps.

\medskip

The observed Stokes $Q$ and $U$ maps are based on products from the third \textit{Planck} data release that we downloaded from the \textit{Planck Legacy Archive}\footnote{\url{http://pla.esac.esa.int/pla/\#home}}.
For our Galactic study, we consider the $Q$ and $U$ 353~GHz maps made from the polarization-sensitive bolometers only, as recommended in \cite{PlaIII2018} and in \cite{PlaXII2018}. We smooth them to a resolution of 80\arcmin.

The modeled $Q$ and $U$ maps are computed from Eq.~\ref{eq:QandU_model} and adjusted to the observations through a linear fit that accounts for the pixel  uncertainties\footnote{If $m = \alpha \, \tilde{m}$ with $\tilde{m} = \left\lbrace Q,\,U \right\rbrace$ directly from Eq.\ref{eq:QandU_model}, the normalization factor $\alpha$ is computed as $(\sum_i(d_i \tilde{m}_i / \sigma_i^2) / \sum_i (\tilde{m}_i^2 / \sigma_i^2))$ where $\sigma_i^2 = \{C_{QQ},\,C_{UU}\}_i$ with $i$ running through all the indices of the concatenated maps.}.
The different parameters entering Eq.~\ref{eq:QandU_model} are obtained as follows:
the last factor, which depends only on the normalized (ordered) magnetic field vector in the shell, is directly taken from our magnetic field model described in Sect.~\ref{subsec:GMF_model}.
The factor $I_{\rm{d}}$ is approximated by the dust total intensity, $I$, and for $I$ we use the map that results from the GNILC component separation algorithm (\citealt{Rem2011}).
Following the recommendation from \cite{PlaXII2018} (see their Sect.~2), we subtract from this intensity map the contribution from the cosmic infrared background monopole ($452\, \mu\rm{K_{CMB}}$) and add back a fiducial Galactic offset ($63\, \mu\rm{K_{CMB}}$).
This map has a uniform resolution of 80\arcmin. 
Finally, the effective polarization fraction, $p_0$, is a scaling factor computed from a linear fit for each set of free-parameter values.
As in \cite{Pel2018}, this choice allows for the optimization of the computation time and reduces by one the number of free parameters of the model.

\begin{figure*}
\begin{center}
\begin{tabular}{ccc}
& $Q$ & $U$     \\

\rotatebox{90}{ \hspace{4em} \textit{Planck} data} &
\includegraphics[trim={0cm .5cm 0cm .7cm},clip,width=.34\linewidth]{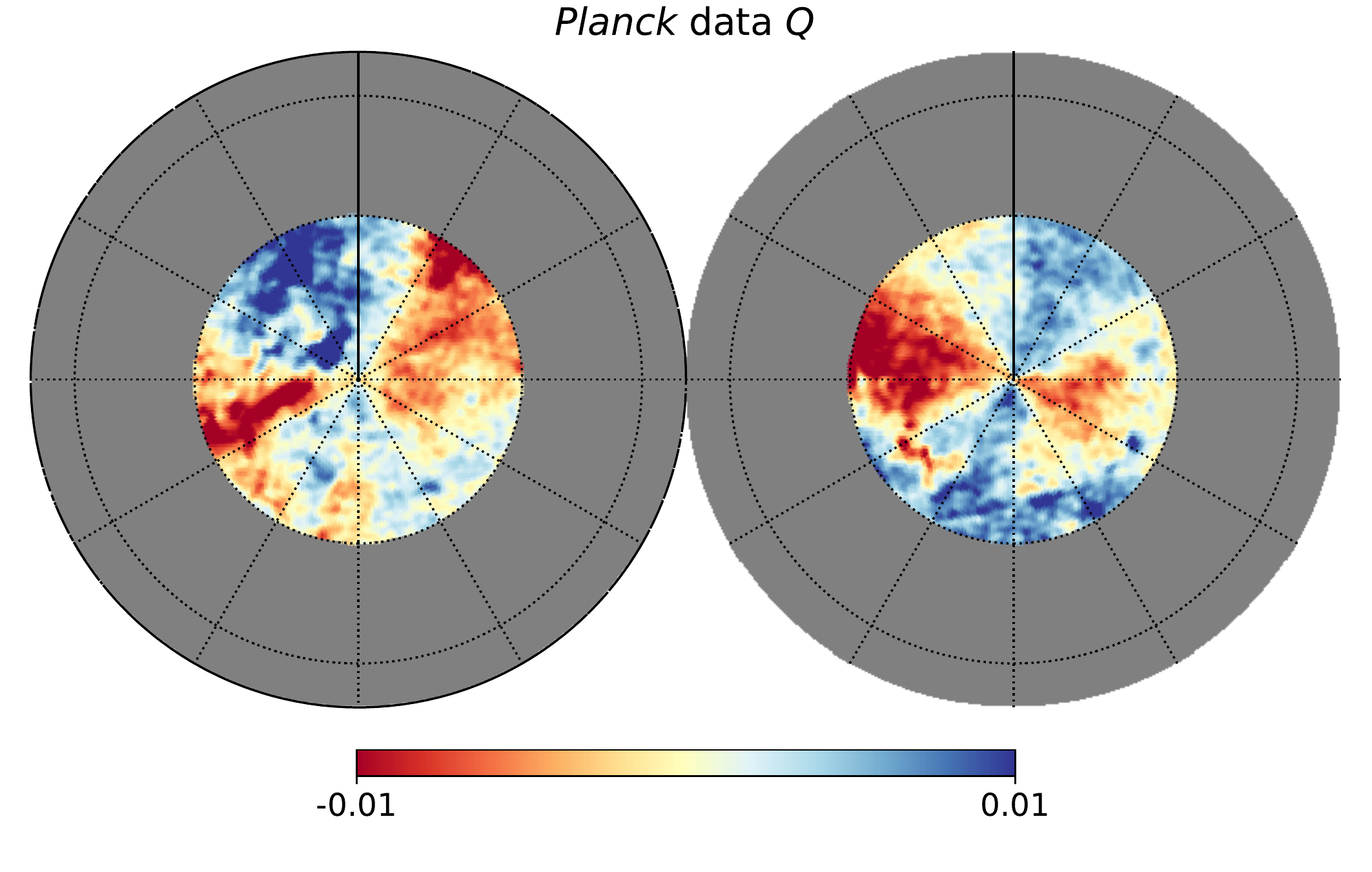}
        &       \includegraphics[trim={0cm .5cm 0cm .7cm},clip,width=.34\linewidth]{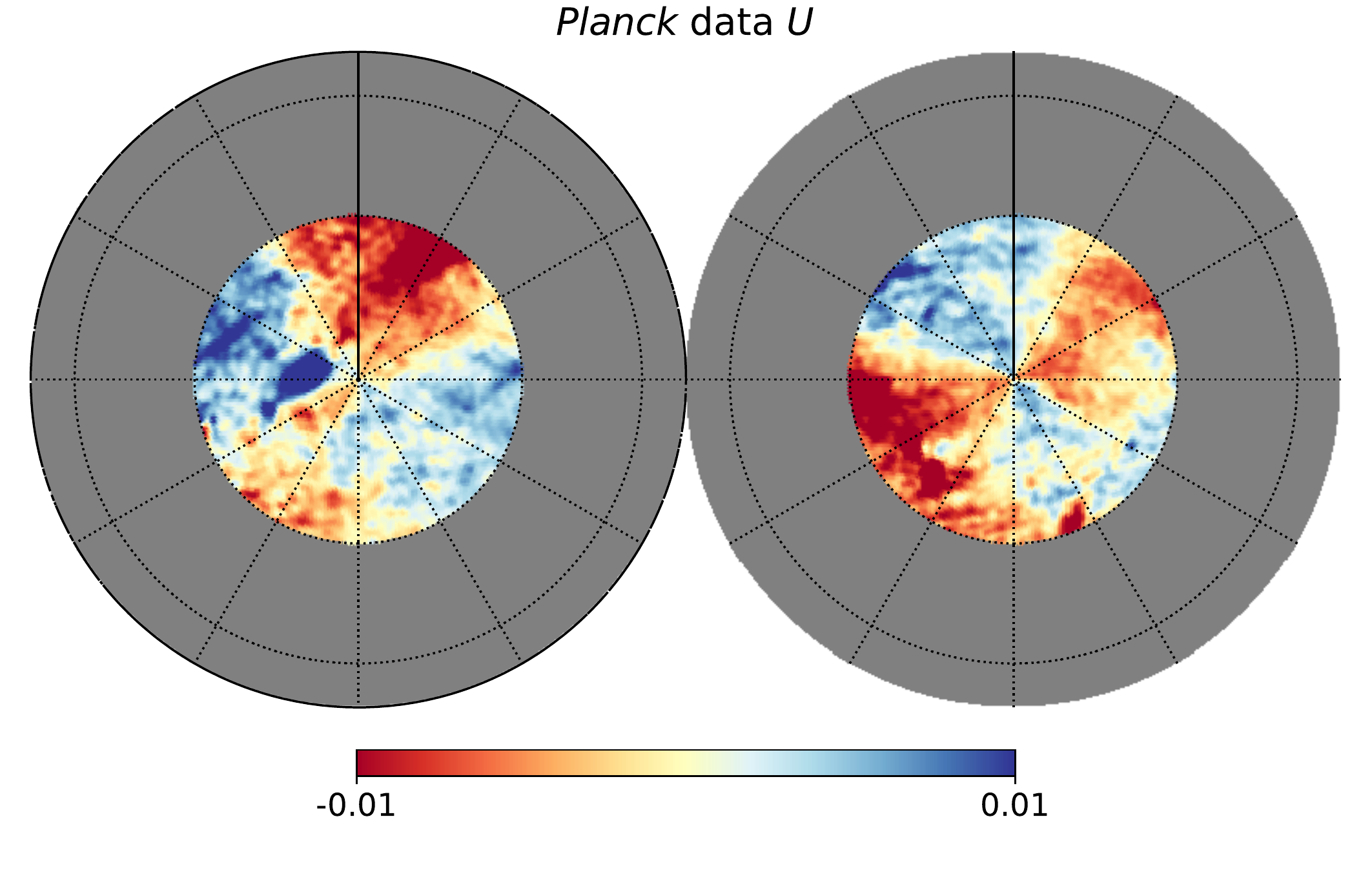}
\\

\rotatebox{90}{ \hspace{4em} $l_{\rm{max}}=2$} &
\includegraphics[trim={0cm .5cm 0cm .7cm},clip,width=.34\linewidth]{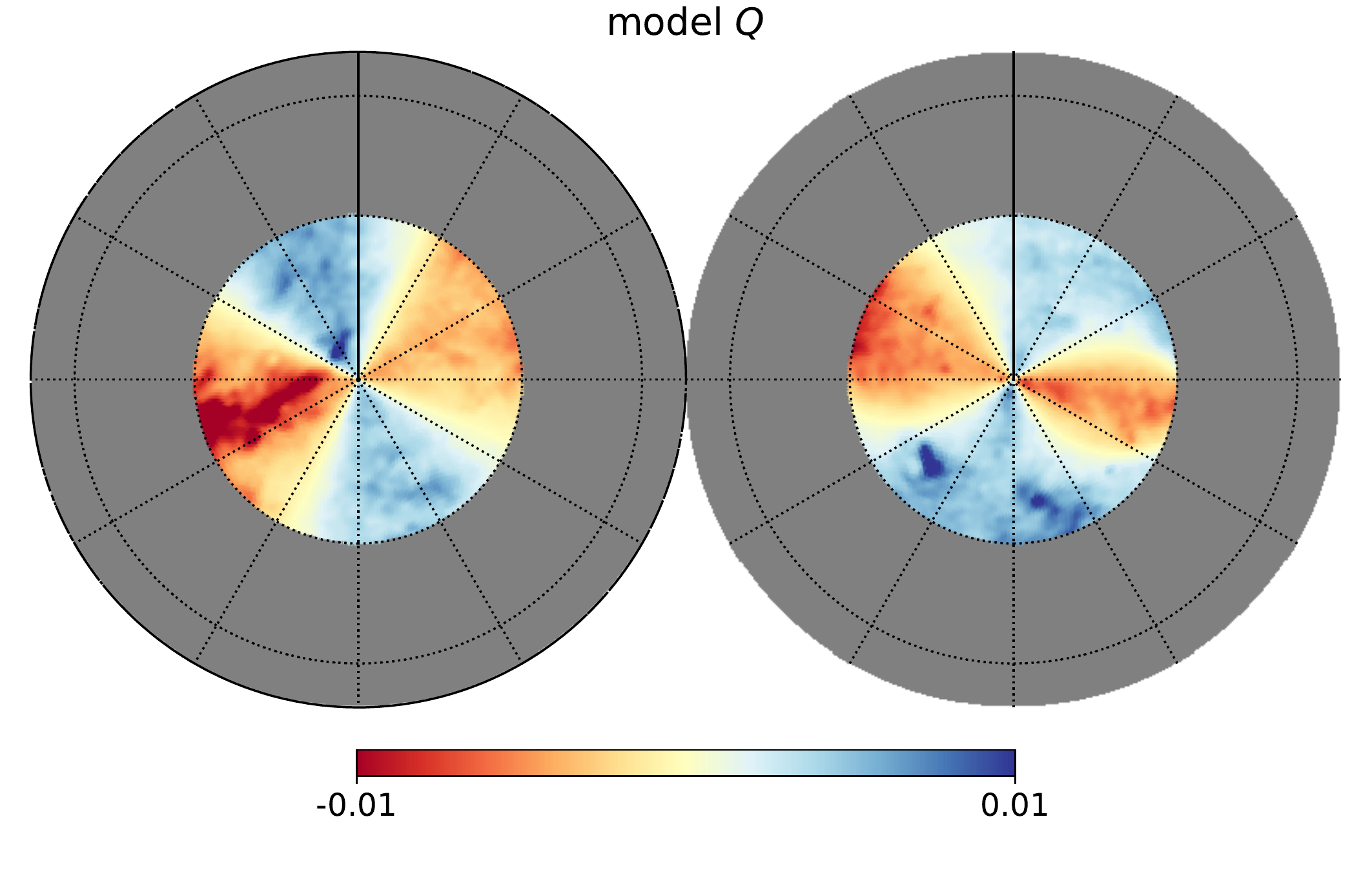}
        &       \includegraphics[trim={0cm .5cm 0cm .7cm},clip,width=.34\linewidth]{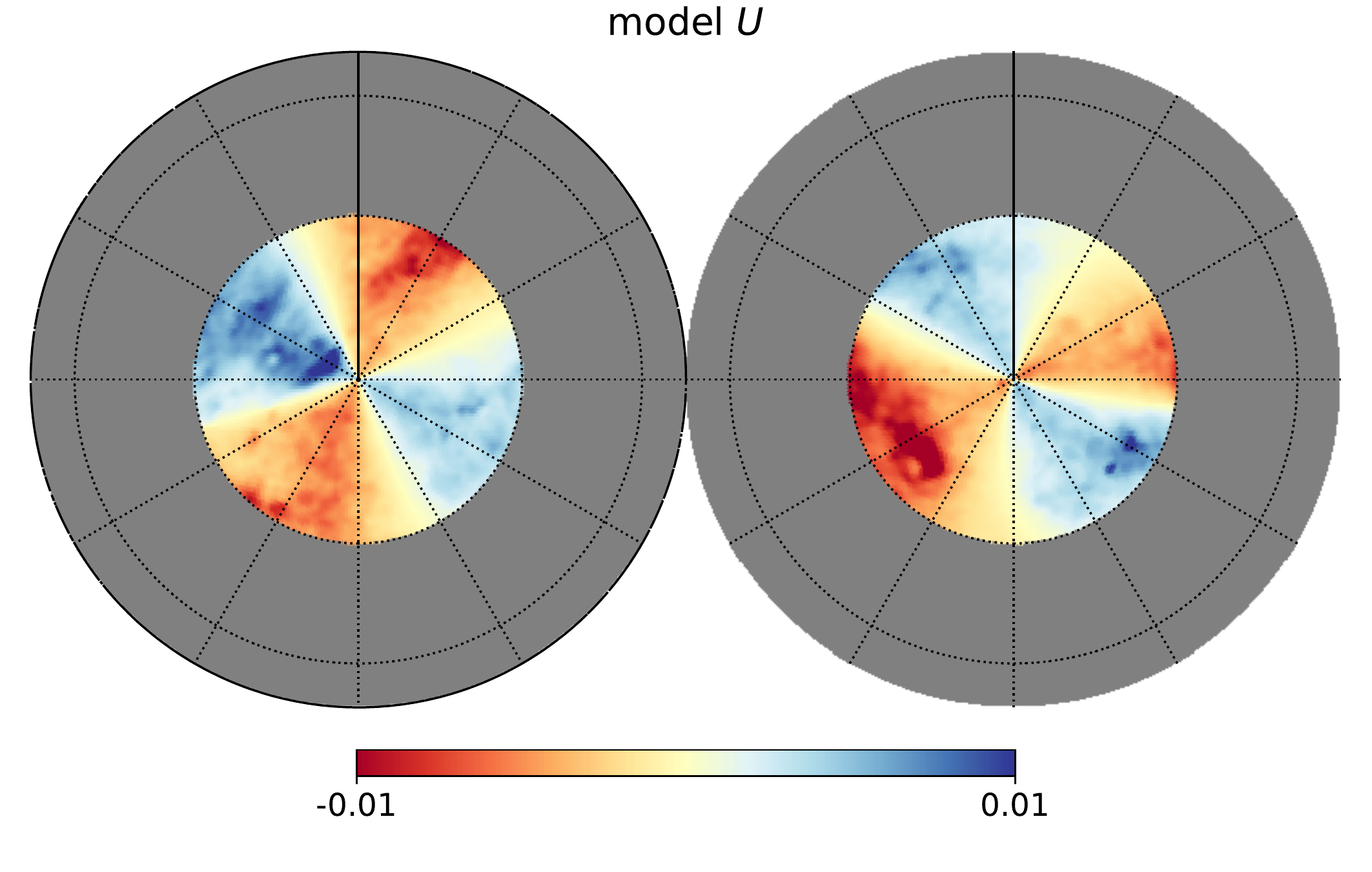}
\\

\rotatebox{90}{ \hspace{4em} $l_{\rm{max}}=4$} &
\includegraphics[trim={0cm .5cm 0cm .7cm},clip,width=.34\linewidth]{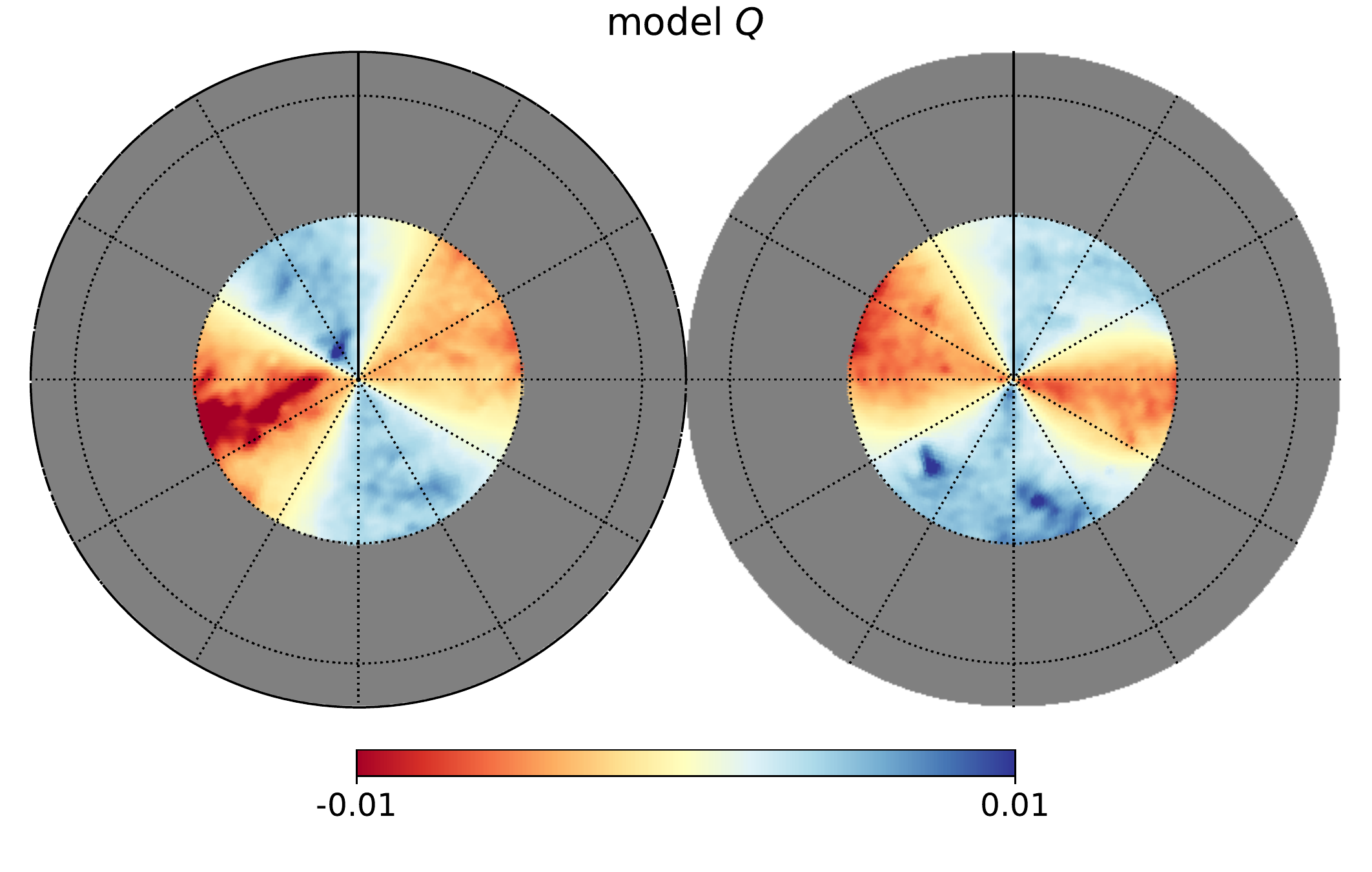}
        &       \includegraphics[trim={0cm .5cm 0cm .7cm},clip,width=.34\linewidth]{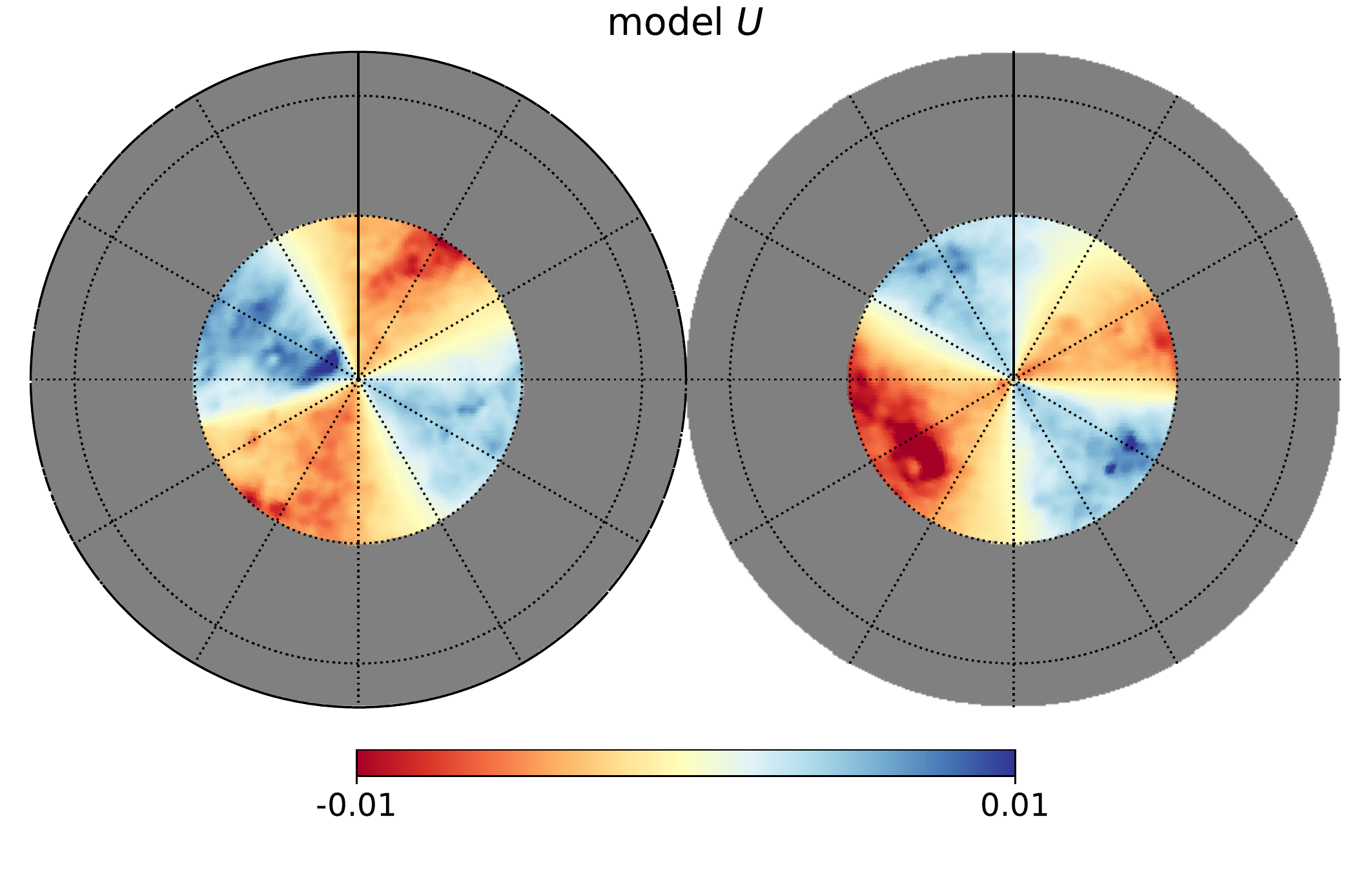}
\\

\rotatebox{90}{ \hspace{4em} $l_{\rm{max}}=6$} &
\includegraphics[trim={0cm .5cm 0cm .7cm},clip,width=.34\linewidth]{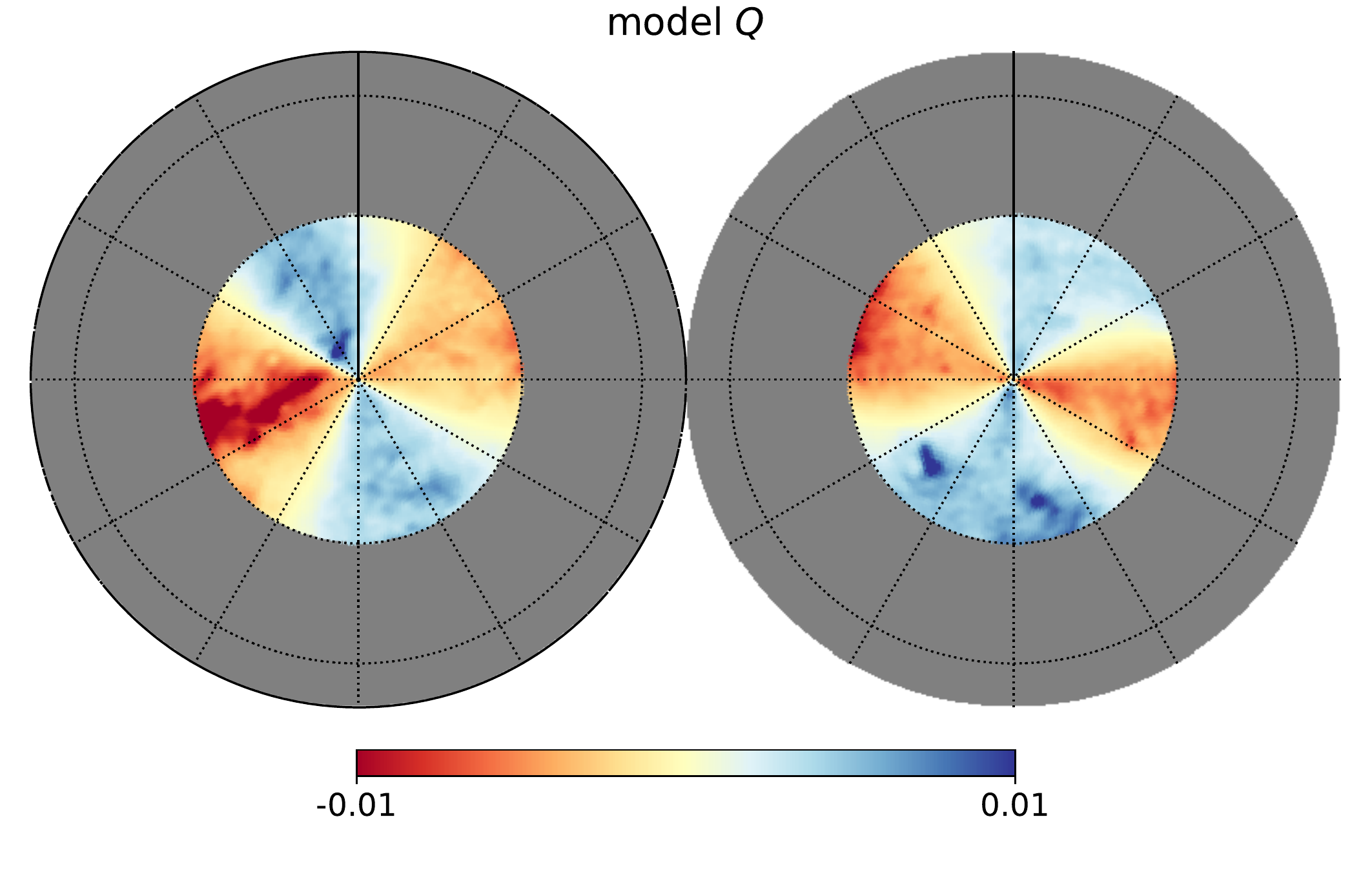}
        &       \includegraphics[trim={0cm .5cm 0cm .7cm},clip,width=.34\linewidth]{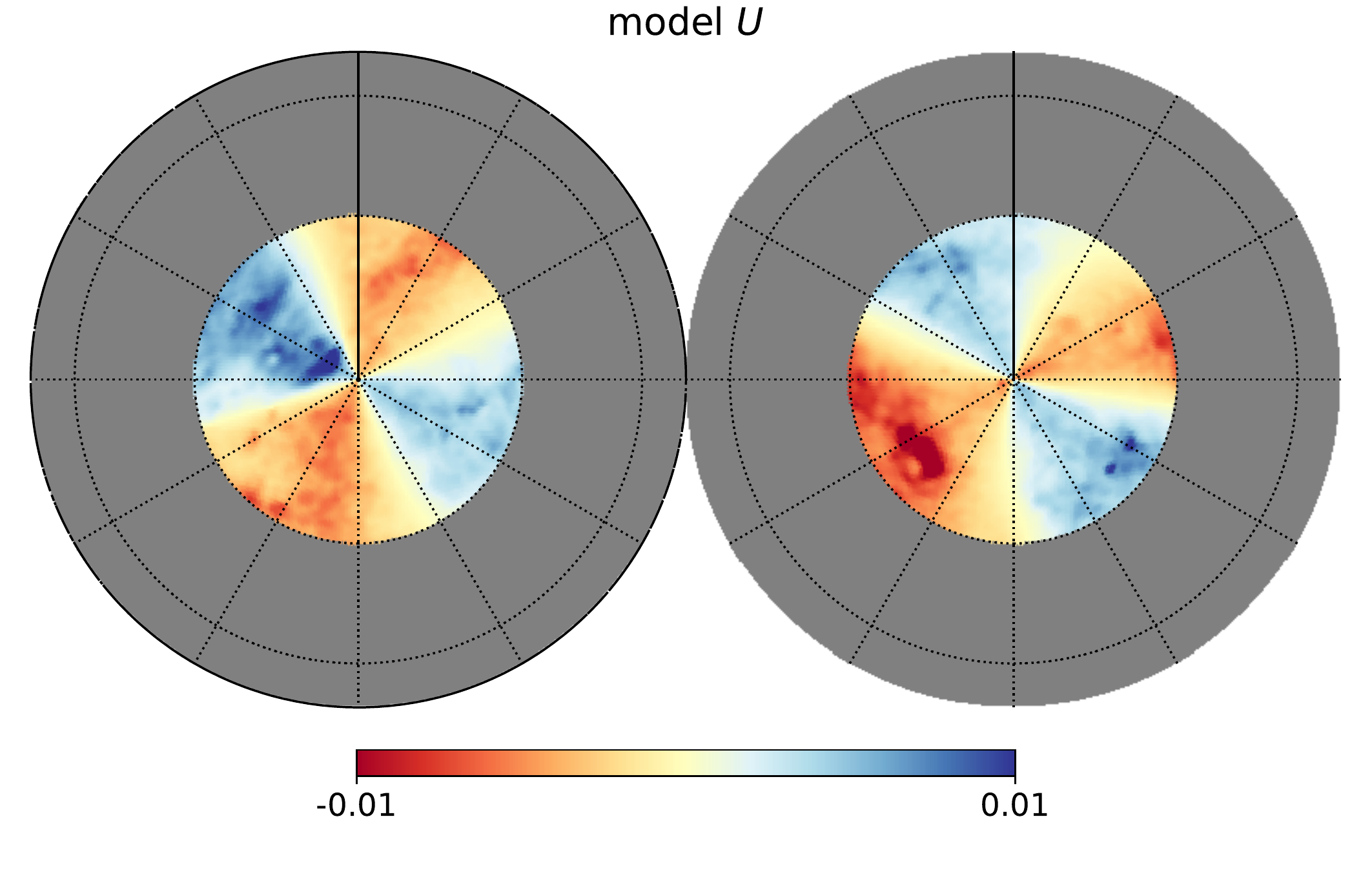}
\\

\rotatebox{90}{ \hspace{4em} $l_{\rm{max}}=8$} &
\includegraphics[trim={0cm .5cm 0cm .7cm},clip,width=.34\linewidth]{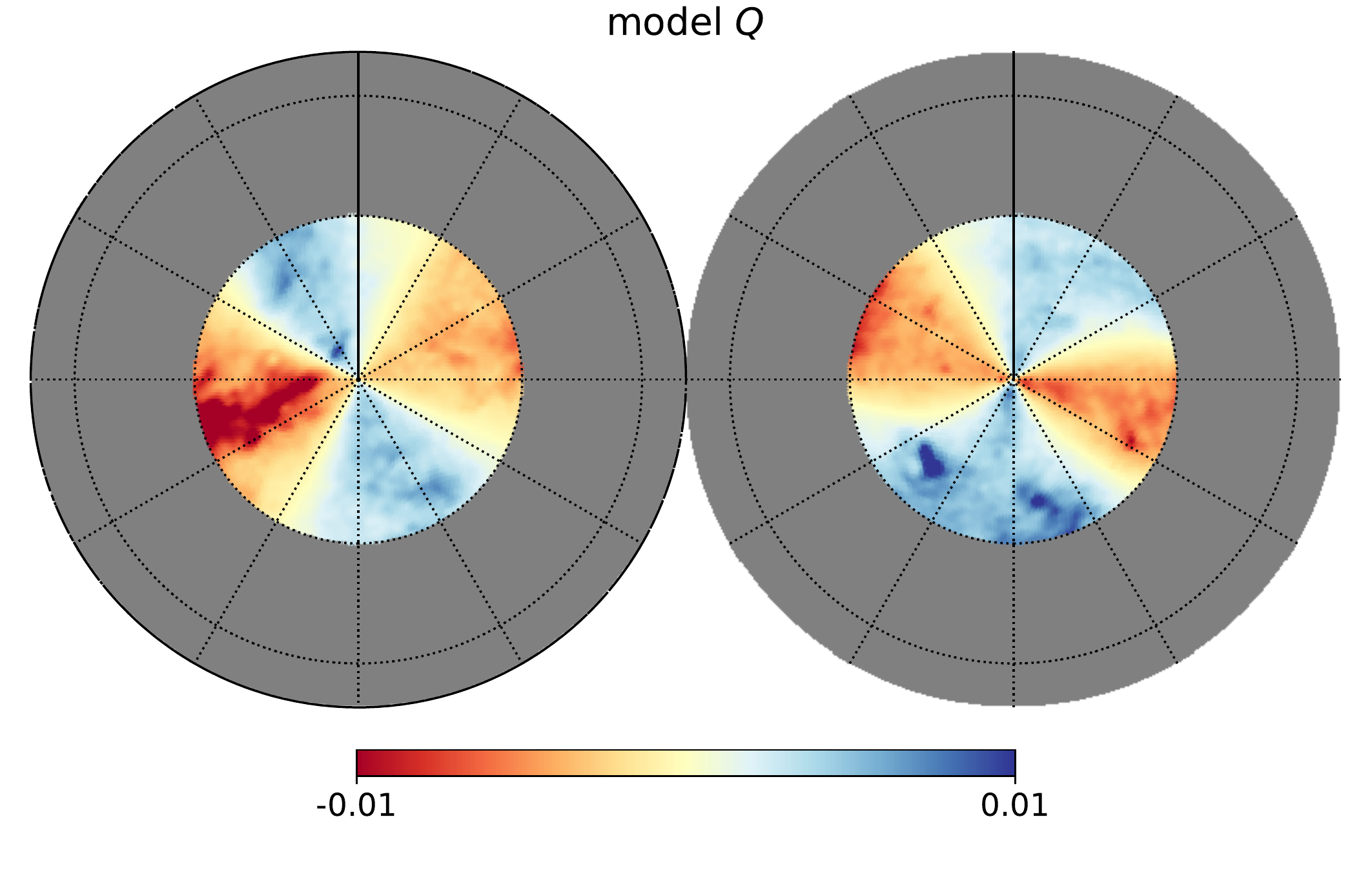}
        &       \includegraphics[trim={0cm .5cm 0cm .7cm},clip,width=.34\linewidth]{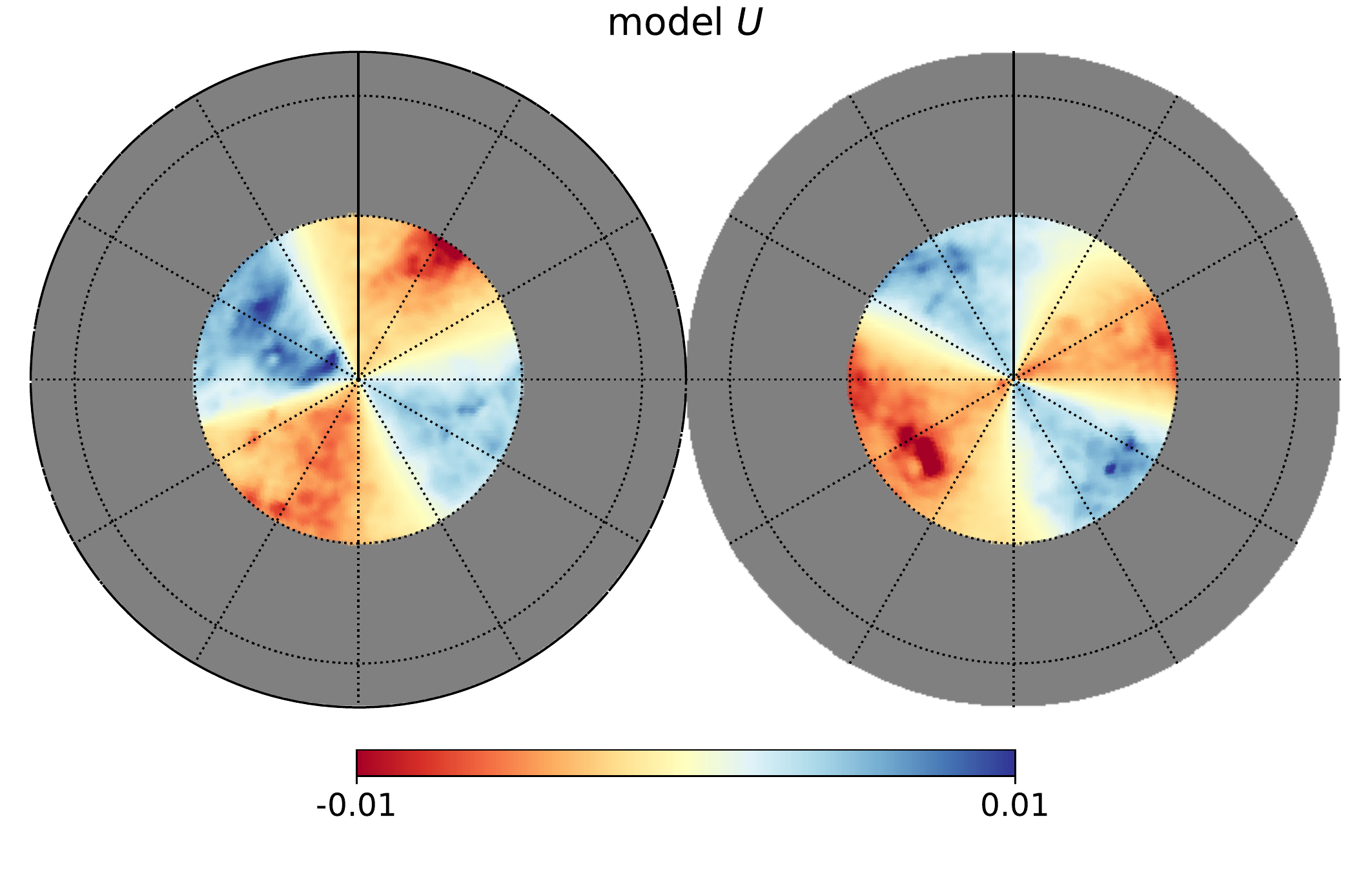}
\\

\rotatebox{90}{ \hspace{4em} $l_{\rm{max}}=10$} &
\includegraphics[trim={0cm .5cm 0cm .7cm},clip,width=.34\linewidth]{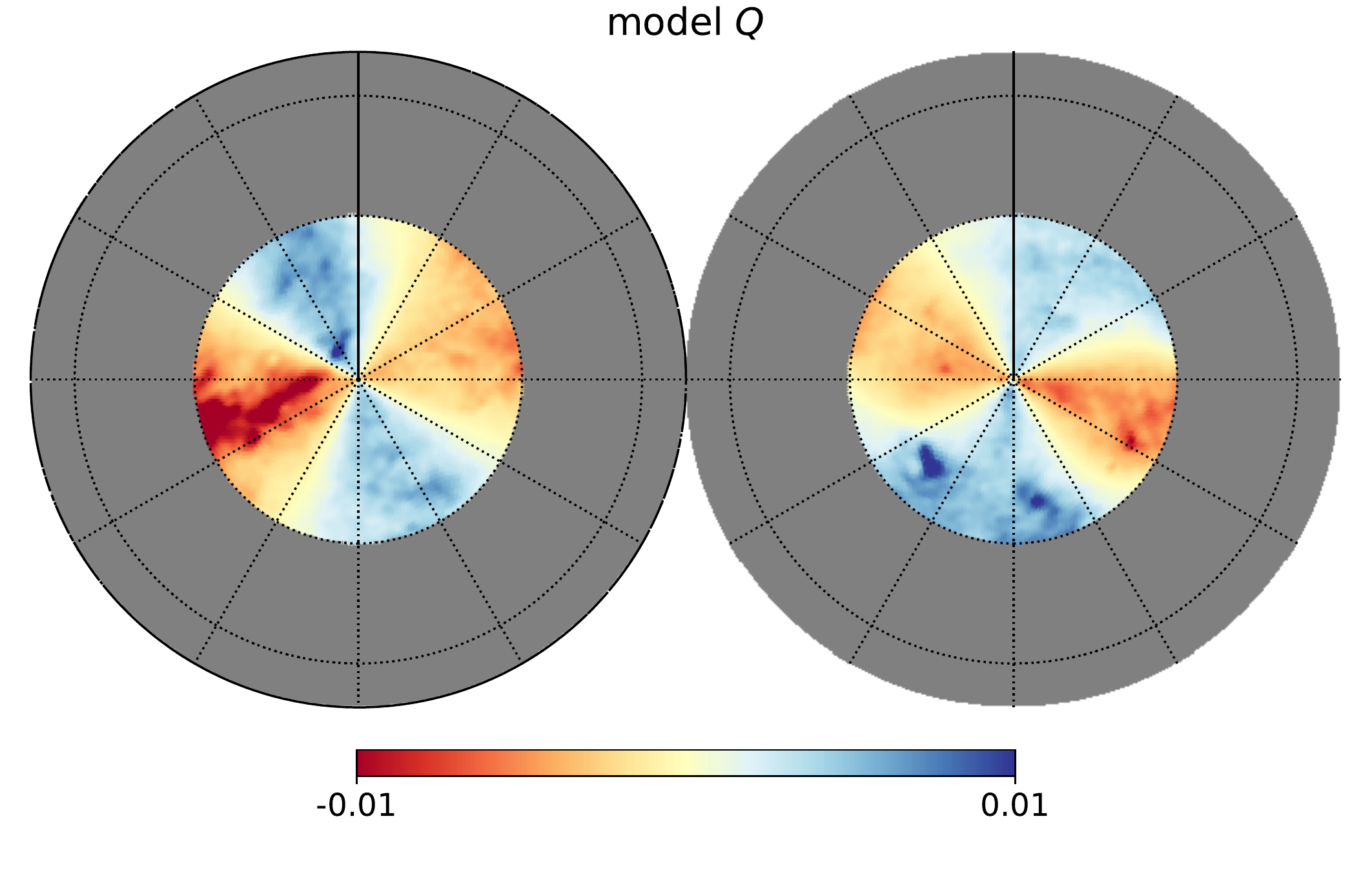}
        &       \includegraphics[trim={0cm .5cm 0cm .7cm},clip,width=.34\linewidth]{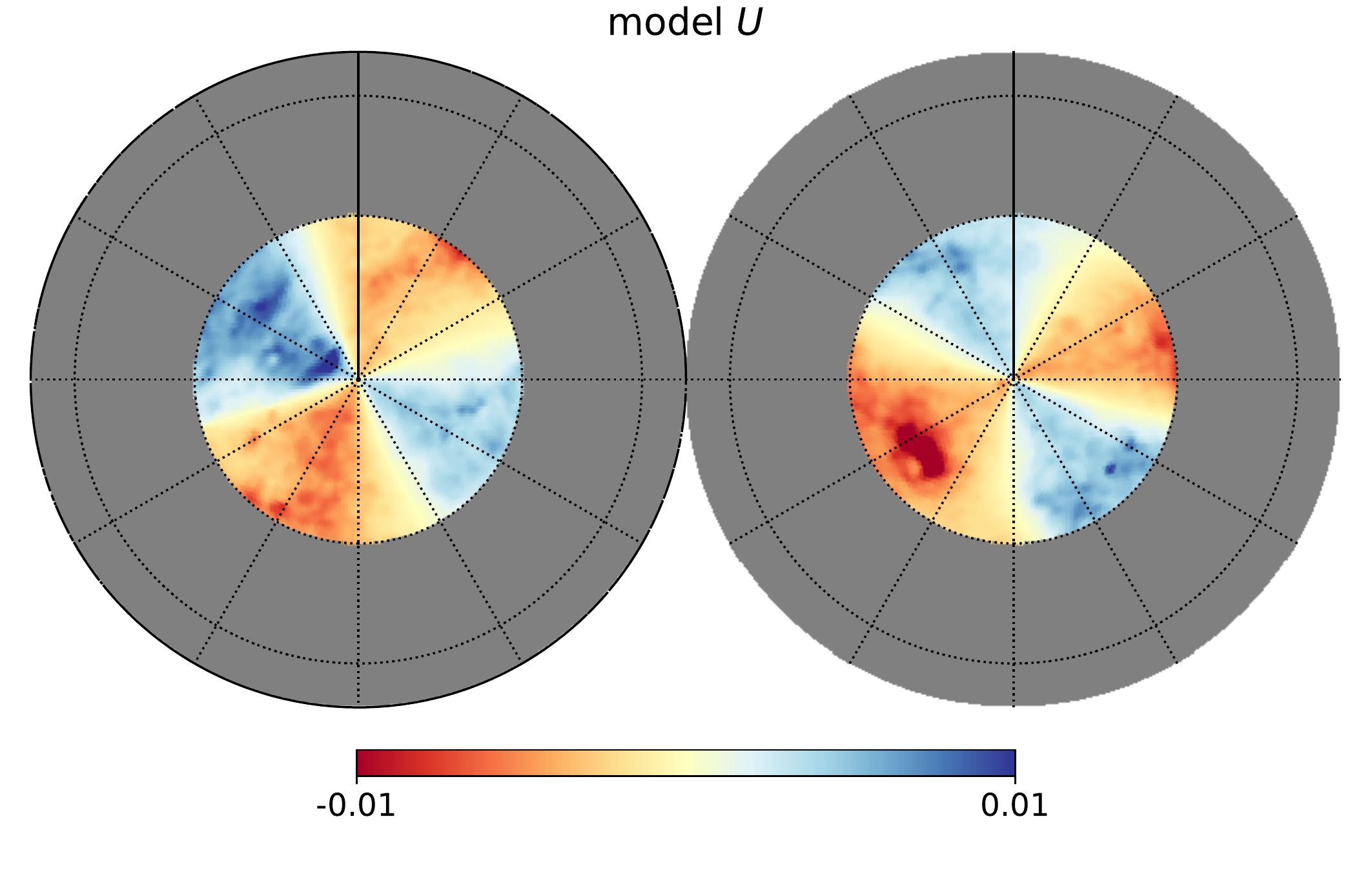}
\\
\end{tabular}
\caption{Orthographic views of the $Q$ (left) and $U$ (right) maps at 353~GHz.
From top to bottom: \textit{Planck} PR3 data and best-fit maps corresponding to the modeled shapes of the LB shell obtained
from the L19 3D extinction map, with $l_{\rm{max}} = 2,\,4,\,6,\,8,\,\text{and}\,10$.
The background structures underlying the maps come from the dust column density taken from the data. Units are MJy/sr.
In orthographic views, the North and South Galactic Poles lie at the center of the left and right circles, respectively,
the vertical solid radius shows longitude $0^\circ$, and the left and right panels touch at $l = 90^\circ$.
The gray area corresponds to the region $|b| < 60^\circ$, which is not considered in the fitting procedure.
}
\label{fig:LBShape_QUmaps}
\end{center}
\end{figure*}

We downgrade the observed maps of $Q$, $U$, and $I$ to the HEALPix grid of $N_{\rm{side}} = 128$ and convert them to MJy~sr$^{-1}$ using the unit conversion factor of $287.5$ MJy sr$^{-1}$ K$^{-1}_{\rm{CMB}}$ given in \cite{PlaIII2018}.
The resulting $Q$ and $U$ maps used as observational reference for our fits are shown in the top row of Fig.~\ref{fig:LBShape_QUmaps}.

The covariance matrix $C$ entering Eq.~\ref{eq:log_LL}, assumed diagonal, takes into account the noise in the \textit{Planck} $Q$ and $U$ data ($\sigma^{\rm noise}_{Q,U}$) and a contribution from the turbulent magnetic field component ($\sigma^{\rm turb}_{Q,U}$) that is otherwise not accounted for in the model and which is added in quadrature. As in A18, the latter is estimated
using modeled $Q$ and $U$ maps from \cite{Van2017}, which fit the \textit{Planck} dust power spectra at 353~GHz. 
The dispersion of $Q/I$ and $U/I$ in these maps, measured over the northern and southern Galactic polar caps separately, yields $\sigma^{\rm turb}_{Q,U} = 0.055 \, \times I_{\rm{d}}$. We note that this estimate is based on a model where the ordered magnetic field is assumed to have a uniform orientation. The corresponding value for the more elaborate model derived in this work could be smaller.
For the statistical noise, we use the covariance matrix of the $Q$ and $U$ GNILC maps, which are already delivered at $80\arcmin$ resolution.
We convert them using the conversion factors reported in Table~B.1. of \cite{PlaXII2018} so that they correspond to the polarization-sensitive bolometers Stokes maps.

\subsection{MCMC fit}
\label{subsec:MCMC_fit}
When modeling complex data, Markov Chain Monte Carlo (MCMC) methods have the strong advantage that they provide direct insight into the correlations and degeneracies between the different model parameters.
They also make it possible to fully explore the parameter space and to monitor the exploration up to completion.

In order to explore the parameter space, find the best-fit values of the parameters and sample their posterior distributions, 
we use the \texttt{emcee} MCMC Python software written by \cite{For2013}, who implemented the Affine-Invariant sampler proposed by \cite{Goo2010}.
Considering a non-informative prior, we require that the explosion center be located within the present-day LB cavity, namely, within the volume interior to our modeled LB shell.
We emphasize that this common-sense requirement
remains consistent with the conclusion of \cite{Mai2001} that the supernovae  contributing to the LB over the past 10 Myrs exploded just outside the boundary of the present cavity;
indeed, the LB and its surrounding ISM
probably drifted as a whole with respect to the Local Standard of Rest (by $\simeq 150$~pc in 10~Myrs; see Sect.~\ref{sec:intro})
and this drift may easily have brought the LB entirely past the explosion sites.
With our simplifying assumption of a single explosion center, the LB drift can simply be seen as a translation of the coordinate
reference frame.
We also note that the above prior sets limits on the location of the explosion center in the direction of the initial magnetic field, which otherwise is not constrained at all by our fit to the \textit{Planck} data (see discussion at the end of Sect.~\ref{subsec:GMF_model}).

To optimize the exploration of the parameter space, we proceed in two stages.
In the first stage, we identify the region of parameter space that maximizes the log-likelihood. 
In the second stage, we determine the set of best-fit parameter values and properly sample the posterior distributions.
Thus, in the first stage, 500 Markov chains are initialized with uniform distributions over a restricted parameter space defined as:
\begin{itemize}
\item $\{\delta x,\, \delta y,\, \delta z \}  \in [-70 ,\,70 ] $ pc
\item $b_0 \in [-90,\,90]$ $^\circ$
\quad \& \quad 
$l_0 \in [0,\,180]$ $^\circ$ \ ,
\end{itemize}
and the MCMC algorithm is run for 1000 steps. 
This first stage can be considered as a burn-in phase of the MCMC experiment.
In the second stage, we retain the best 250 chains obtained at the last MCMC step of the first stage; these chains are initialized at their last positions in parameter space, and the MCMC algorithm is run until the convergence criteria proposed by \cite{Gel1992} are fulfilled for all the model parameters, with a threshold value of $1.03$. 
We test for convergence every 100 MCMC steps. 
For all the fits presented in this paper, convergence is reached within 5000 steps. 
We verified on one of the fits that the same result is obtained when initializing ten times more Markov chains at the first stage within a wider volume of parameter space.

\subsection{Results}
\label{subsec:GMF_Result}
We use our MCMC procedure to fit the \textit{Planck} $Q$ and $U$ maps and thereby constrain the five free parameters of our magnetic field model.
We consider the cases when the spherical harmonic expansion of the inner surface of the LB shell in Sect.~\ref{subsubsec:mod_LBedges} is truncated at $l_{\rm{max}} = 2,\,4,\,6,\,8$, and $10$.
We present the results of the fits in Sect.~\ref{subsec:model_Bfield} and discuss systematic uncertainties in Sect.~\ref{subsec:uncertainties_Bfield}.

Since dust polarization only gives the orientation of the magnetic field, not its direction, each solution for $(l_0,\,b_0)$ actually corresponds to the pair of solutions $(l_0,\,b_0)$ and $(l_0+180^\circ,\,-b_0)$. Among these two solutions, we select the one that is closer to the large-scale magnetic field direction derived from rotation measures of nearby pulsars, which is almost equivalent to selecting the value of $l_0$ that lies in the range $[0^\circ,\,180^\circ]$ (e.g., \citealt{Fer2015}).

\subsubsection{Magnetic field model}
\label{subsec:model_Bfield}
\begin{figure}
\centering
\begin{tabular}{cc}
\rotatebox{90}{ \hspace{9em} $l_{\rm{max}}=2$} &
\includegraphics[trim={0cm .0cm 0cm .0cm},clip,width=.8\linewidth]{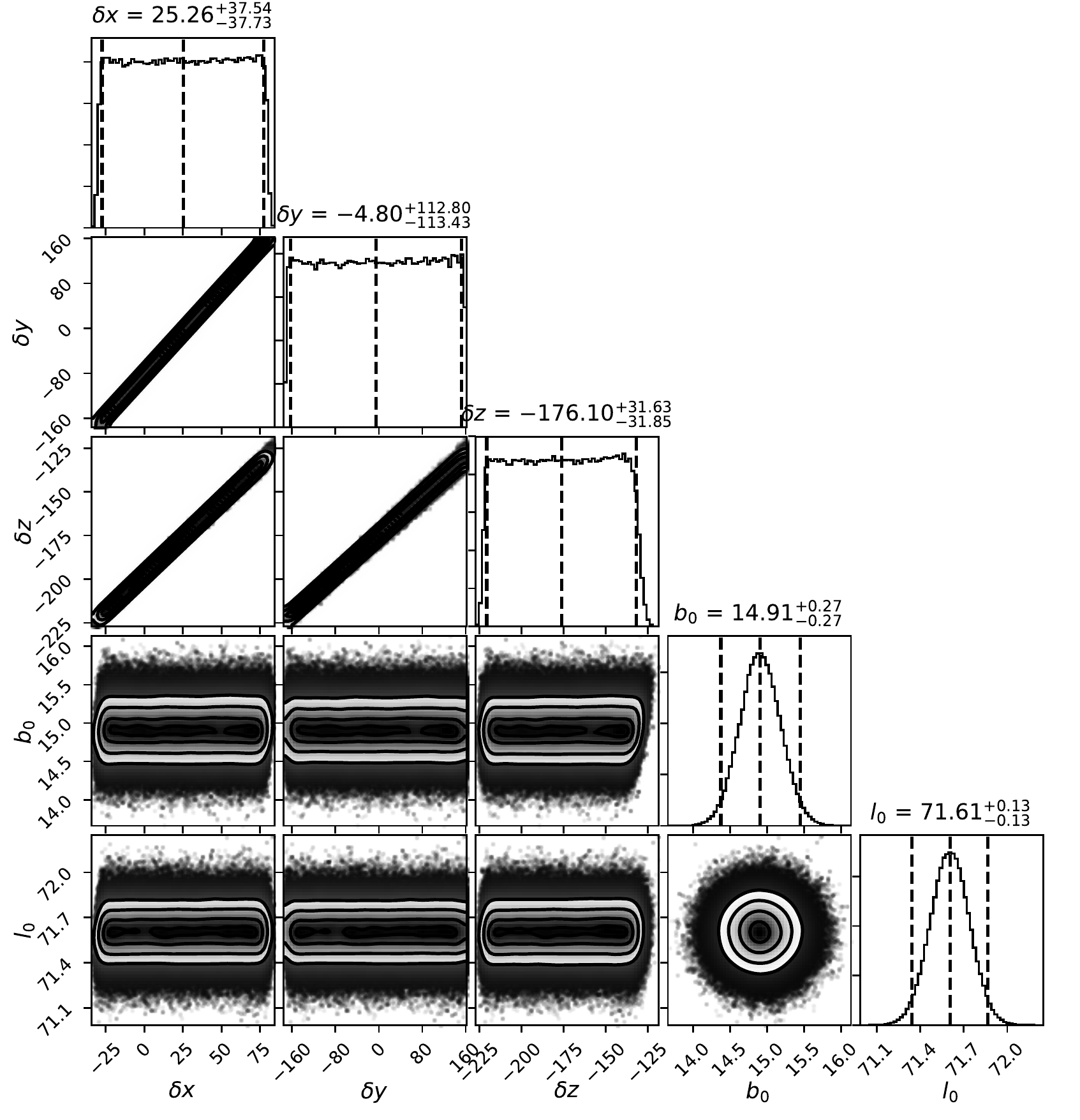} \\
\rotatebox{90}{ \hspace{9em} $l_{\rm{max}}=6$} &
\includegraphics[trim={0cm .0cm 0cm .0cm},clip,width=.8\linewidth]{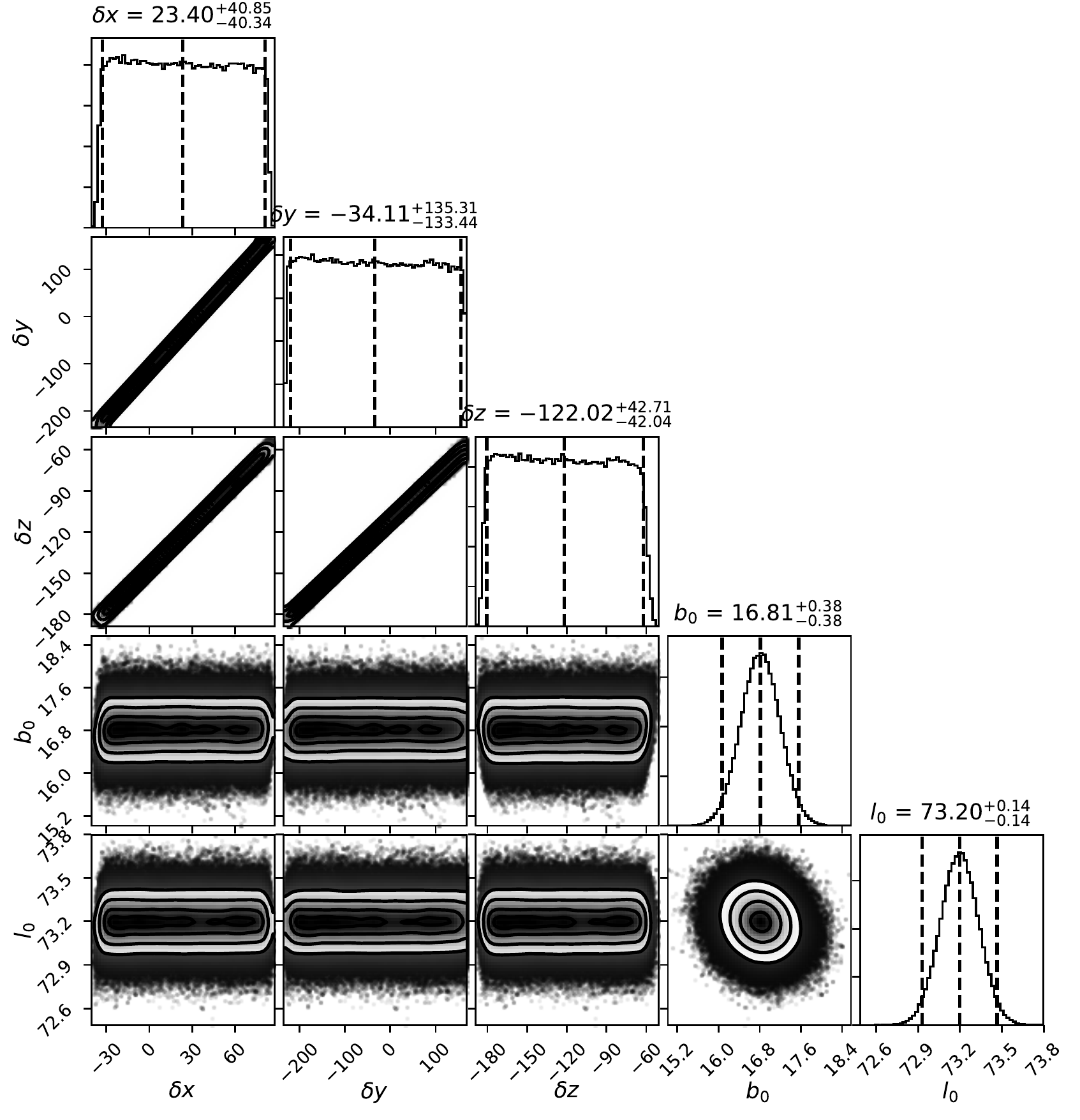} \\
\rotatebox{90}{ \hspace{9em} $l_{\rm{max}}=10$} &
\includegraphics[trim={0cm .25cm 0cm .0cm},clip,width=.8\linewidth]{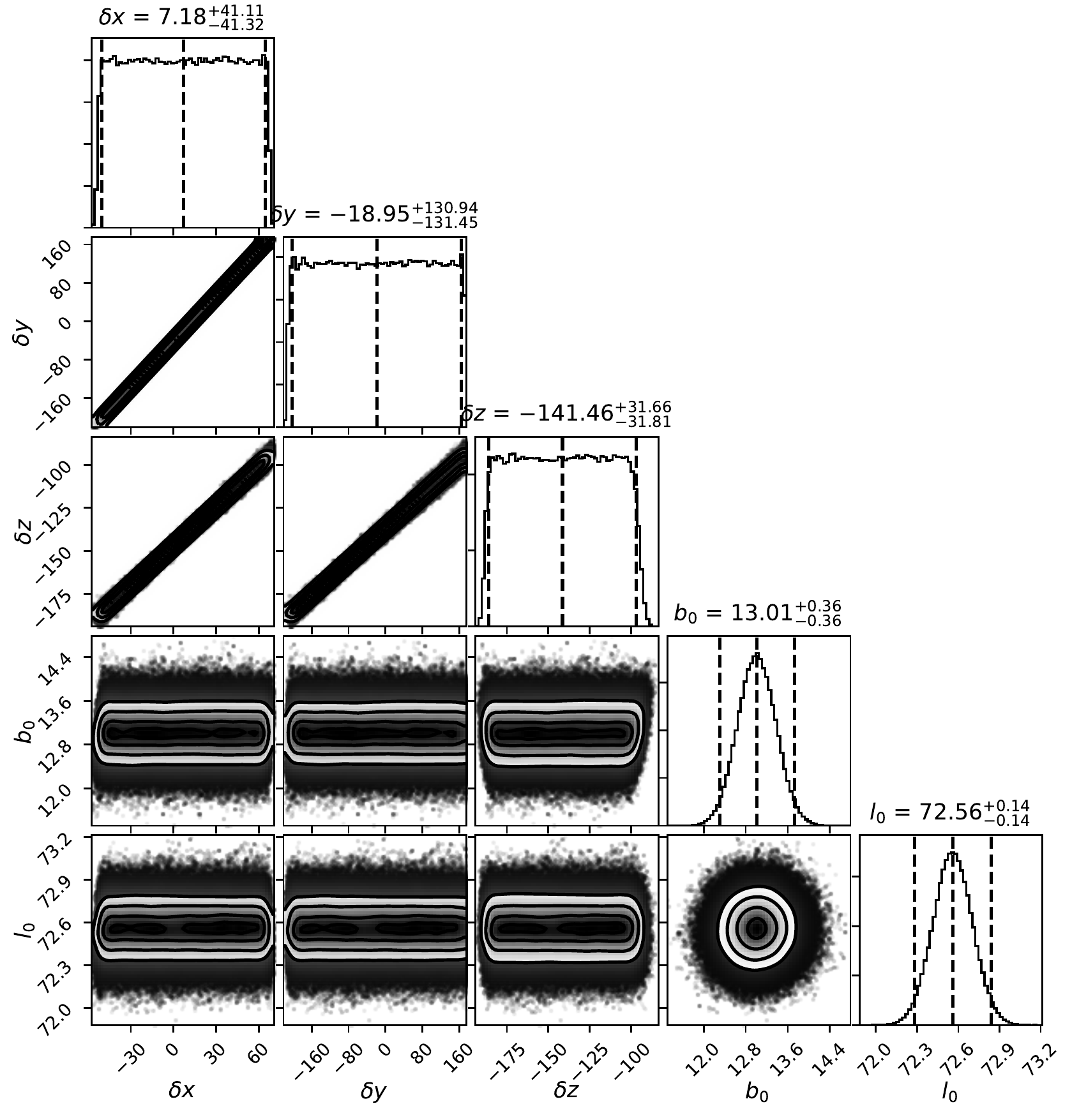}\\
\end{tabular}
\caption{1D and 2D marginalized posterior distributions of the model free parameters.
The modeled shapes of the LB shell are obtained from the L19 3D extinction map, with $l_{\rm{max}} = 2,\,6,\,\text{and}\,10$. The vertical dashed lines show the 2.5, 50 and 97.5 percentiles.}
\label{fig:GMFfit_L19-06_corner}
\end{figure}

The best-fit $Q$ and $U$ maps obtained with the five values of $l_{\rm{max}}$
are shown in Fig.~\ref{fig:LBShape_QUmaps}, below the observed \textit{Planck} maps.
For the purposes of visualization, Fig.~\ref{fig:GMFfit_L19-06_corner} displays the 1D and 2D marginalized posterior distributions of the fits obtained with $l_{\rm{max}} = 2,\,6,\,\text{and}\,10$. 
The posterior distributions are produced from converged fractions of the MCMC chains.

In these corner plots, it appears that the coordinates of the explosion center, ($\delta x$, $\delta y$, $\delta z$), and those of the initial magnetic field direction, ($l_0$, $b_0$), are not correlated around the best-fit values.
Similar results are obtained with $l_{\rm{max}} = 4$ and $8$.
The 2D marginalized posterior distributions of $(\delta x,\,\delta y,\,\delta z)$ reflect the model degeneracy discussed at the end of Sect.~\ref{subsec:GMF_model}:
the orientation of the present-day magnetic field in the shell is insensitive to the position of the explosion center along the direction of the initial magnetic field, $\mathbf{B}_0$.

Table~\ref{tab:Table1} lists the best-fit values of the five model parameters.
The quoted uncertainties correspond to the standard deviation of the 1D marginalized posterior distribution of each parameter. 
They only account for the data noise and the turbulent component of the magnetic field.

Using the 2015 \textit{Planck} data release, A18 found that the dominant contribution to the uncertainty budget on their model parameters is from the \textit{Planck} residual systematics.
Here, to provide a full error budget, we need to assess the impact on our model results of both residual systematics in the \textit{Planck} data and uncertainties in the 3D extinction map.

\begin{table*}
\caption{Best-fit parameter values of the adjusted magnetic field model for the values of $l_{\rm{max}}$ used to model the shape of the LB shell based on the L19 3D extinction map.
The fit was performed on the joined $Q$ and $U$ maps from the \textit{Planck} PR3 353~GHz data.
The error bars correspond to the standard deviations of the 1D marginalized posterior distributions.
The last two columns give the reduced $\chi^2$ and the effective polarization fraction obtained for the best fits (see text).
}
\label{tab:Table1}
\centering
\small{
\begin{tabular}{l c c c c c | c c}
\hline\hline
\\[-1.5ex]
$l_{\rm{max}}$ & $\delta x$ [pc] & $\delta y$ [pc] & $\delta z$ [pc] & $b_0$ [$^\circ$]      & $l_0$ [$^\circ$]
        & $\tilde{\chi^2}$      & $p_0$ [\%]    \\
\\[-1.5ex]
\hline \\[-1.5ex]
2 &     $32.0 \pm 31.9$ &       $15.4 \pm 95.9$ &       $-170.6 \pm 27.0 $       &       $14.9 \pm 0.3$  &       $71.6 \pm 0.1$  &       $0.68$  &       $8.18$  \\

4 &     $-16.9 \pm 29.3$        &       $ -184.5\pm 96.5$       &       $-195.5 \pm  28.5$      &       $15.8 \pm  0.3$ &       $73.1 \pm 0.1$  &       $0.69$  &       $8.27$  \\

6 &     $57.6 \pm 34.5$ &       $79.2 \pm 114.1$        &       $-86.3 \pm 36.1$   &       $16.8 \pm 0.4$  &       $73.2 \pm 0.1$  &       $0.75$  &       $8.17$  \\

8 &     $-9.0 \pm 35.4$ &       $-96.7 \pm 115.0$       &       $-150.2 \pm 30.7$           &       $14.3 \pm 0.4$  &       $72.9 \pm 0.1$  &       $0.74$  &       $8.45$  \\

10 &    $51.2 \pm 35.0$ &       $121.3 \pm 111.5$       &       $-107.6 \pm 27.1$           &       $13.0 \pm 0.4$  &       $72.6 \pm 0.1$  &       $0.78$  &       $8.29$  \\
       
\hline
\end{tabular}
}
\end{table*}

\begin{table*}
\caption{Best-fit parameter values and corresponding results of the E2E simulations,
for the modeled shapes of the LB shell obtained from the L19 (top) and LE19 (bottom) 3D extinction maps, with $l_{\rm{max}} = 6$.
The fits were performed on the joined $Q$ and $U$ maps from the \textit{Planck} PR3 353~GHz data.
The error bars on the best-fit values correspond to the standard deviations of the 1D marginalized posterior distributions.
The E2E results are the mean values and the standard deviations of the best fits to
the 10 mock maps with \textit{Planck} residual systematics
(see text).
}
\label{tab:Table2}
\centering
\small{
\begin{tabular}{l c c c c c | c c }
\hline\hline
\\[-1.5ex]
3D map & $\delta x$ [pc] & $\delta y$ [pc] & $\delta z$ [pc] & $b_0$ [$^\circ$] & $l_0$ [$^\circ$] & $\tilde{\chi^2}$  & $p_0$ [\%]       \\
\\[-1.5ex]
\hline \\[-1.5ex]
L19 \\
\hspace{.2cm} best fit  &       $57.6 \pm 34.5$ &       $79.2 \pm 114.1$        &       $-86.3 \pm 36.1$       &       $16.8 \pm 0.4$  &       $73.2 \pm 0.1$  & $0.75$        & $8.17$  \\
\hspace{.2cm} E2E               &    $22.5 \pm 35.8$    &   $-22.3 \pm 99.0$    &   $-120.1 \pm 30.9$     &   $16.2 \pm 1.2$      &   $72.3 \pm 1.2$    &   $-$ &   $8.16 \pm 0.02$ \\
\\[-1.5ex]
LE19 \\
\hspace{.2cm} best fit & $56.4 \pm 41.6$ & $166.6 \pm 104.9$ &   $-98.5 \pm 6.5$   &   $3.2 \pm 0.3$  &   $68.4 \pm 0.1$ &  $0.61$  &   $10.86$ \\
\hspace{.2cm} E2E &     $31.0 \pm 33.7$ &       $99.0 \pm 81.3$ &   $-96.0 \pm 9.3$    &   $5.1 \pm 2.3$ & $67.8 \pm 0.7$ &   $-$ & $10.89 \pm 0.02$\\
\\
\hline
\end{tabular}
}
\end{table*}

\subsubsection{Systematic uncertainties}
\label{subsec:uncertainties_Bfield}
In this section, we assess the uncertainties associated with first the \textit{Planck} data systematics and second the 3D extinction map used to compute the inner surface of the LB shell (see also Sect.~\ref{subsec:comp3Dmap}).

For the \textit{Planck} residual systematics, we follow the  following three steps.
First, we produce mock $Q$ and $U$  maps based on the model maps computed for the best-fit parameters, to which we add independent realizations of the \textit{Planck} systematics. 
Here, we use the end-to-end (E2E) simulations available in the \textit{Planck Legacy Archive} (see Appendices~A.1 and~A.2 in \cite{PlaXI2018}).
\footnote{The E2E maps were downloaded from \url{https://wiki.cosmos.esa.int/planck-legacy-archive/index.php/Simulation\_data\#Noise\_and\_instrumental\_effect\_residual\_maps}, smoothed at 80\arcmin ~and downgraded at $N_{\rm{side}} = 128$ before being co-added to the best-fit maps.}
Next, we fit each set of $Q$ and $U$ mock maps with our MCMC code in the same way as we fitted the \textit{Planck} maps, using the same covariance matrix and $I_{\rm{d}}$ map.
Last, we compare the best-fit parameter values obtained for 10 mock samples with the input model values to quantify the uncertainties associated with the \textit{Planck} residual systematics.
To estimate the uncertainties associated with the 3D extinction map, we repeat the analysis of the mock maps and the fit to the \textit{Planck} maps, using the LB shape derived from the LE19 map instead of the L19 map.

\smallskip

Table~\ref{tab:Table2} summarizes the results of this data analysis.
For each of the 3D extinction maps, we first report the best-fit parameter values and the standard deviations of the 1D marginalized posterior distributions obtained by fitting to the \textit{Planck} data. In the second line, we report the mean values and the standard deviations of the best fits to the 10 mock maps.
Here are the conclusions we draw from this analysis.\\
(\textit{i}) The \textit{Planck} residual systematics do not induce bias in the best-fit parameter values since the input parameter values are always found within one standard deviation from the mean values measured in the mock maps. \\
(\textit{ii}) The difference between input and output parameter values is slightly larger than the uncertainties from the fit to the \textit{Planck} data, which shows that the residual systematics are a significant source of uncertainty. 
This conclusion is substantiated by the dispersions in the best-fit values of $l_0$ and $b_0$, which are larger than those derived in the MCMC data fit for the \textit{Planck} noise and turbulence. We note that, as expected, the uncertainties in $l_0$ and $b_0$ are smaller for the PR3 maps than those obtained in A18 for the previous version of the \textit{Planck} maps.\\
(\textit{iii}) The best-fit parameter values depend significantly on the 3D dust maps used to model the geometry of the LB shell. 
Indeed, for most of the model parameters, the posterior distributions obtained with the LE19 map are significantly different from the corresponding distributions obtained with the L19 map.
At this stage, we recommend that  both solutions be considered, as neither one is actually better than the other (however, see the discussion in Sects.~\ref{subsec:comp3Dmap} and~\ref{subsec:radio_pol}).
The dispersion between the different models is more representative of the margins of error.

\subsection{Discussion}
This work extends the analytical modeling of the local Galactic magnetic field in A18 into a consistent model where the shape of the LB shell is derived from a 3D extinction map (L19 or LE19), rather than  approximated with an ad-hoc geometry.
To assess and discuss the validity of our model, we compare it to two other models that we fitted to the same \textit{Planck} PR3 353~GHz data and with the same MCMC code.
One of these models is the spheroid model of A18, the other one assumes, as in \cite{PlaXLIV2016}, that the magnetic field orientation is uniform over each Galactic polar cap.
In all cases, the models were fitted over both polar caps with a single value of $p_0$.
As a basis for the model comparison (discussed below), Table~\ref{tab:Table3} provides a few representative quantities corresponding to the best fit of each model.

\begin{table*}
\caption{Comparison of the best-fit magnetic fields for
four different models discussed in the text:
the `Uniform' model assumes a uniform magnetic field
orientation over each Galactic polar cap,
`A18' is the spheroid model presented in A18, and
`L19' and `LE19' are the models developed in this paper with the shape of the LB shell extracted from the L19 and LE19 3D extinction maps, respectively, each modeled with $l_{\rm{max}} = 6$.
The four models were fitted to the \textit{Planck} PR3 353~GHz data.
Listed here are the reduced $\chi^2$, the effective polarization fraction, $p_0$, the orientation of the magnetic field averaged
over each of the northern and southern polar caps, $(l,\,b)_{\rm N,S}$,
and $\cos^2\gamma$ averaged over each polar cap, with $\gamma$ the angle of the magnetic field to the plane of the sky.
}
\label{tab:Table3}
\centering
\small{
\begin{tabular}{lcccc}
\hline\hline
\\[-1.5ex]
    & Uniform & A18 & L19   & LE19 \\
    & & & $l_{\rm{max}} = 6$ & $l_{\rm{max}} = 6$\\
\\[-1.5ex]
\hline \\[-1.5ex]
$\tilde{\chi^2}$ & $0.58$ &  $0.60$ & $0.75$    & $0.61$  \\
$p_0$ [\%]   & $10.33$  & $11.83$  & $8.17$     & $10.86$\\
$(l,\,b)_{\rm{N}}$ [$^\circ$] & $(67.9,\,-26.4)$  & $(67.4,\, 34.3)$  & $(71.0,\,-10.9)$  & $(72.5,\,0.06)$\\
$(l,\,b)_{\rm{S}}$ [$^\circ$] & $(74.9,\,24.2)$ & $(69.8,\,-6.7)$ & $(74.0,\, 5.8)$  & $(76.8,\,-15.2)$\\
$\left\langle \cos^2 \gamma \right\rangle_{\rm{N}}$
&  $0.78$ & $0.61$  & $0.91$ & $0.64$ \\
$\left\langle \cos^2 \gamma \right\rangle_{\rm{S}}$
&  $0.80$ & $0.71$  &  $0.95$ & $0.79$ \\
\\[-1.5ex]
\hline
\end{tabular}
}
\end{table*}

\subsubsection{Goodness of fit}
Our model fits the data with good reduced $\chi^2$.
The values of $\tilde{\chi^2}$ in Table~\ref{tab:Table1} increase with increasing $l_{\rm{max}}$,
from $\tilde{\chi^2} = 0.68$ for $l_{\rm{max}} = 2$ to $\tilde{\chi^2} = 0.78$ for $l_{\rm{max}} = 10$.
The variation is small but systematic. This trend suggests that the increasingly detailed structure of the LB surface that the model captures as $l_{\rm{max}}$ increases does not match structure in the \textit{Planck} polarization maps.
Our models appear to be mainly successful at modeling the magnetic field orientation in the LB shell on very large angular scales, that is, the first few multipoles accounting for the variation from the northern and southern polar caps.

This conclusion is supported by the comparison with the reduced $\chi^2$
obtained when fitting either a model with uniform magnetic field orientation over each polar cap ($\tilde{\chi^2} = 0.58$) or the spheroid model of A18 ($\tilde{\chi^2} = 0.60$) to the same \textit{Planck} PR3 data.
It is interesting to note that the reduced $\chi^2$ is only slightly smaller for the A18 spheroid model ($\tilde{\chi^2} = 0.60$) than for our model with $l_{\rm{max}} = 2$ ($\tilde{\chi^2} = 0.68$),
even though our model has three fewer free parameters.
This is quite satisfying, especially as the best-fit spheroid obtained by A18 poorly matches the LB shape derived from extinction and X-ray data.

For $l_{\rm{max}} = 6$, the reduced $\chi^2$ is smaller when the LB shell is modeled using the LE19 3D extinction map
($\chi^2 = 0.61$) than using the L19 map ($\chi^2 = 0.75$).
This indicates that the former is somewhat favored by
the fit of our magnetic field model to the \textit{Planck} data.
However, this conclusion has to be confronted to the caveats drawn in Sect.~\ref{subsec:comp3Dmap} regarding the use of the LE19 map to model the shape of the LB shell (see also our discussion in Sect.~\ref{subsec:radio_pol}).

The values of the reduced $\chi^2$ obtained for the three tested models of the magnetic field in the solar
neighborhood are all below unity. This suggests an overestimation of the uncertainties entering the log-likelihood that we are maximizing. It is the contribution from the turbulent component of the magnetic field ($\sigma_{Q,U}^{\rm turb}$) that dominates the uncertainty budget. It is, therefore, likely that the degree of turbulence that we adopt for our magnetic field modeling, and which we take from \cite{Van2017}, is globally overestimated. This does not impact the results of our fit, which depend only on the relative weighting of pixels, not on the actual value of the uncertainties.

\subsubsection{Model parameters}
Here, we discuss the best-fit values obtained for our free parameters, namely, the angular coordinates ($l_0$, $b_0$) of the initial magnetic field, $\mathbf{B}_0$, and the Cartesian coordinates ($\delta x$, $\delta y$, $\delta z$) of the explosion center.

The best-fit values of $l_0$
are all $\sim 70^\circ$: the values obtained from the L19 map (for different values of $l_{\rm{max}}$; see Table~\ref{tab:Table1}) lie in the range $\simeq 72^\circ - 73^\circ$, while the value obtained from the LE19 map (for $l_{\rm{max}} = 6$; see Table~\ref{tab:Table2}) is $\simeq 68^\circ$.
These values correspond to a magnetic pitch angle
$\sim 20^\circ$ in the solar neighborhood
($\simeq 17^\circ - 18^\circ$ from the L19 map and $\simeq 22^\circ$ from the LE19 map).
This pitch angle is consistent, within the error bars, both with the A18 value and with the values obtained by \cite{Pel2018b} upon fitting large-scale Galactic magnetic field models to full-sky \textit{Planck} dust polarized emission maps.

The best-fit values of $b_0$
appear to be much more sensitive to the chosen value of $l_{\rm{max}}$ and to the adopted 3D extinction map:
the values obtained from the L19 map lie in the range $\simeq 13^\circ - 17^\circ$, while the value obtained  from the LE19 map is $\simeq 3^\circ$.
These values
indicate that the magnetic field in the solar neighborhood points upwards, crossing the Galactic plane at
a small angle.
For comparison, the spheroid model of A18 fitted to the \textit{Planck} PR2 353~GHz data predicts a magnetic field pointing downwards, with an angle to the Galactic plane $\simeq -16^\circ$.
However, when re-fitting the spheroid model of A18 to the \textit{Planck} PR3 353~GHz data,
we find $b_0 \simeq -2^\circ$, which
is almost half-way between the value reported in A18 and the values obtained here from the L19 map.
Hence, part of the discrepancy
between A18 and the present study
can be attributed to the different data sets
(PR2 versus PR3).

The best-fit position of the explosion center depends quite sensitively on the chosen value of $l_{\rm{max}}$ and on the adopted 3D extinction map.
Moreover, the error bars are always large, especially in the $Y$-direction.
This is because the position of the explosion center cannot be constrained in the direction of the initial magnetic field, $\mathbf{B}_0$ (see end of Sect.~\ref{subsec:GMF_model}), except for the requirement that it be contained within the present-day cavity.
Since $\mathbf{B}_0$ is found to be nearly parallel to the Galactic plane ($b_0$ close to $0^\circ$) and at a small angle to the $Y$-direction ($l_0$ close to $90^\circ$), $\delta y$ is very poorly determined, with error bars comparable to the LB radius in the $Y$-direction.
The error bars on $\delta z$ are smaller,
with the result that
none of the best-fit positions is found close to the Galactic plane ($\delta z = 0$).
Only the solution obtained from the L19 map, with
$l_{\rm{max}} = 6$, is compatible with $|\delta z| \lesssim 50$~pc within the uncertainties.

\subsubsection{Mean orientation of $\mathbf{B}$ and effective polarization fraction $p_0$}
\label{subsec:GMF_compmean}

We compute the mean magnetic field direction for our best-fit models by averaging the Cartesian coordinates of $\mathbf{B}$ over each Galactic polar cap.
For the shape of the LB shell obtained from the L19 3D extinction map,
with $l_{\rm{max}} = 6$,
we find that the mean magnetic field points towards Galactic coordinates $(l, b) = (71^\circ \pm  1.3^\circ, -10.9^\circ \pm 0.1^\circ )$ and $(74^\circ \pm 1.4^\circ, 5.8^\circ \pm 0.7^\circ)$ in the northern and southern polar caps, respectively.
The corresponding coordinates of the mean magnetic field obtained with the LE19 map, still with $l_{\rm{max}} = 6$, are $(l, b) = (72.5^\circ \pm  0.6^\circ, -0.1^\circ \pm 1.9^\circ )$ and $(76.8^\circ \pm 1.4^\circ, -15.2^\circ \pm 2.5^\circ)$ in the northern and southern polar caps, respectively.
The error bars are derived from the results obtained on the mock observations described in Sect.~\ref{subsec:uncertainties_Bfield}.
The mean magnetic field coordinates do not depend much on $l_{\rm{max}}$.
However, they depend significantly on the adopted geometric model (see Table~\ref{tab:Table3}).

The mean dust polarization fraction over the Galactic polar caps involves the product of the effective polarization fraction,
$p_0$, and the mean value $\left\langle \cos^2\gamma \right\rangle$, where $\gamma$ is the angle between $\mathbf{B}$ and the plane of the sky.
Thus, in the fit to the data, the value of $p_0$ depends on $\left\langle \cos^2\gamma \right\rangle$.
In our best-fit model based on the L19 extinction map, with $l_{\rm{max}} = 6$, the value averaged over both polar caps is $\left\langle \cos^2 \gamma \right\rangle = 0.93$, with a small scatter,
which implies that the mean magnetic field has a small angle to the plane of the sky ($\gamma \approx 15^\circ$).
For comparison, in the best-fit model based on the LE19 extinction map, still with $l_{\rm{max}} = 6$,
and in the "uniform" and A18 spheroid models applied to the \textit{Planck} PR3 data,
$\left\langle \cos^2 \gamma \right\rangle = 0.715$, $0.79$, and $0.66$, respectively,
which means that the mean magnetic field is more significantly tilted to the plane of the sky
($\gamma \approx 32^\circ$, $27^\circ$, and $36^\circ$, respectively).
These differences in the averaged value of $\left\langle \cos^2 \gamma \right\rangle$ explain the differences in the value of $p_0$.
It is interesting to note
that the product $p_0 \, \left\langle \cos^2 \gamma \right\rangle$ is roughly conserved between the
different models listed in Table~\ref{tab:Table3}.

\subsubsection{Comparison with radio polarization data}
\label{subsec:radio_pol}
The analysis of the \textit{Parkes} survey of the southern sky ($\delta < 20^\circ$) at frequencies of 300–480\,MHz (\citealt{Wol2019}) presented by \cite{Dic2019} provides an additional constraint, which may be used to discriminate between the different models listed in Table~\ref{tab:Table3}.
The intensity-weighted mean of the Faraday depth 
$\bar{\Phi}$, has average values $\langle \bar{\Phi} \rangle_{\rm{N}}=1.0\,$rad~m$^{-2}$ and $\langle \bar{\Phi} \rangle_{\rm{S}}=1.7\,$rad~m$^{-2}$ over
the northern and southern polar caps, respectively, with an uncertainty $\sim 0.3\,$rad~m$^{-2}$ (see Fig.~8 in \citealt{Dic2019}).
We note that these values have the same signs as the corresponding rotation measures (RMs), 
$\langle {\rm RM} \rangle_{\rm{N}}=5.4\pm 0.2\,$rad~m$^{-2}$ and $\langle {\rm RM} \rangle_{\rm{S}}=4.5\pm 0.2 \,$rad~m$^{-2}$  (see Fig.~11 in \citealt{Dic2019}), measured in the map of Galactic RMs built from observations of extragalactic sources by \cite{Oppermann15}. The positive signs of the two sets of values indicate that the line-of-sight magnetic field points towards us from both the North and South Galactic poles.
This is in agreement with the results obtained here from the L19 map, 
since we find that the magnetic field points towards negative latitudes for the northern polar cap ($b_{\rm N} < 0^\circ$) and to positive latitudes for the southern cap ($b_{\rm S} > 0^\circ$).
The positive signs of $\langle\bar{\Phi}\rangle_{\rm{N}}$ and $\langle\bar{\Phi}\rangle_{\rm{S}}$ are also in agreement with the predictions of the "uniform" model, whereas they are in opposition to the signs predicted by both the A18 spheroid model and the shell model based on the LE19 map.

Within the simple model presented by \cite{Sok1998}, the mean Faraday depth, towards 
either polar cap is given by
$|\bar{\Phi}|= \frac{1}{2} K \, {\rm DM} \, |B_Z|$, where $K =0.81 \, {\rm{rad \, m^{-2} \, cm^{3} \,}}\mu{\rm{G^{-1} \, pc^{-1}}}$, ${\rm DM}$ is the dispersion measure, and $B_Z$ is the magnetic field component along the $Z$-axis perpendicular to the Galactic plane. This is an approximative model, which we use for qualitative comparison. For example, the ratio between the full RM through the Galaxy and the mean Faraday depth is 2 in the model, while the observed ratio is $\sim 4$.
From pulsar measurements, we know that ${\rm DM} \simeq 26\, {\rm{pc \, cm^{-3}}}$ (\citealt{Gae2008}), and thus
$|\bar{\Phi}$(rad~m$^{-2})| \simeq 10.5 \, |B_Z(\mu$G$)|$.
To match the observed values of $\langle\bar{\Phi}\rangle_{\rm{N}}$ and $\langle\bar{\Phi}\rangle_{\rm{S}}$, $|B_Z|$ must be $\sim 0.15\,\mu$G, that is, a small fraction of the total strength (a few $\mu$G, \citealt{Bec2003}) of the mean magnetic field in the solar neighborhood.
In other words, the magnetic field in each polar cap must be nearly in the plane of the sky,
that is, the angle $\gamma$ must be small.
This, too, is in agreement with the results obtained from the L19 map,
which lead to values of $\left\langle \cos^2 \gamma \right\rangle_{\rm{N}}$ and $\left\langle \cos^2 \gamma \right\rangle_{\rm{S}}$ close to 1.
The agreement is not as good  with the predictions of the other models listed in Table~\ref{tab:Table3}, which give smaller values of $\left\langle \cos^2 \gamma \right\rangle_{\rm{N}}$ and $\left\langle \cos^2 \gamma \right\rangle_{\rm{S}}$.
Altogether, our comparison with the mean Faraday depths derived by \cite{Dic2019} leads us to give preference to the shell model based on the L19 map.

\section{Summary \& Perspective}
\label{sec:end}
In this paper, we pursued a physically motivated approach to model interstellar polarization data at high Galactic latitudes in a self-consistent way, which involved modeling the magnetic field in that part of the sky.
We relied on observational evidence showing that the dust polarized emission is dominated by a nearby dusty and magnetized structure.
Associating this structure with the shell of the Local Bubble (LB) led us to model the magnetic field in the LB shell, for which polarization data can be obtained.
We used a dedicated analytical model that relates the structure of the magnetic field in the LB shell to the shell geometry and we extracted the latter from actual 3D maps of the local dusty ISM.

This work is, therefore, two-fold. First, we extracted and modeled the geometry of the LB shell.
Second, we inserted the resulting shell model into our analytical model of the magnetic field in the shell, which we then fitted to the \textit{Planck} PR3 353~GHz polarization data at high Galactic latitudes. The key steps of this process and the main results are summarized below.

\smallskip

The first part of the paper focuses on the shape of the LB shell.
We first we developed an ad-hoc method to extract the LB shell from 3D extinction maps
through the identification of its inner and outer surfaces.
We applied this method to the most recent 3D extinction maps that
cover the full sky, namely, the maps from \cite{Lal2019} (L19) and from \cite{Lei2019} (LE19).
We then expanded the shell inner surface in spherical harmonics, up to a variable maximum multipole degree, $l_{\rm{max}}$, which enabled us to model its shape with the desired level of complexity.
We compared the results obtained from the two extinction maps
both with each other and with
the shape of the Local Hot Bubble derived by \cite{Liu2017} from X-ray emission data.
Both comparisons indicated overall agreement, with, however, noticeable differences in the details.
The overall agreement between the LB shapes inferred from extinction maps and from X-ray emission data lends credence to our multi-phase view of the LB.
We hope that our modeling of the dusty LB and the agreement found with the Local Hot Bubble will motivate further investigations and will help understand and physically model the multi-phase ISM in the solar vicinity.

\smallskip

The second part of the paper focuses on the magnetic field in the LB shell.
Following up on the work of \cite{Alv2018} (A18), we used the shape of the shell inner surface derived in the first part of the paper as input to the A18 analytical model of the magnetic field in the LB shell.
Fixing the shape of the shell inner surface left us with only five free parameters:
the three Cartesian coordinates of the explosion center and the two angular coordinates of the initial magnetic field.
Using a MCMC method, we fitted our magnetic field model to the dust polarized emission at high Galactic latitudes ($|b|>60^{\circ}$)
as measured by \textit{Planck}.
The fit was performed for different values of $l_{\rm{max}}$.

Our model fits the \textit{Planck} data with good reduced $\chi^2$, close to the value obtained for the spheroid model of A18,
which however has three more free parameters.
Moreover, the best-fit spheroid of A18 does not match the observed geometry of the LB, whereas our
modeled shell, which is directly derived from 3D extinction data, automatically does.

The derived orientation of the initial magnetic field, $\mathbf{B_0}$, appears to be stable for the different models
of the shell inner surface that we tested  (models with different values of $l_{\rm{max}}$) or different underlying extinction maps).
It is also consistent with models of the large-scale Galactic magnetic field.
However, our models of the present-day magnetic field in the shell show more complexity in their geometry than those large-scale Galactic magnetic field models, including a North-South asymmetry.
The position of the explosion center is only constrained in a plane perpendicular to $\mathbf{B_0}$,
and even there it is only loosely constrained.
None of the best-fit positions are found within less than $\sim 100$~pc from the Galactic plane.
It is unclear whether this is clashes with existing models of the origin of the LB.

\smallskip

For the shell model derived from the L19 extinction map, the mean magnetic field in each polar cap
has a small angle to the plane of the sky,
and (assuming $l_0 \in [0^\circ,\,180^\circ]$) its line-of-sight component points towards us. Both results are
in agreement with Faraday spectra of the Galactic diffuse synchrotron emission
from the nearby ISM.
The corresponding results obtained for either the shell model derived from the LE19 map, the "uniform" model, or the A18 spheroid model (see Table~\ref{tab:Table3}) do not show the same agreement with the Faraday spectra. 
It is, therefore, the LB shell model based on the L19 map that we consider to be the best.

\smallskip

We further investigated the sources of uncertainty in our approach to model the dust polarized emission in the Galactic polar caps. We considered the impact on our fits of the \textit{Planck} residual systematics at 353~GHz and of the choice of the 3D extinction map used to extract and model the shape of the LB shell.
We found that the largest uncertainty in our modeling came from the choice of the 3D extinction map, which, at this stage, was found to strongly depend on the
underlying data set.
Significant improvements are expected in that research area.

\smallskip

In principle, the modeling of the magnetic field in the LB shell and its contribution to the dust polarized sky can be evaluated for the full sky, corresponding to a first layer of the Galactic dust polarized foregrounds.
In future studies, we will extend our modeling towards lower Galactic latitudes, where it will be required to connect the local magnetic field to the large-scale Galactic magnetic field.
In a way, with this paper, we are setting the stage for the next generation of Galactic magnetic field models which would integrate external data sets or specific models derived from them. As a consequence, we expect this paper and our results to be useful both for modeling the local ISM as traced by its different components and modeling the Galactic dust polarized emission, a long-awaited input for studies of the polarized foregrounds to the CMB and to the optical polarization of extragalactic sources (\citealt{Pel19}).

\begin{acknowledgements}
We would like to warmly thank Marta Alves for all the inspiring discussions and motivations that took place during the early phases of this project. We thank Kostas Tassis for comments that help improving the presentation of the work. We also thank J. van Loon for his fruitful review.
This work has received funding from the European Research Council (ERC) under the European Unions Horizon 2020 research and innovation programme under grant agreement No. 771282, and by the Agence Nationale de la Recherche (project BxB: ANR-17-CE31-0022).
We acknowledge the use of data from the Planck/ESA mission,
downloaded from the Planck Legacy Archive, and of the Legacy
Archive for Microwave Background Data Analysis (LAMBDA).
Support for LAMBDA is provided by the NASA Office of Space
Science. Some of the results in this paper have been derived
using the HEALPix (G{\'o}rski et al. 2005) package.
\end{acknowledgements}

%
\bibliographystyle{aa} 
\interlinepenalty=10000
\bibliography{myLBbiblio} 

\begin{thebibliography}{49}
\expandafter\ifx\csname natexlab\endcsname\relax\def\natexlab#1{#1}\fi

\bibitem[{{Alves} {et~al.}(2018){Alves}, {Boulanger}, {Ferri{\`e}re}, \&
  {Montier}}]{Alv2018}
{Alves}, M.~I.~R., {Boulanger}, F., {Ferri{\`e}re}, K., \& {Montier}, L. 2018,
  \aap, 611, L5

\bibitem[{{Andrae} {et~al.}(2018){Andrae}, {Fouesneau}, {Creevey}, {Ordenovic},
  {Mary}, {Burlacu}, {Chaoul}, {Jean-Antoine-Piccolo}, {Kordopatis}, {Korn},
  {Lebreton}, {Panem}, {Pichon}, {Th{\'e}venin}, {Walmsley}, \&
  {Bailer-Jones}}]{And2018}
{Andrae}, R., {Fouesneau}, M., {Creevey}, O., {et~al.} 2018, \aap, 616, A8

\bibitem[{{Beck} {et~al.}(2003){Beck}, {Shukurov}, {Sokoloff}, \&
  {Wielebinski}}]{Bec2003}
{Beck}, R., {Shukurov}, A., {Sokoloff}, D., \& {Wielebinski}, R. 2003, \aap,
  411, 99

\bibitem[{{Berdyugin} {et~al.}(2014){Berdyugin}, {Piirola}, \&
  {Teerikorpi}}]{Ber2014}
{Berdyugin}, A., {Piirola}, V., \& {Teerikorpi}, P. 2014, \aap, 561, A24

\bibitem[{{Breitschwerdt} {et~al.}(2016){Breitschwerdt}, {Feige}, {Schulreich},
  {Avillez}, {Dettbarn}, \& {Fuchs}}]{Bre2016}
{Breitschwerdt}, D., {Feige}, J., {Schulreich}, M.~M., {et~al.} 2016, \nat,
  532, 73

\bibitem[{{Capitanio} {et~al.}(2017){Capitanio}, {Lallement}, {Vergely},
  {Elyajouri}, \& {Monreal-Ibero}}]{Cap2017}
{Capitanio}, L., {Lallement}, R., {Vergely}, J.~L., {Elyajouri}, M., \&
  {Monreal-Ibero}, A. 2017, \aap, 606, A65

\bibitem[{{Chen} {et~al.}(2019){Chen}, {Huang}, {Yuan}, {Wang}, {Fan}, {Xiang},
  {Zhang}, {Tian}, \& {Liu}}]{Che2019}
{Chen}, B.~Q., {Huang}, Y., {Yuan}, H.~B., {et~al.} 2019, \mnras, 483, 4277

\bibitem[{{Cox} \& {Reynolds}(1987)}]{Cox1987}
{Cox}, D.~P. \& {Reynolds}, R.~J. 1987, \araa, 25, 303

\bibitem[{{de Avillez} \& {Breitschwerdt}(2009)}]{deA2009}
{de Avillez}, M.~A. \& {Breitschwerdt}, D. 2009, \apjl, 697, L158

\bibitem[{{Dickey} {et~al.}(2019){Dickey}, {Landecker}, {Thomson}, {Wolleben},
  {Sun}, {Carretti}, {Douglas}, {Fletcher}, {Gaensler}, {Gray}, {Haverkorn},
  {Hill}, {Mao}, \& {McClure-Griffiths}}]{Dic2019}
{Dickey}, J.~M., {Landecker}, T.~L., {Thomson}, A. J.~M., {et~al.} 2019, \apj,
  871, 106

\bibitem[{{Farhang} {et~al.}(2019){Farhang}, {van Loon}, {Khosroshahi},
  {Javadi}, \& {Bailey}}]{Far2019}
{Farhang}, A., {van Loon}, J.~T., {Khosroshahi}, H.~G., {Javadi}, A., \&
  {Bailey}, M. 2019, Nature Astronomy, 3, 922

\bibitem[{{Ferri{\`e}re}(2015)}]{Fer2015}
{Ferri{\`e}re}, K. 2015, in Journal of Physics Conference Series, Vol. 577,
  Journal of Physics Conference Series, 012008

\bibitem[{{Foreman-Mackey} {et~al.}(2013){Foreman-Mackey}, {Conley},
  {Meierjurgen Farr}, {Hogg}, {Lang}, {Marshall}, {Price-Whelan}, {Sanders}, \&
  {Zuntz}}]{For2013}
{Foreman-Mackey}, D., {Conley}, A., {Meierjurgen Farr}, W., {et~al.} 2013,
  {emcee: The MCMC Hammer}, Astrophysics Source Code Library

\bibitem[{{Frisch} {et~al.}(2012){Frisch}, {Andersson}, {Berdyugin}, {Piirola},
  {DeMajistre}, {Funsten}, {Magalhaes}, {Seriacopi}, {McComas}, {Schwadron},
  {Slavin}, \& {Wiktorowicz}}]{Fri2012}
{Frisch}, P.~C., {Andersson}, B.~G., {Berdyugin}, A., {et~al.} 2012, \apj, 760,
  106

\bibitem[{{Gaensler} {et~al.}(2008){Gaensler}, {Madsen}, {Chatterjee}, \&
  {Mao}}]{Gae2008}
{Gaensler}, B.~M., {Madsen}, G.~J., {Chatterjee}, S., \& {Mao}, S.~A. 2008,
  \pasa, 25, 184

\bibitem[{{Galeazzi} {et~al.}(2014){Galeazzi}, {Chiao}, {Collier}, {Cravens},
  {Koutroumpa}, {Kuntz}, {Lallement}, {Lepri}, {McCammon}, {Morgan}, {Porter},
  {Robertson}, {Snowden}, {Thomas}, {Uprety}, {Ursino}, \& {Walsh}}]{Gal2014}
{Galeazzi}, M., {Chiao}, M., {Collier}, M.~R., {et~al.} 2014, \nat, 512, 171

\bibitem[{{Gelman} \& {Rubin}(1992)}]{Gel1992}
{Gelman}, A. \& {Rubin}, D.~B. 1992, Statistical Science, 7, 457

\bibitem[{{Gontcharov} \& {Mosenkov}(2019)}]{Gon2019}
{Gontcharov}, G.~A. \& {Mosenkov}, A.~V. 2019, \mnras, 483, 299

\bibitem[{{Goodman} \& {Weare}(2010)}]{Goo2010}
{Goodman}, J. \& {Weare}, J. 2010, Communications in Applied Mathematics and
  Computational Science, Vol.~5, No.~1, p.~65-80, 2010, 5, 65

\bibitem[{{G{\'o}rski} {et~al.}(2005){G{\'o}rski}, {Hivon}, {Banday}, {Wand
  elt}, {Hansen}, {Reinecke}, \& {Bartelmann}}]{Gor2005}
{G{\'o}rski}, K.~M., {Hivon}, E., {Banday}, A.~J., {et~al.} 2005, \apj, 622,
  759

\bibitem[{{Green} {et~al.}(2019){Green}, {Schlafly}, {Zucker}, {Speagle}, \&
  {Finkbeiner}}]{Gre2019}
{Green}, G.~M., {Schlafly}, E., {Zucker}, C., {Speagle}, J.~S., \&
  {Finkbeiner}, D. 2019, \apj, 887, 93

\bibitem[{{Gry} \& {Jenkins}(2014)}]{Gry2014}
{Gry}, C. \& {Jenkins}, E.~B. 2014, \aap, 567, A58

\bibitem[{{Heiles}(1998)}]{Hei1998}
{Heiles}, C. 1998, {The magnetic field near the local bubble}, ed.
  D.~{Breitschwerdt}, M.~J. {Freyberg}, \& J.~{Truemper}, Vol. 506, 227

\bibitem[{{Lallement} {et~al.}(2019){Lallement}, {Babusiaux}, {Vergely},
  {Katz}, {Arenou}, {Valette}, {Hottier}, \& {Capitanio}}]{Lal2019}
{Lallement}, R., {Babusiaux}, C., {Vergely}, J.~L., {et~al.} 2019, \aap, 625,
  A135

\bibitem[{{Lallement} {et~al.}(2018){Lallement}, {Capitanio}, {Ruiz-Dern},
  {Danielski}, {Babusiaux}, {Vergely}, {Elyajouri}, {Arenou}, \&
  {Leclerc}}]{Lal2018}
{Lallement}, R., {Capitanio}, L., {Ruiz-Dern}, L., {et~al.} 2018, \aap, 616,
  A132

\bibitem[{{Lallement} {et~al.}(2014){Lallement}, {Vergely}, {Valette},
  {Puspitarini}, {Eyer}, \& {Casagrande}}]{Lal2014}
{Lallement}, R., {Vergely}, J.~L., {Valette}, B., {et~al.} 2014, \aap, 561, A91

\bibitem[{{Lee} \& {Draine}(1985)}]{Lee85}
{Lee}, H.~M. \& {Draine}, B.~T. 1985, \apj, 290, 211

\bibitem[{{Leike} \& {En{\ss}lin}(2019)}]{Lei2019}
{Leike}, R.~H. \& {En{\ss}lin}, T.~A. 2019, \aap, 631, A32

\bibitem[{{Leroy}(1999)}]{Ler1999}
{Leroy}, J.~L. 1999, \aap, 346, 955

\bibitem[{{Liu} {et~al.}(2017){Liu}, {Chiao}, {Collier}, {Cravens}, {Galeazzi},
  {Koutroumpa}, {Kuntz}, {Lallement}, {Lepri}, {McCammon}, {Morgan}, {Porter},
  {Snowden}, {Thomas}, {Uprety}, {Ursino}, \& {Walsh}}]{Liu2017}
{Liu}, W., {Chiao}, M., {Collier}, M.~R., {et~al.} 2017, \apj, 834, 33

\bibitem[{{Ma{\'\i}z-Apell{\'a}niz}(2001)}]{Mai2001}
{Ma{\'\i}z-Apell{\'a}niz}, J. 2001, \apjl, 560, L83

\bibitem[{{Oppermann} {et~al.}(2015){Oppermann}, {Junklewitz}, {Greiner},
  {En{\ss}lin}, {Akahori}, {Carretti}, {Gaensler}, {Goobar}, {Harvey-Smith},
  {Johnston-Hollitt}, {Pratley}, {Schnitzeler}, {Stil}, \&
  {Vacca}}]{Oppermann15}
{Oppermann}, N., {Junklewitz}, H., {Greiner}, M., {et~al.} 2015, \aap, 575,
  A118

\bibitem[{{Pelgrims}(2019)}]{Pel19}
{Pelgrims}, V. 2019, \aap, 622, A145

\bibitem[{{Pelgrims} \& {Mac{\'\i}as-P{\'e}rez}(2018)}]{Pel2018b}
{Pelgrims}, V. \& {Mac{\'\i}as-P{\'e}rez}, J.~F. 2018, arXiv e-prints,
  arXiv:1807.10516

\bibitem[{{Pelgrims} {et~al.}(2018){Pelgrims}, {Mac{\'\i}as-P{\'e}rez}, \&
  {Ruppin}}]{Pel2018}
{Pelgrims}, V., {Mac{\'\i}as-P{\'e}rez}, J.~F., \& {Ruppin}, F. 2018, arXiv
  e-prints, arXiv:1807.10515

\bibitem[{{Planck Collaboration III}(2018)}]{PlaIII2018}
{Planck Collaboration III}. 2018, arXiv e-prints, arXiv:1807.06207

\bibitem[{{Planck Collaboration Int. XLIV}(2016)}]{PlaXLIV2016}
{Planck Collaboration Int. XLIV}. 2016, \aap, 596, A105

\bibitem[{{Planck Collaboration XI}(2018)}]{PlaXI2018}
{Planck Collaboration XI}. 2018, arXiv e-prints, arXiv:1801.04945

\bibitem[{{Planck Collaboration XII}(2018)}]{PlaXII2018}
{Planck Collaboration XII}. 2018, arXiv e-prints, arXiv:1807.06212

\bibitem[{{Planck Collaboration XX}(2015)}]{PlaXX2015}
{Planck Collaboration XX}. 2015, \aap, 576, A105

\bibitem[{{Planck Collaboration XXX}(2016)}]{PlaXXX2016}
{Planck Collaboration XXX}. 2016, \aap, 586, A133

\bibitem[{{Puspitarini} {et~al.}(2014){Puspitarini}, {Lallement}, {Vergely}, \&
  {Snowden}}]{Pus2014}
{Puspitarini}, L., {Lallement}, R., {Vergely}, J.~L., \& {Snowden}, S.~L. 2014,
  \aap, 566, A13

\bibitem[{{Remazeilles} {et~al.}(2011){Remazeilles}, {Delabrouille}, \&
  {Cardoso}}]{Rem2011}
{Remazeilles}, M., {Delabrouille}, J., \& {Cardoso}, J.-F. 2011, \mnras, 418,
  467

\bibitem[{{Santos} {et~al.}(2011){Santos}, {Corradi}, \& {Reis}}]{San2011}
{Santos}, F.~P., {Corradi}, W., \& {Reis}, W. 2011, \apj, 728, 104

\bibitem[{{Skalidis} \& {Pelgrims}(2019)}]{Ska2019}
{Skalidis}, R. \& {Pelgrims}, V. 2019, \aap, 631, L11

\bibitem[{{Snowden} {et~al.}(2015){Snowden}, {Koutroumpa}, {Kuntz},
  {Lallement}, \& {Puspitarini}}]{Sno2015}
{Snowden}, S.~L., {Koutroumpa}, D., {Kuntz}, K.~D., {Lallement}, R., \&
  {Puspitarini}, L. 2015, \apj, 806, 120

\bibitem[{{Sokoloff} {et~al.}(1998){Sokoloff}, {Bykov}, {Shukurov},
  {Berkhuijsen}, {Beck}, \& {Poezd}}]{Sok1998}
{Sokoloff}, D.~D., {Bykov}, A.~A., {Shukurov}, A., {et~al.} 1998, \mnras, 299,
  189

\bibitem[{{Vansyngel} {et~al.}(2017){Vansyngel}, {Boulanger}, {Ghosh},
  {Wandelt}, {Aumont}, {Bracco}, {Levrier}, {Martin}, \& {Montier}}]{Van2017}
{Vansyngel}, F., {Boulanger}, F., {Ghosh}, T., {et~al.} 2017, \aap, 603, A62

\bibitem[{{Wolleben} {et~al.}(2019){Wolleben}, {Landecker}, {Carretti},
  {Dickey}, {Fletcher}, {McClure-Griffiths}, {McConnell}, {Thomson}, {Hill},
  {Gaensler}, {Han}, {Haverkorn}, {Leahy}, {Reich}, \& {Taylor}}]{Wol2019}
{Wolleben}, M., {Landecker}, T.~L., {Carretti}, E., {et~al.} 2019, \aj, 158, 44

\end{thebibliography}
%

\begin{appendix}
\section{Supplementary material}
\begin{figure*}
\begin{center}
\begin{tabular}{ccc}
& $Q$ & $U$     \\

\rotatebox{90}{ \hspace{4em} $\sigma$} &
\includegraphics[trim={0cm .5cm 0cm .7cm},clip,width=.34\linewidth]{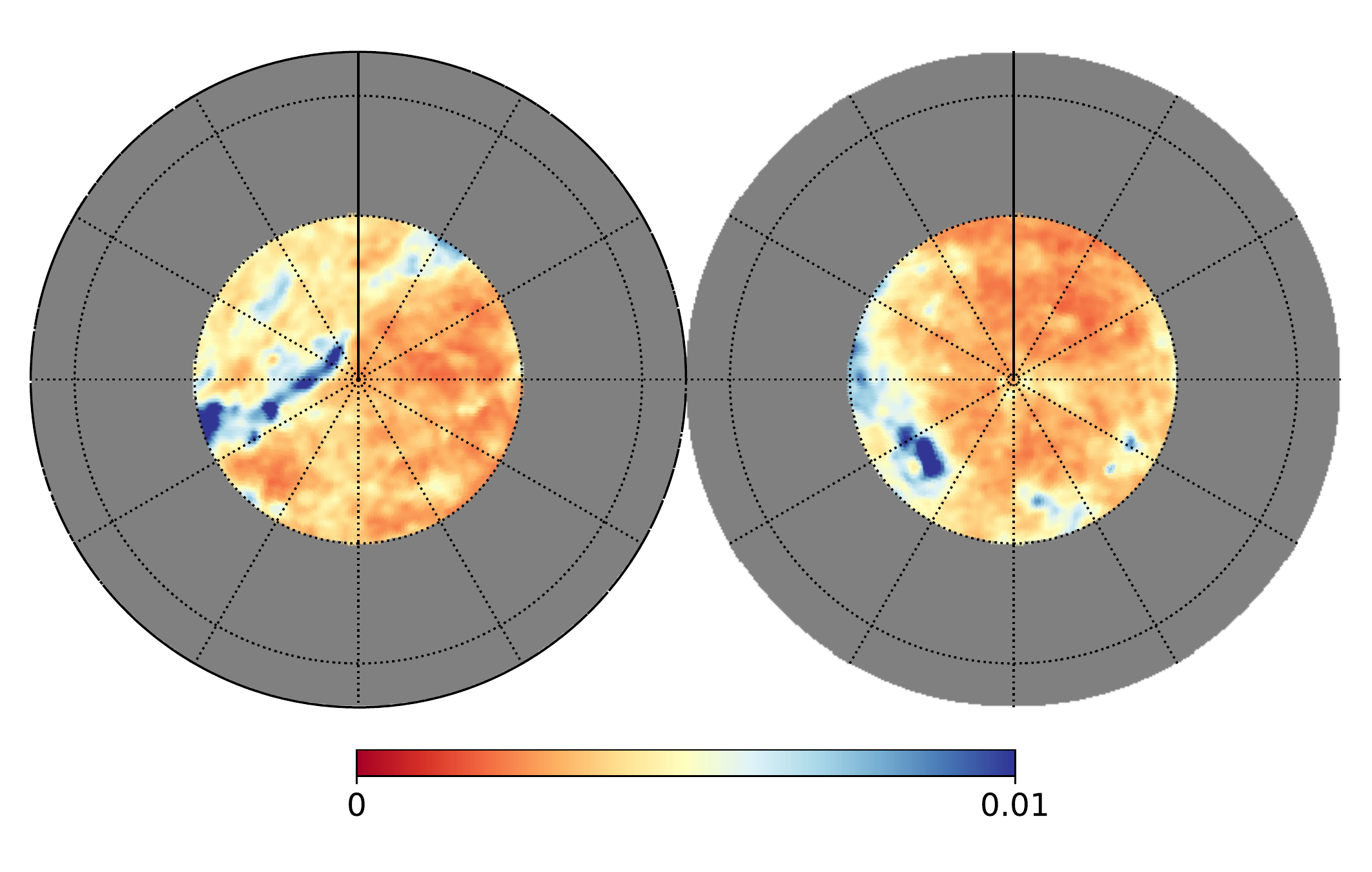}
        &       \includegraphics[trim={0cm .5cm 0cm .7cm},clip,width=.34\linewidth]{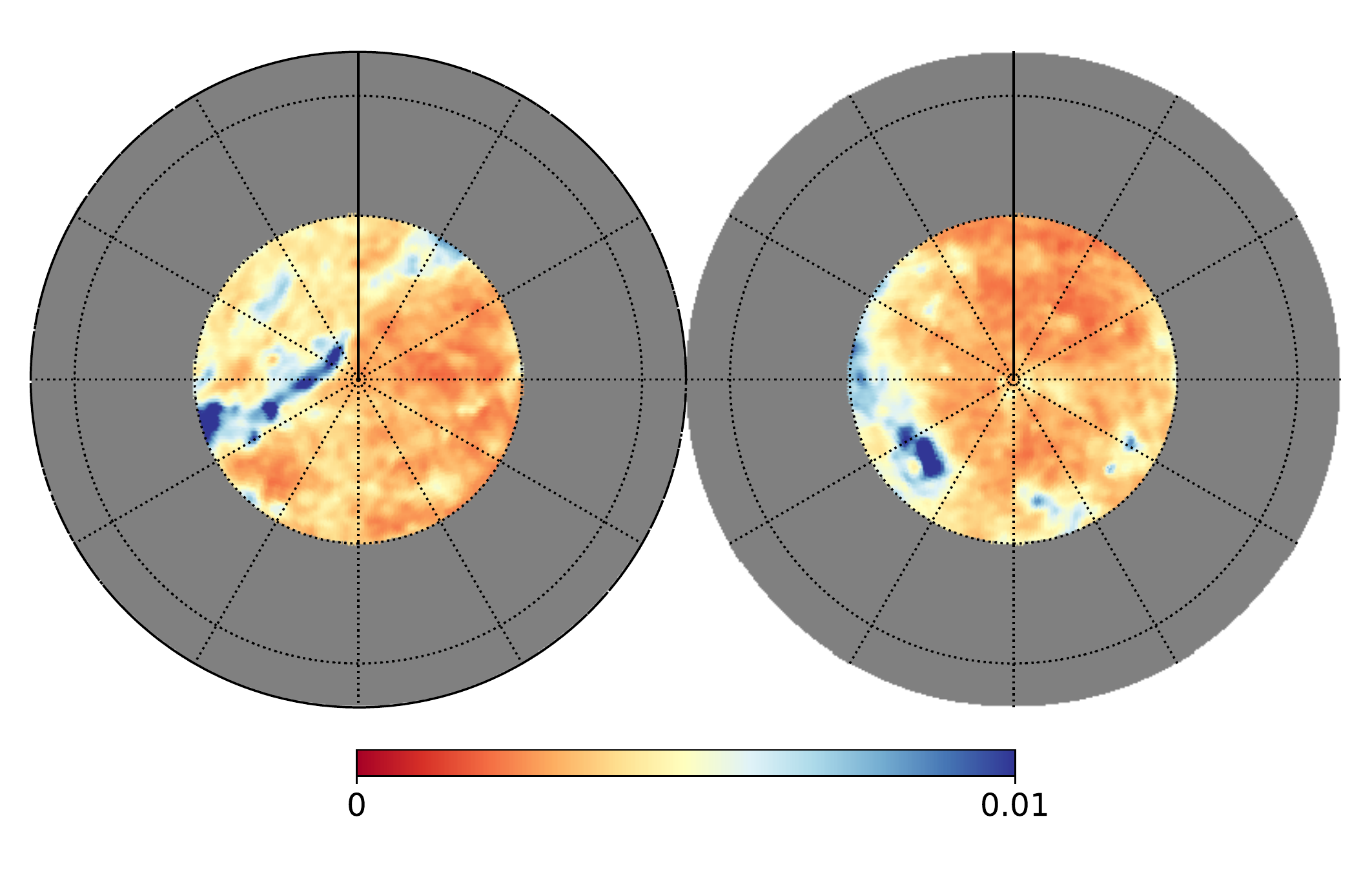}
\\

\rotatebox{90}{ \hspace{4em} $l_{\rm{max}}=2$} &
\includegraphics[trim={0cm .5cm 0cm .7cm},clip,width=.34\linewidth]{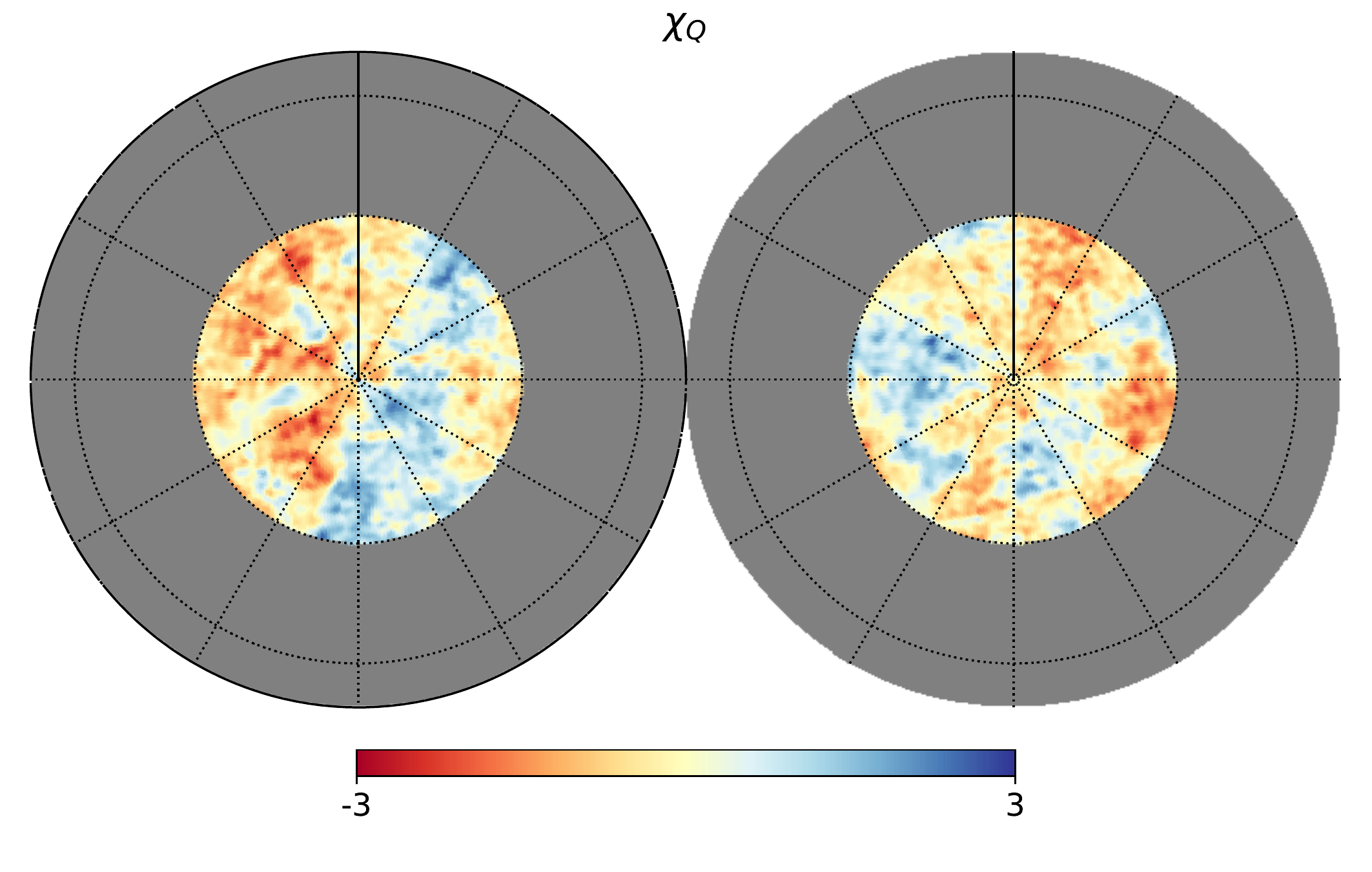}
        &       \includegraphics[trim={0cm .5cm 0cm .7cm},clip,width=.34\linewidth]{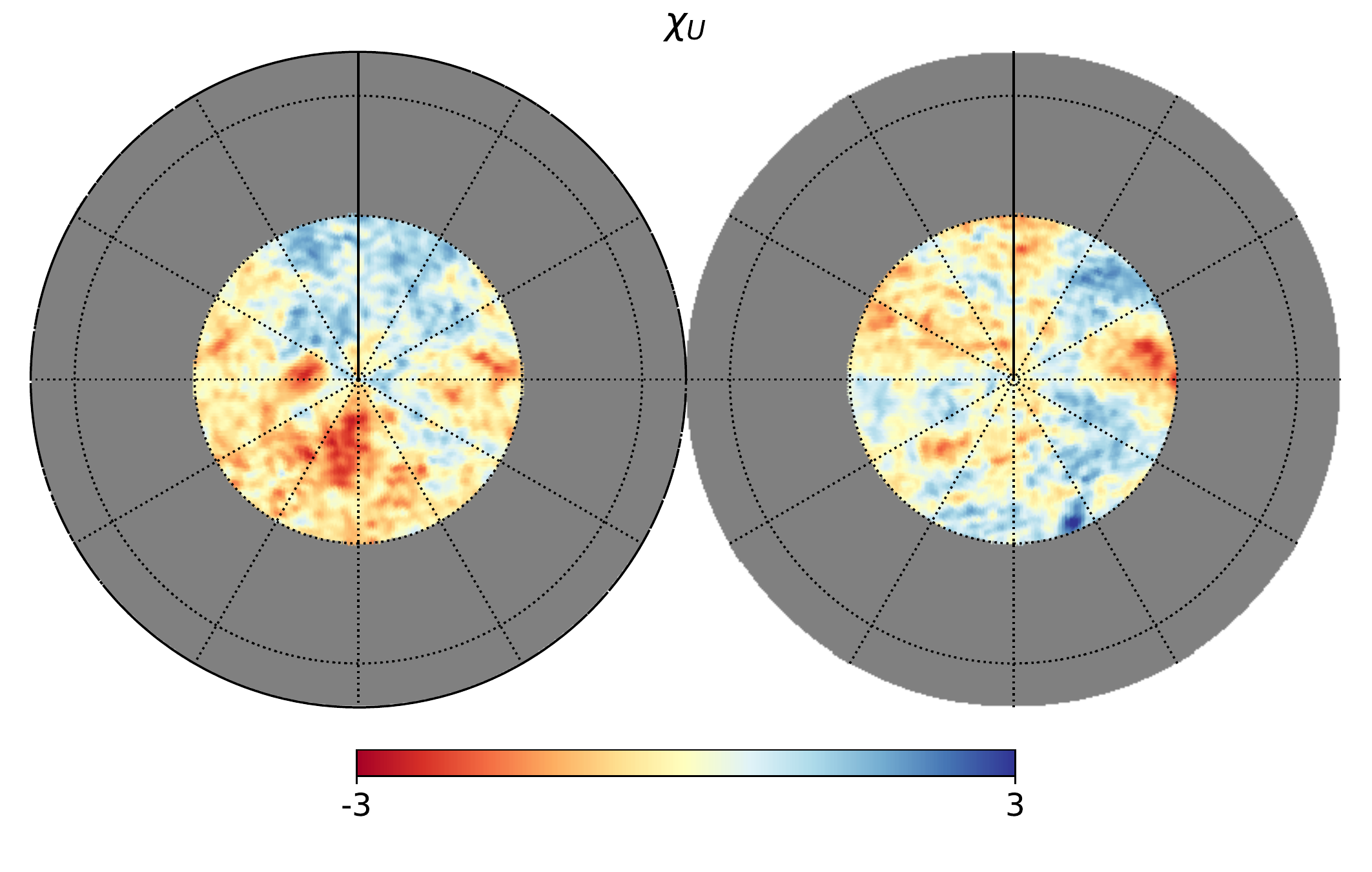}
\\

\rotatebox{90}{ \hspace{4em} $l_{\rm{max}}=4$} &
\includegraphics[trim={0cm .5cm 0cm .7cm},clip,width=.34\linewidth]{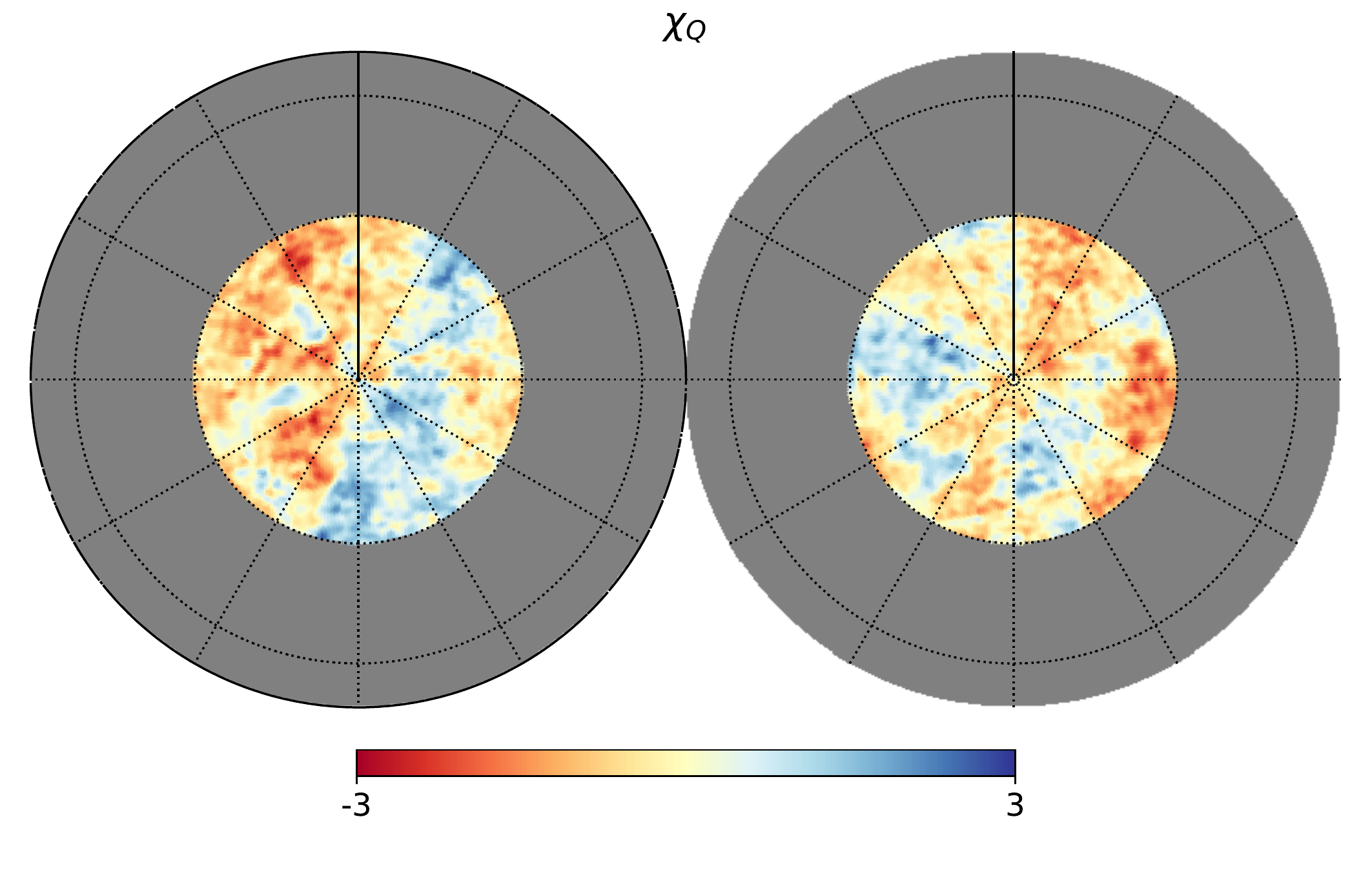}
        &       \includegraphics[trim={0cm .5cm 0cm .7cm},clip,width=.34\linewidth]{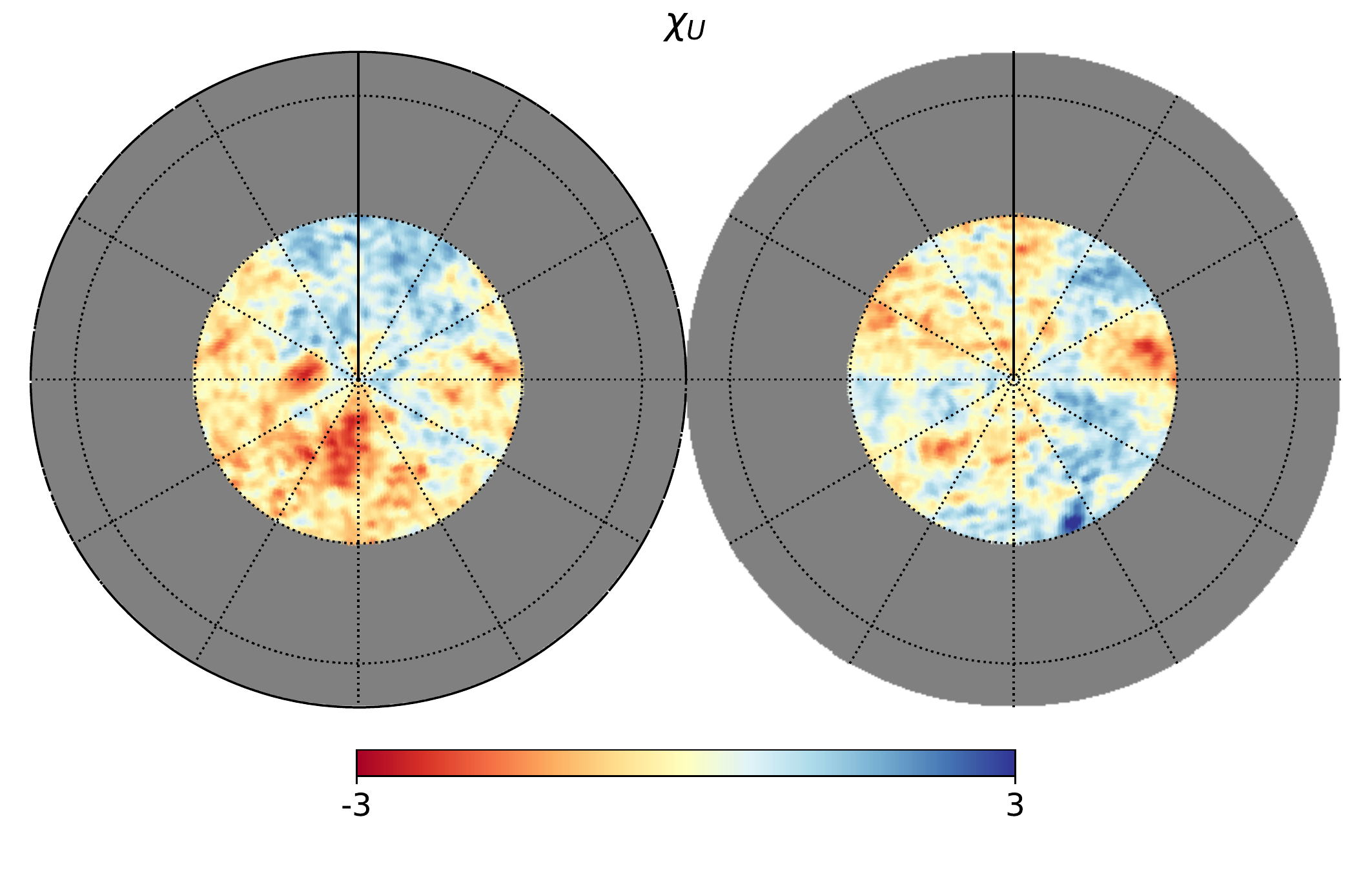}
\\

\rotatebox{90}{ \hspace{4em} $l_{\rm{max}}=6$} &
\includegraphics[trim={0cm .5cm 0cm .7cm},clip,width=.34\linewidth]{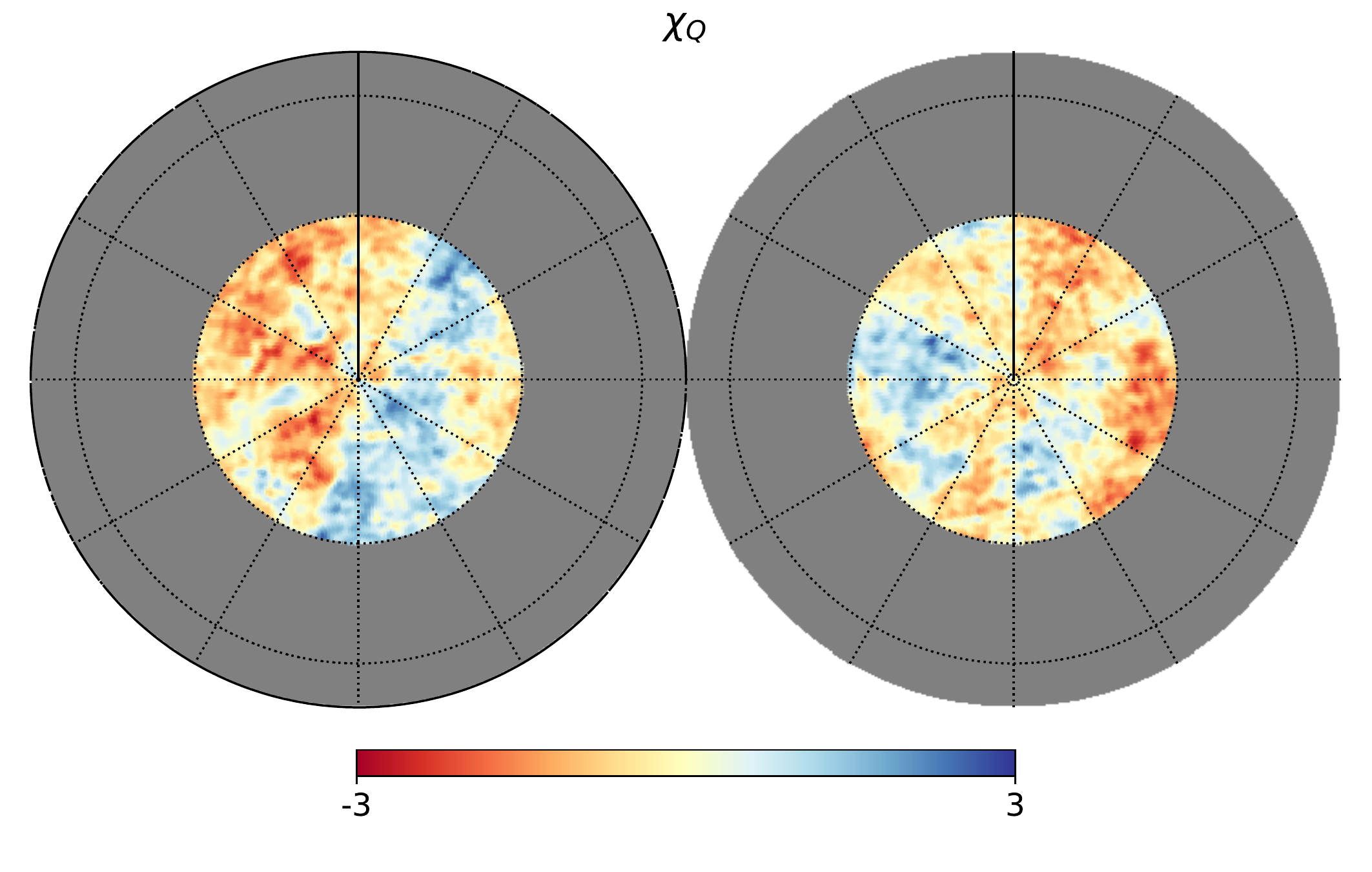}
        &       \includegraphics[trim={0cm .5cm 0cm .7cm},clip,width=.34\linewidth]{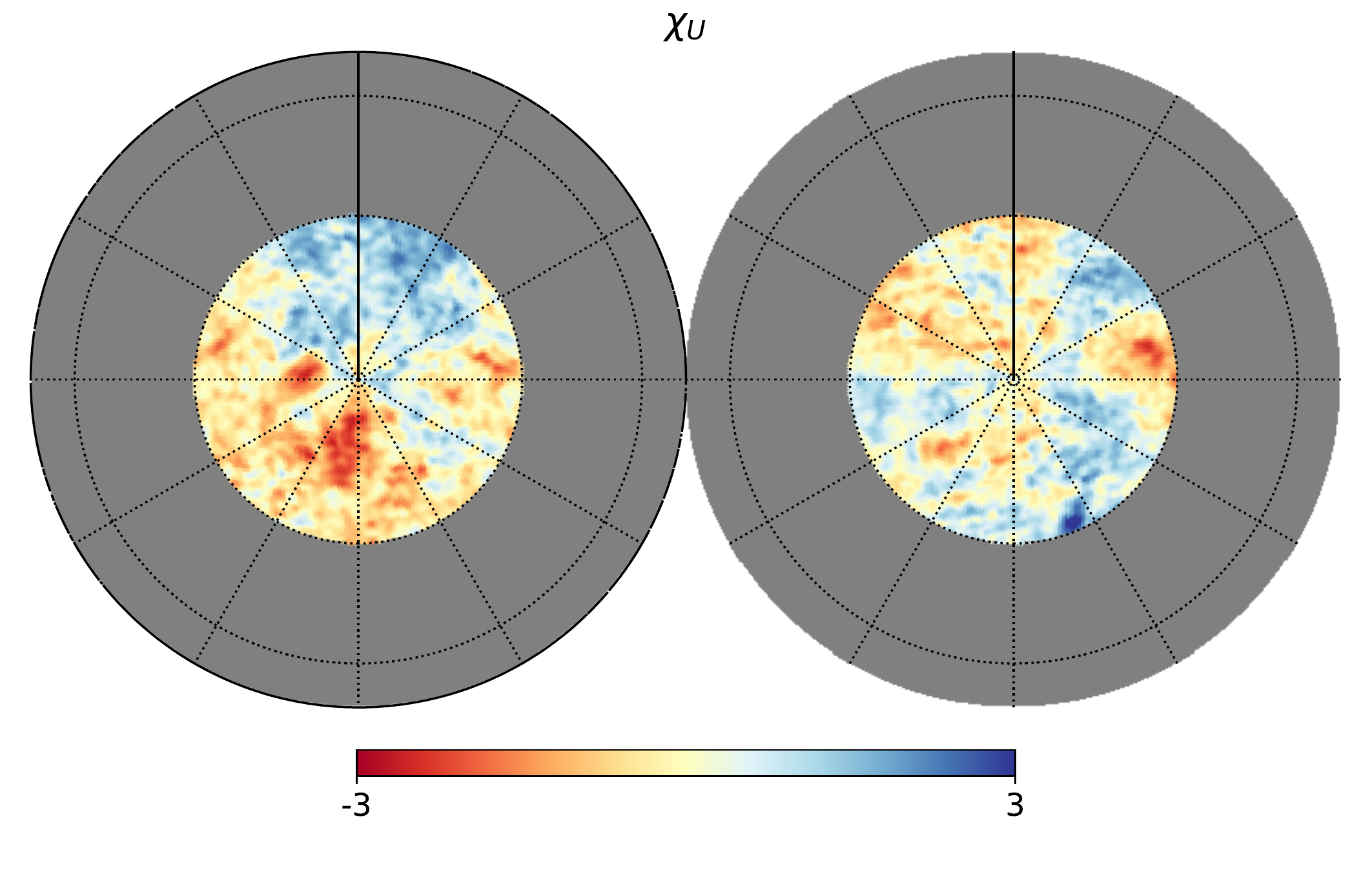}
\\

\rotatebox{90}{ \hspace{4em} $l_{\rm{max}}=8$} &
\includegraphics[trim={0cm .5cm 0cm .7cm},clip,width=.34\linewidth]{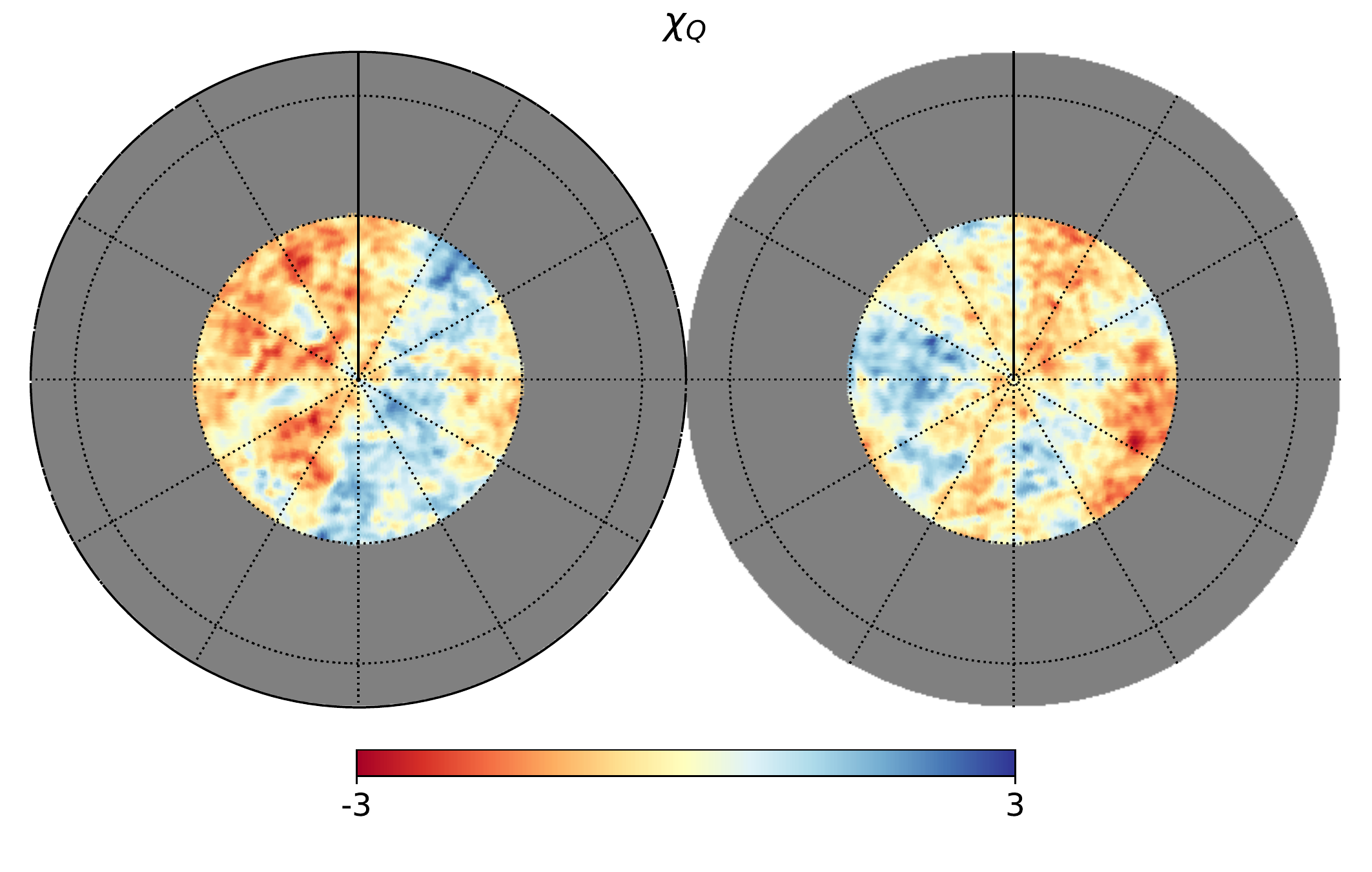}
        &       \includegraphics[trim={0cm .5cm 0cm .7cm},clip,width=.34\linewidth]{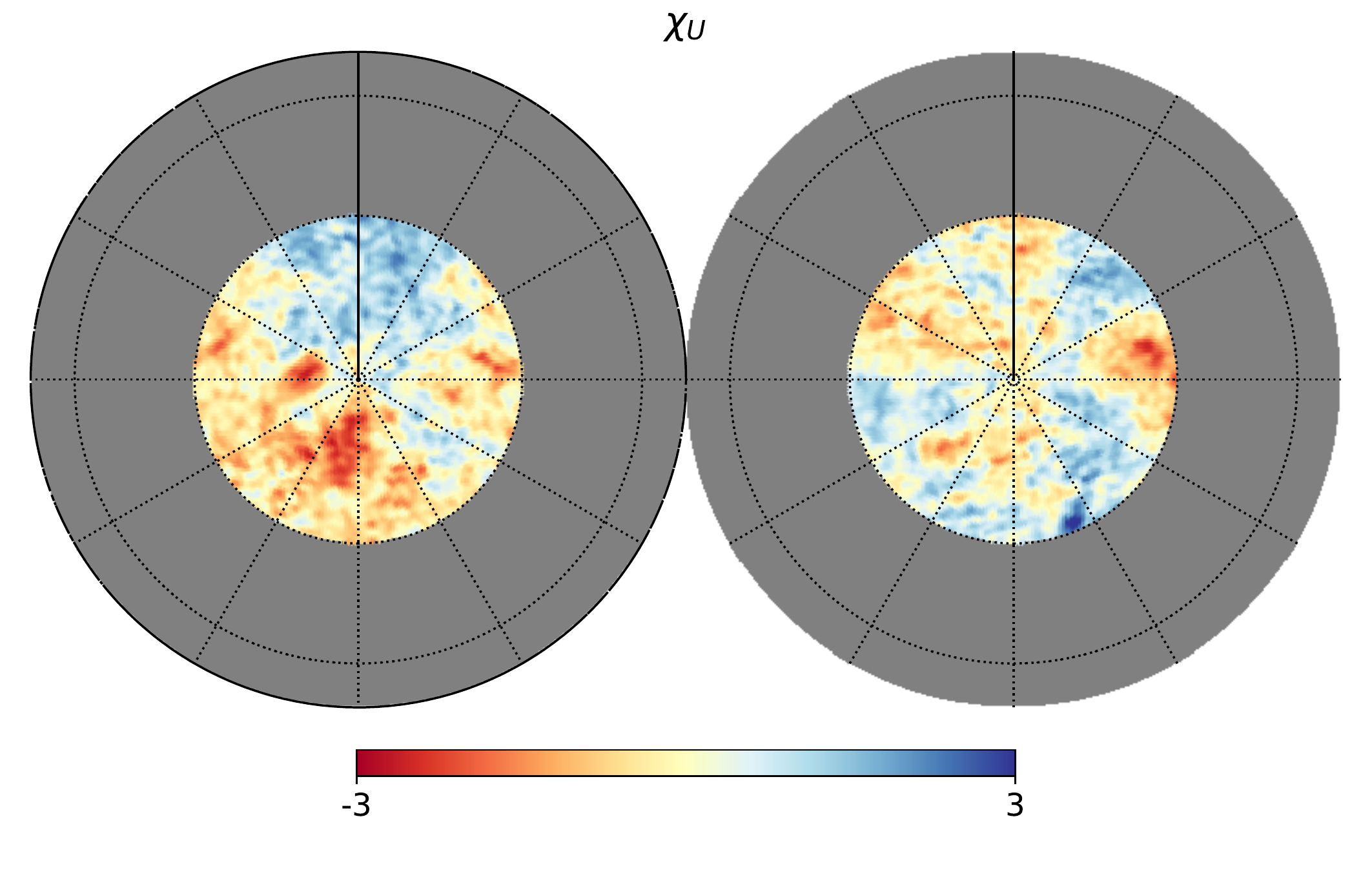}
\\

\rotatebox{90}{ \hspace{4em} $l_{\rm{max}}=10$} &
\includegraphics[trim={0cm .5cm 0cm .7cm},clip,width=.34\linewidth]{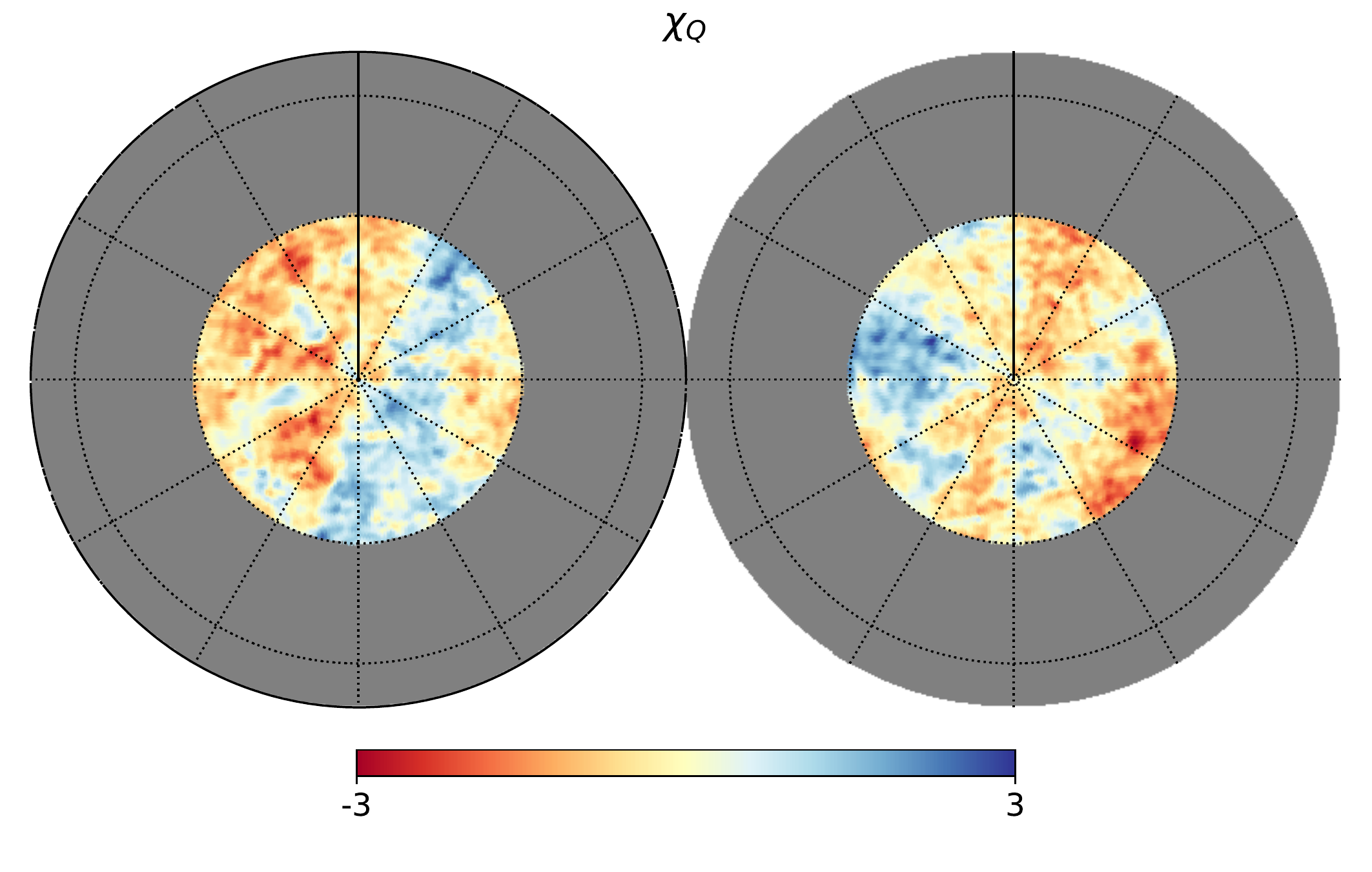}
        &       \includegraphics[trim={0cm .5cm 0cm .7cm},clip,width=.34\linewidth]{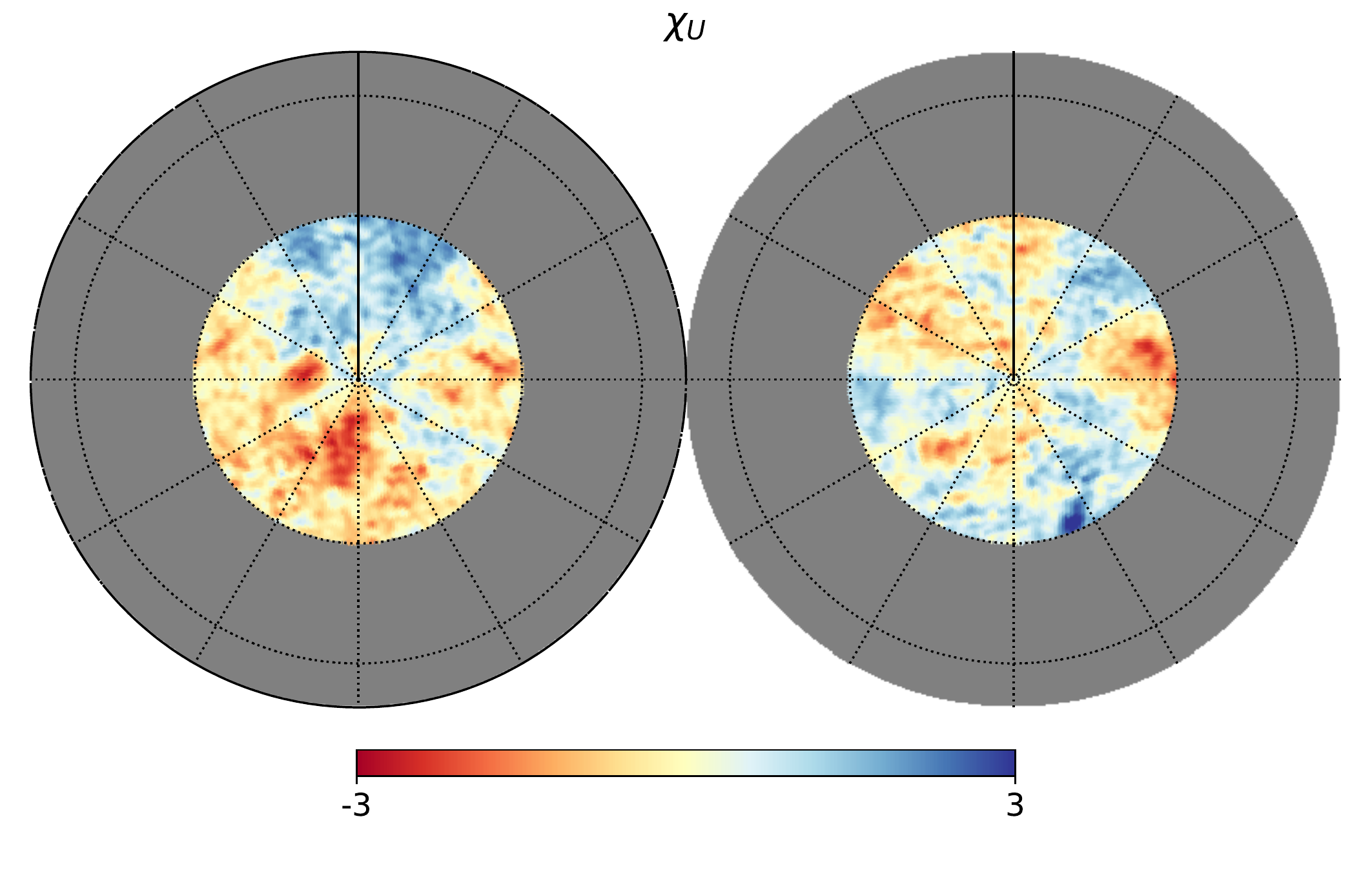}
\\
\end{tabular}
\caption{Orthographic views of the uncertainty (top row) and significance (rows 2 to 6) maps of $Q$ (left) and $U$ (right) at 353~GHz.
The uncertainty maps are in MJy/sr. The significance maps show the significance of the residuals, defined as $(m - d)/\sigma$ per pixel, for the best-fit maps presented in Fig.~\ref{fig:LBShape_QUmaps}.
Conventions are the same as in Fig.~\ref{fig:LBShape_QUmaps}.}
\label{fig:LBShape_QUmaps_res}
\end{center}
\end{figure*}

In Fig.~\ref{fig:LBShape_QUmaps_res} we show the maps of the significance of the residuals corresponding to the best-fit maps obtained in Sect.~\ref{subsec:GMF_Result} for all investigated values of $l_{\rm{max}}$. The significance of the residuals are computed as $(m_i - d_i)/\sigma_i$ for each pixel $i$ where $d_i$ and $m_i$ are either the Stokes $Q$ or $U$ from the observation or from the model, respectively, and $\sigma_i$ is the corresponding uncertainty.

\end{appendix}

\end{document}